\documentclass[nofootinbib,preprint,noshowpacs,noshowkeys,floatfix,
aps]{revtex4-1}

\usepackage{graphicx}
\usepackage[caption=false]{subfig}
\usepackage{dcolumn} 
\usepackage{multirow} 
 \usepackage{tabularx}
\usepackage{rotating} 
\usepackage{array}
\usepackage{amsmath}
\usepackage{esint}
\usepackage{amsfonts} 
\usepackage{dsfont}  
\usepackage{amsthm} 
	\theoremstyle{plain}
	
	\theoremstyle{definition}
\usepackage{mathrsfs} 
\usepackage{tabularx,multirow,float}
\usepackage{verbatim}
\usepackage{xcolor}
\usepackage{url}

\usepackage{xpatch} 
\makeatletter
	\xpatchcmd{\@ssect@ltx}{\@xsect}{\edef\@currentlabelname{#8}\@xsect}{}{}
	\xpatchcmd{\@sect@ltx}{\@xsect}{\edef\@currentlabelname{#8}\@xsect}{}{}
\makeatother
\usepackage[
	pdfstartview=FitBV,
	bookmarks=false,
	colorlinks =true,
 	linkbordercolor=blue,
 	linkcolor = blue,
	citecolor = blue,
	urlcolor=blue 
]{hyperref} 

\newcommand{\ownref}[2]{\hyperref[#2]{#1~\ref*{#2}}}
\newcommand{\ownrefSimple}[2]{\hyperref[#2]{#1}}
\DeclareMathOperator{\csch}{csch}

\begin{document}

\title{Symmetric Tops Subject to Combined Electric Fields:\\
Conditional Quasi-Solvability \\ via the Quantum Hamilton-Jacobi
 Theory}

\author{Konrad Schatz}
\author{Bretislav Friedrich}
\email{bretislav.friedrich@fhi-berlin.mpg.de}
\affiliation{
Fritz-Haber-Institut der Max-Planck-Gesellschaft \\ Faradayweg 4-6, D-14195 Berlin, Germany}

\author{Simon Becker}
\author{Burkhard Schmidt}
\email{burkhard.schmidt@fu-berlin.de}
\affiliation{
Institute for Mathematics, Freie Universit\"{a}t Berlin \\ Arnimallee 6, D-14195 Berlin, Germany}

\date{\today}

\begin{abstract} We make use of the Quantum Hamilton-Jacobi (QHJ) theory to investigate conditional quasi-solvability of the quantum symmetric top subject to combined electric fields (symmetric top pendulum). We derive the conditions of quasi-solvability of the time-independent Schr\"odinger equation as well as the corresponding finite sets of exact analytic solutions. We do so for this prototypical trigonometric system as well as for its anti-isospectral hyperbolic counterpart.  An examination of the algebraic and numerical spectra of these two systems reveals mutually closely related patterns.  The QHJ approach allows to retrieve the closed-form solutions for the spherical and planar pendula and the Razavy system that had been obtained in our earlier work via Supersymmetric Quantum Mechanics as well as to find a cornucopia of additional exact analytic solutions.
\end{abstract}

\maketitle


\section{Inroduction}

A realization of a quantum symmetric top is a molecule that possesses at least a threefold rotational symmetry axis. The symmetric top rotational states, $|J,K,M\rangle$, are characterized by the angular momentum quantum number, $J$, and the projections, $K$ and $M$, of the angular momentum on the body- and space-fixed axis, respectively. Polar symmetric top molecules, i.e., those with a body fixed electric dipole moment, exhibit a unique behavior in electric fields: in their precessing states, in which $J$, $K$, and $M$ are all nonzero, the body fixed electric dipole does not average out in first order, as a result of which such states are inherently oriented in the field \cite{CPL-1998-Aoiz-Fri-Rab-Ver-Herr-H-DCl,haertelt2008}. Like other polar molecules, symmetric tops are also amenable to pendular orientation, a higher-order effect that involves hybridization over both even and odd $J$'s of the top's pure rotational states. Pendular states of a different kind -- that exhibit alignment rather than orientation -- can be created via the induced-dipole interaction of an external electric or optical field with the anisotropic polarizability of the symmetric top; in this case, the rotational hybrids comprise {\it either} even {\it or} odd $J$'s.  We note that {\it orientation} is like a single-headed arrow pointing {\it in} a certain direction whereas {\it alignment} is like a double-headed arrow directed {\it along} a certain direction. A combination of the permanent and induced-dipole interactions provides versatile means to orient  symmetric top molecules, often with only a small admixture of the parity-breaking orienting permanent dipole interaction to the parity-conserving aligning induced-dipole interaction \cite{friedrich1999, haertelt2008}.


\begin{table}[h]
\caption{\label{tab:applications} Examples of problems and applications in chemistry and physics where the trigonometric symmetric top pendulum and its special cases  make a prominent appearance. Cf. also Ref. \cite{LemKreDoyKais:MP2013}. We note that the hyperbolic symmetric top potential resembles the generalized P\"oschl-Teller potential \cite{SUSYQM_Cooper_Khare_Sukhatme1995}. }

\begin{tabular}{l l }
\hline
Problems \& Applications &Reference\\
\hline \hline
Molecular alignment/orientation 
		& 	\cite{Brooks1976Science,LoeschRem1990,FriHer1991Nature,Ortigoso1999,Seideman1999,Larsen:00a,CCCC2001,Averbukh:01a,Leibscher:03a,Leibscher:04a}  \\
Deflection, focusing, trapping 
		& 	\cite{Toennies1964,StolteBerichte1982,StapCorkumPRA1997,BumSukFriPRA2016,TruppeTarbutt2017NaturePhysics} \\
Reaction stereodynamics 
		&\cite{BernHerschLevine1989,KremsStwalleyFriedrich2009,DulieuOsterwalder2018} \\
Stark spectroscopy 
		& 	\cite{Liptay1974,FriSlenHer1994,Slenczka1999} \\
Molecules in combined fields 
		& 	\cite{JCP1999FriHer,CaiFri2001,PRL2003Sakai,JMO2003-Fri,RosarioPRL2012,schmidt2014b,SchmiFri2015a} \\
Photoelectron angular distributions
		& 	\cite{Bisgaard2009,Holmegaard2010,Hansen2011} \\
Diffraction-from-within
		&	\cite{Landers2001} \\
High-order harmonic generation and orbital imaging
		& 	\cite{Itatani:04a,CorkumHHG,BandraukHHG,Smirnova2009,Woerner} \\
Quantum simulation and computing
		& 	\cite{Bar:12,Man:13,DeMille2002,de6,de7,de8,QiKaisFrieHer2011a,QiKaisFrieHer2011b,Karra2016} \\

\hline
\end{tabular}
\end{table}

At the same time, a symmetric top molecule under the combined permanent and induced dipole interaction represents a realization of a {\it symmetric top quantum pendulum} \cite{haertelt2008}. As such it is a prototypical quantum system that lurks behind numerous applications, some of which are summarized in Table \ref{tab:applications}. Unlike other key quantum prototypes, such as the harmonic oscillator, the quantum pendulum lacks, in general, {\it exact} eigenproperties, and so its energies and wavefunctions have to be obtained by solving the corresponding Schr\"{o}dinger equation numerically \cite{SchmiFri2015a}. However, as exemplified in our previous work on the three-dimensional (3D) quantum pendulum, realized by a polar and polarizable $^1\Sigma$ molecule -- a special case of a symmetric top with $K=0$\footnote{We note that linear molecules in states with non-zero electronic angular momentum that fall under Hund's cases (a) and (c) are genuine symmetric tops, with $K\ne0$ \cite{Levefre-Field:2004}.}, there are classes of eigenproblems in quantum mechanics that are neither  exactly solvable nor unsolvable, but lie somewhere in between, i.e., possess a finite number of exact solutions for specific values of the interaction or other defining parameters, cf. Table \ref{tab:solvability}. The analytic, closed-form, and algebraic solutions that we present below and that are defined in Table \ref{tab:solvability} are always exact solutions, i.e., no approximation is involved in obtaining them. 

Herein, we explore quasi-solvability of the symmetric top quantum pendulum within the framework of the Quantum Hamilton-Jacobi (QHJ) theory \cite{leacock1983,geojo2003} and find the conditions for its quasi-solvability as well as the closed-form solutions themselves. In the process, we retrieve the conditions of quasi-solvability and the closed-form solutions derived previously for the planar (2D) and  spherical (3D) quantum pendulum within the framework of Supersymmetric Quantum Mechanics (SUSY QM) \cite{LemMusKaisFriPRA2011, LemMusKaisFriNJP2011, schmidt2014b, SchmiFri2015a, b_friedrich2017} as well as identify a cornucopia of new closed-form solutions and the conditions under which they obtain.

In the case of SUSY QM, the closed-form solutions follow from a suitable {\it Ansatz} for the superpotential, which requires an ``educated guess'' that is quite hard to make. In contrast, the QHJ theory offers a generic way of constructing such solutions, but calls for an ``educated guess'' concerning the choice of appropriate coordinates in which to express them; this is an easier task than coming up with an Ansatz for the superpotential.  

Moreover, we show that both the exact (in fact, algebraic, cf. Table \ref{tab:solvability}) and numerical spectra of the symmetric top pendulum exhibit patterns that are intrinsically related to those of the eigenproblem obtained by the anti-isospectral transformation
 \cite{Krajewska1997a, b_friedrich2017} that converts the {\it trigonometric} symmetric top to a {\it hyperbolic} one.
 
\begin{table}[h]
\footnotesize
\caption{\label{tab:solvability} Classification of quantum systems according to their solvability \cite{sato2002}. If the solutions within a solvability class for a given system only obtain under certain conditions imposed on the system's parameters, the system is termed {\it conditionally} QS, QES, or QPS \cite{dutra1993,lopez1993}. Furthermore, we use the following terminology: an {\it analytic solution} is one in terms of elementary functions (the set of elementary functions is not closed under limits and infinite sums) and of (some) special functions (except for those that are infinite sums); a {\it closed-form solution/expression} is an analytic solution obtained via a finite number of operations (a narrower class than analytic solutions that in practice excludes some special functions); an {\it algebraic solution} is a closed-form solution built up from integer constants and algebraic operations (addition, multiplication, exponentiation). A comprehensive overview is given in Ref. \cite{wiki-Closed-form_expression}.}

\begin{tabularx}{1.02\linewidth}{l l l l}
	\hline 
Class &Spectrum&Normalizability&Number of solutions\\
\hline\hline 
Exactly solvable, ES&entire& normalizable& infinite \\
Quasi-solvable, QS& part&normalizable \& non-normalizable& finite\\
Quasi-exactly solvable, QES&part& normalizable& finite\\
Quasi-perturbatively solvable, QPS&part& non-normalizable& finite\\
\hline
\end{tabularx}
\end{table}

The present paper is structured as follows: In Section \ref{sec:hams-intro}, we lay out the Hamiltonians of the trigonometric as well as hyperbolic symmetric tops and the corresponding Schr\"{o}dinger equations whose quasi-solvability we investigate. In Section \ref{sec:polar_constr-alg-sols}, we derive for either top the QHJ
equation which is solved by the quantum momentum function (QMF). The quantum momentum function is closely related to the 
SUSY QM superpotential which is, in turn, linked to the ground-state wavefuntion. We then derive the conditions of quasi-solvability by analyzing the singularity structure of the tops' potential and subsequently construct the excited-state solutions algebraically. The physical relevance of the closed-form solutions found is evaluated by making use of the limit-point and limit-circle classifications. In Section \ref{sec:examples}, we provide a sampling of the closed-form solutions obtained from the QHJ theory for the trigonometric and hyperbolic symmetric top as well as derive their special cases, the planar and spherical pendula and the Razavy system.


\section{Symmetric top Hamiltonian} \label{sec:hams-intro}

The Hamiltonian of a symmetric top molecule subject to collinear
electric fields with interaction strengths $\eta$ and $\zeta$ is given by
\begin{align}\label{eq:ham_org-phys_symm-top}
	\mathcal{H}_t = B \textbf{J}^2 -  B \rho J_3^2 +  \mathcal{V}_t(\theta) \, 
\end{align}
where 
\begin{align} \label{eq:ext-pot}
	\mathcal{V}_t(\theta)
			={}&
			 - \eta \cos{\theta} 
			 - \zeta 	
			\cos^2{\theta} \, 
\end{align}
is the potential. The trigonometric character of the Hamiltonian and its potential is emphasized by the subscript $t$.
The interaction strengths $\eta$ and $\zeta$ arise, respectively, from the fields $\varepsilon_1$ and $\varepsilon_2$ such that
\begin{align}
	\eta = \mu \varepsilon_1  \quad  \text{and} \quad
	\zeta = \zeta_{\|}-\zeta_{\bot} \quad \text{with} \quad
	\zeta_{\|,\bot} = \frac{\alpha_{\|,\bot} \varepsilon_2^2}{2}  \, 
\end{align}
where $\mu$ and $\alpha_{\|,\bot}$ denote the body-fixed electric dipole moment and 
principal polarizability components parallel and perpendicular to the body-fixed
axis $3$ (the figure axis) \cite{haertelt2008} and the parameter $\rho$
determines whether the  inertia tensor of the symmetric top is prolate or oblate: 
\begin{align}
	\rho = \begin{cases}
               A/B - 1 > 0 \quad \text{prolate}\\
                C/B - 1 < 0 \quad \text{oblate}
            \end{cases}
\end{align}
Here $A = \frac{\hbar^2}{2 I_A}$, $B = \frac{\hbar^2}{2 I_B}$ and $C =
\frac{\hbar^2}{2 I_C}$ are the rotational constants defined via the principal moments of inertia $I_A$, $I_B$,
$I_C$.

In terms of the Euler angles $(\varphi,\theta,\chi)$, the body-fixed components $(1,2,3)$ of the angular momentum operator, $\textbf{J}$, are given by
\begin{align}
	J_1 
	={}& 
		i \left(- \sin \chi \, \partial_\theta 
		+
		\frac{\cos \chi}{\sin \theta} \, \partial_\varphi
		-
		\cot \theta \cos \chi\, \partial_\chi\right) \,\\
	J_2
	={}&
		i \left(-\cos \chi \, \partial_\theta 
		- 
		\frac{\sin \chi}{\sin \theta}\, \partial_\varphi
		+
		\cot \theta \sin \chi\, \partial_\chi\right)\,\\
	J_3
	={}&
		-i \, \partial_\chi \, 
\end{align}
and so the square of the angular momentum operator becomes
\begin{align}
	\textbf{J}^2 
	=
			-\partial_\theta^2 
			-
			\cot \theta\, \partial_\theta
			-
			\frac{1}{\sin^2 \theta} \left(\partial_\varphi^2 + \partial_\chi^2 \right) 
			+ 
			2 \cot \theta \csc \theta\,
			\partial_\varphi\, \partial_\chi \, 
\end{align} 
Since  potential (\ref{eq:ext-pot}) only depends on the polar angle $\theta$,  we can separate variables and write the solution, $\psi_{3D,t}(\theta,\varphi,\chi)$, to the Schr\"odinger equation
\begin{align} \label{eq:ham_org-phys_symm-top_3-dim}
	\mathcal{H}_t\,  \psi_{3D,t}(\theta,\varphi,\chi) = E_t\,
	\psi_{3D,t}(\theta,\varphi,\chi)
\end{align}
as $\psi_{3D,t}(\theta,\varphi,\chi) = \hat{\psi}_t (\theta)\,
e^{-i M \varphi} e^{-i K \chi}$, with $M$ and $K$ the (constant) integer projections of $\textbf{J}$ on the space- and body-fixed axis, respectively. Substitution of the wavefunction $\psi_{3D,t}(\theta,\varphi,\chi)$ into Eq. 
\eqref{eq:ham_org-phys_symm-top_3-dim} then leads to the Schr\"odinger equation 
\begin{align} \notag
	\hat{\mathcal{H}}_t \hat{\psi}_t(\theta) 
	={}&
		B \left(
			-\partial_\theta^2 
			-
			\cot \theta\, \partial_\theta
			+
			\left(M^2 + K^2  \right)\csc^2 \theta 
			- 
			2 M K \csc \theta \, \cot \theta 
			-
			\rho K^2
			\right) \hat{\psi}_t(\theta)\\\notag
			&+
			\left(
			 -
			  \eta \cos{\theta} 
			 - \zeta 	
			\cos^2{\theta}
			\right) \hat{\psi}_t(\theta) 
		\\\label{eq:ham_org-phys_symm-top_3-dim_separ}
	={}&
		E_t\, \hat{\psi}_t(\theta) \,  
\end{align} 
for $\hat{\psi}_t (\theta)$ which, when gauged\footnote{For the origin of this gauge transformation
 see, e.g., \cite{vilenkin1995,sezgin1998,haertelt2008}.} as
\begin{align} \label{eq:wavefunc-3d-gauge}
 \hat{\psi}_t (\theta) = \psi_t (\theta)/\sqrt{\sin \theta},
\end{align} 
allows to recast our initially 3-dimensional eigenproblem as a 1-dimensional one
\begin{align} \label{eq:ham_org-phys_symm-top_1-dim}
		H_t \psi_t(\theta)
	=
		- B\, \partial_\theta^2 \psi_t(\theta)
		+
		V_{t}(\theta)\psi_t(\theta)
	=
		E_t\, \psi_t(\theta)
\end{align}
for an effective potential
\begin{align} \notag
			V_{t}(\theta)
			={}&
			B\left[
				\left(
					M^2 +K^2 -\frac{1}{4}				
				\right) \csc^2{\theta}
				- 2  M K \cot{\theta}\csc{\theta}
				- \rho K^2 
				- \frac{1}{4}
			\right]
			 \\\label{eq:org-effective-pot_1-dim}
			  &- \eta \cos{\theta} 
			 - \zeta \cos^2{\theta} \,  
\end{align}

We note that for $K=0$, Eq. \eqref {eq:org-effective-pot_1-dim} yields the  spherical pendulum Hamiltonian
\cite{LemMusKaisFriPRA2011, LemMusKaisFriNJP2011,SchmiFri2014a, SchmiFri2015a}
and for $(K,M)=(0,1/2)$, the planar pendulum Hamiltonian
\cite{schmidt2014b,b_friedrich2017}.  

As in our previous work \cite{b_friedrich2017}, we will also consider the ``hyperbolic counterpart'' of the above trigonometric symmetric top, obtained by the  coordinate transformation $\theta \mapsto i
\theta$ and gauging 
\begin{align}  
\label{eq:wavefunc-3d-gauge_hyperb}
 \psi_h (\theta) = \hat{\psi}_h (\theta) \sqrt{\sinh \theta}
\end{align}
where the subscript $h$ stands for hyperbolic.
This leads to the Schr\"odinger equation 
\begin{align} \label{eq:ham_org-phys_symm-top_1-dim_hyperb}
		H_h \psi_h(\theta) 
	=
		- B\, \partial_\theta^2 
		+
		V_{h}(\theta) \psi_h(\theta)
	=
		- E_t\, \psi_h(\theta) \,  
\end{align}
with the effective potential
\begin{align} \notag
			V_{h}(\theta)
			={}&
			B \left[
				\left(
					M^2 +K^2 -\frac{1}{4}
				\right) \csch^2{\theta}
				- 2 M K \csch{\theta}\, \coth{\theta}
				+ \rho K^2 
				+ \frac{1}{4}
			\right]
			 \\\label{eq:org-effective-pot_hyperb_1-dim} 
			  &+  \eta \cosh{\theta} 
			 + \zeta \cosh^2{\theta} \,  
\end{align}
We note that the transformation $\theta \mapsto i\theta$ is anti-isospectral
\cite{Krajewska1997a,b_friedrich2017}, as $E_t \mapsto E_h := - E_t$.

In what follows, we will refer to the tops described by Schr\"odinger equations
\eqref{eq:ham_org-phys_symm-top_1-dim} and
\eqref{eq:ham_org-phys_symm-top_1-dim_hyperb} as the {\it trigonometric} and {\it hyperbolic top}, respectively.


\section{Conditional quasi-solvability}
\label{sec:polar_constr-alg-sols}

In this section, we apply the Quantum Hamilton-Jacobi  theory
\cite{gangopadhyaya2011,ranjani2004} to the  trigonometric and hyperbolic top eigenproblems for Hamiltonians
\eqref{eq:ham_org-phys_symm-top_1-dim} and 
\eqref{eq:ham_org-phys_symm-top_1-dim_hyperb}, respectively,
and construct the closed-form solutions. 

Since the derivations for the two types of top are analogous to one another, we show the derivation for the trigonometric top only and subsequently provide a summary of the results for the hyperbolic one.


\subsection{Quantum Hamilton-Jacobi Equation}\label{subsec:singularity_structure}  

The Schr\"odinger equation \eqref{eq:ham_org-phys_symm-top_1-dim} can be recast as a Ricatti equation 
\begin{align} \label{eq:qhj-eq}
	p(\theta)^2 - i  \sqrt{B}p'(\theta) = E -  V(\theta) \, 	
\end{align} 
for the quantum momentum function 
\begin{align} \label{eq:qmf-wavefunc-phys-coord}
	p(\theta) = - i	\sqrt{B}\frac{\psi'(\theta)}{\psi(\theta)}
\end{align}
with $ \psi'(\theta)\equiv \partial_\theta \psi(\theta)$, where we dropped the subscripts $t$ or $h$ on $p$, $E$, $V$, and $\psi$ for simplicity. Crucially, the wavefunction $\psi(\theta)$ is assumed to be meromorphic in QHJ theory, i.e., containing at most isolated singularities (poles). Eq. \eqref{eq:qhj-eq} is termed the quantum Hamilton-Jacobi equation; in the limit $\hbar \rightarrow 0$, i.e., $B \rightarrow 0$, it turns into the classical Hamilton-Jacobi equation. We note that the quantum momentum function is related to the SUSY QM superpotential $W(\theta)$ via $p(\theta) = i W(\theta)$ \cite{gangopadhyaya2011}.

In an appropriate new coordinate $z=z(\theta)$, the quantum Hamilton-Jacobi equation \eqref{eq:qhj-eq} can be transformed into a purely rational form, 
\begin{align} \label{eq:qhj-eq_transf-compact}
	\tilde{p}(z)^2 + \sqrt{B}\, \tilde{p}'(z) 
	+
	{\theta'(z)}^2 \left[
		E - \tilde{V}(z)
	\right]
	=
	0 \, 
\end{align}
with a $p(\theta) \mapsto \tilde{p}(z)$ mapping given by Eq. \eqref{eq:qhj-eq_transf-compact2} (for details see Appendix \ref{subsec:rational}). This can be considered a \textit{normal form} \cite{bank1981} of the Riccati equation \eqref{eq:qhj-eq}.
This will prove key for finding its solutions  algebraically via Laurent series expansion of the quantum momentum function in Subsection \ref{subsec:fixed-qmf-constr}. For the explicit choice of $z=(\cos \theta+1)/2$ as the new variable,\footnote{We note that a more intuitive choice of the new variable, $z=\cos{\theta}$, would have no effect on the solution spaces, as it is just a M\"obius transform of $z = (\cos{\theta}+1)/2$, see Appendix \ref{subsec:rational}. However, it would result in four diagonal elements, whereas our above choice has only three, which is of computational advantage (see Subsection 
\ref{subsec:step_matrix-els}).} the corresponding purely rational potential of the trigonometric top becomes:
\begin{align} \notag
		\tilde{V_t}(z) 
	={}&
		-\eta  (2 z-1)- \zeta(2 z-1)^2
		\\\label{eq:qhj-pot_transformed}
		&+
		B \left(
			\frac{
				-4 K^2 \rho z^2
				+4 K^2 \rho z
				-K^2
				+4 K M z
				-2 K M 
				- M^2
				+\frac{1}{4}
			}{
				4 (z-1) z
			} 
			-\frac{1}{4}
		\right)\, 
\end{align}
Note that its singularity structure differs from that of the original potential \eqref{eq:org-effective-pot_1-dim}: whereas the original potential $V_t(\theta)$ has double poles at $\theta_m = m \pi$ 
on $\mathbb{R}$ with $m$ integer,
 the transformed potential $\tilde{V}_t(z)$ possesses simple poles at the points $z_{1,2} \in \{0,1\}$ in 
 the physical domain $[0,1]$ and its extension, the complex plane
 $\mathbb{C}$. In addition, it has a double pole at $z_0 = \infty$. The singularity structure is of consequence for determining the quasi-solvability conditions, see Subsection \ref{subsec:qs-conds}.
 
 
\subsection{Construction of the quantum momentum function}\label{subsec:fixed-qmf-constr}

Given that Hamiltonians \eqref{eq:ham_org-phys_symm-top_1-dim} and
\eqref{eq:ham_org-phys_symm-top_1-dim_hyperb} 
belong to conditionally quasi-solvable Hamiltonians in the coordinates
$z\propto \cos \theta$ and $z \propto \cosh \theta$, respectively \cite{lopez1994}, they can
be gauge-transformed into $sl(2)$-algebraizable Sturm-Liouville Operators $T$ which preserve 
the finite-dimensional monomial subspaces\footnote{The monomial subspaces form
the flag $\mathcal{P}_{0} \subset \mathcal{P}_{1} \subset ... \subset ...\subset \mathcal{P}_{n}$. If the Hamiltonian
preserved a complete flag $\mathcal{P}_{0} \subset \mathcal{P}_{1} \subset ...  \subset ... \subset
\mathcal{P}_{n} \subset...$, it would be said to be exactly solvable.} \cite{ullate2013} 
\begin{align}\label{eq:polyn-span}
	 \mathcal{P}_{n} = \mathrm{span}\{1,z,z^2,...,z^{n}\} \, 
\end{align}
As a result, the  wavefunction 
\begin{align} \label{eq:newpsi-interms-of-p}
	\tilde{\psi}(z)= e^{ \frac{1}{\sqrt{B}} \int^z
	\tilde{p}(y) \mathrm{d}y}
\end{align}
that corresponds to the transformed problem, Eq. \eqref{eq:qhj-eq_transf-compact}, and is related to the original wavefunction via 
\begin{align} \label{eq:alg-sols_first-order_transf-new-old}
	\psi(\theta) = 
		\left. \tilde{\psi}(z) \sqrt{ \theta'(z) }
				\right|_{z=z(\theta)}		\, 	
\end{align}
cf. Eq. \eqref{eq:new-schrod-type-eq_general} of the Appendix, 
can be factorized into a product of a {\it seed} function $\phi_f(z)$ and a polynomial $\phi_{m,i}(z)$
\begin{align} \label{eq:new-wavefunc_factoriz} 
	\tilde{\psi}_{i}(z) =\phi_f(z)  \phi_{m,i}(z) \quad \, \text{with} \quad 
	\phi_{m,i}(z) \in \mathcal{P}_{n} \, 
\end{align}
pertaining to eigenenergy $E_i$. For $n=0$, $\tilde{\psi}_{0}(z) \propto \phi_f(z)$, whence we see that the seed function $\phi_f(z)$
 determines the \textit{algebraic sector} of the solutions, i.e., the class of wavefunctions which  
 only differ from one another by the polynomial $\phi_{m,i}(z)$ in Eq.
 \eqref{eq:new-wavefunc_factoriz}. Hence the seed function  can serve as a gauge factor
 which allows the transformation of Hamiltonians \eqref{eq:ham_org-phys_symm-top_1-dim} and
\eqref{eq:ham_org-phys_symm-top_1-dim_hyperb}  to the Sturm-Liouville self-adjoint operators \cite{lopez1993,lopez1994} 
\begin{align} \label{eq:ham_gauged2}
	T_{t,h}(z) = \left.\frac{1}{\phi_f(z)} H_{t,h}(\theta) \phi_f(z)
	\right|_{\theta=\theta(z)}	\, 
\end{align} 
with the subscripts $t$ and $h$ referring to the
\textit{trigonometric} and \textit{hyperbolic top}, respectively\footnote{That the seed function is a meromorphic and algebraic function of $z$ forces $T_{t,h}$ of Eq. \eqref{eq:ham_gauged2} to preserve the span \eqref{eq:polyn-span} and to be self-adjoint}.

By combining Eqs.  \eqref{eq:newpsi-interms-of-p} and \eqref{eq:new-wavefunc_factoriz}, we can  decompose the quantum momentum function as
\begin{align} \label{eq:qmf_decomp2}
	\tilde{p}(z) = \tilde{p}_f(z) + \tilde{p}_m(z)
\end{align} 
with 
\begin{align}\notag
\tilde{p}_f(z) &= \sqrt{B}\,  \partial_z
\phi_f(z)/\phi_f(z)
\end{align}
and
\begin{align}\label{eq:pfpm}
\tilde{p}_m(z) &= \sqrt{B}\,  
\partial_z \phi_{m,i}(z)/\phi_{m,i}(z)
\end{align}
Clearly, if $\tilde{p}_f(z)$ is known, the seed
function can be determined via $\phi_f(z) = e^{\frac{1}{\sqrt{B}} \int^z
\tilde{p}_f(y) \mathrm{d}y}$.

Our notation above, starting with Eq. \eqref{eq:new-wavefunc_factoriz}, anticipates the fact that the quantum momentum function $\tilde{p}(z)$ can have two types of poles: {\it fixed} poles (subscript $f$) that are due to the poles of $\tilde{V}(z)$ and $\theta'(z)^2$ of Eq. \eqref{eq:qhj-eq_transf-compact}, and {\it moving} poles (subscript $m$) that are due to the nodes of the wavefunction. Thus the notation  accounts for the fact that  $\tilde{p}_f(z)$, the part of the quantum momentum function that pertains to the seed function, has fixed poles, and that $\tilde{p}_m(z)$,  the part of the quantum momentum function that pertains to polynomials in $z$, has moving poles. As noted above, the fixed poles occur at $z_j \in
\{\infty,0,1\}$.

We will now determine the closed-form expressions for
the  quantum momentum function term $\tilde{p}_f(z)$ that pertains to the fixed poles.
 
By setting $\phi_{m,i} = const.$ in Eq. \eqref{eq:pfpm}, which corresponds to the lowest-order solution $\tilde{\psi}_0$ with $n=0$, Eq.
\eqref{eq:pfpm} yields $\tilde{p}_{m,i}=0$ and Eq. \eqref{eq:qmf_decomp2} $\tilde{p}=\tilde{p}_f$. As a result, the quantum Hamilton-Jacobi equation \eqref{eq:qhj-eq_transf-compact} reduces to
\begin{align} \label{eq:qhj-eq_transf3b}
	\tilde{p}_f(z)^2 + \sqrt{B}\, \tilde{p}_f\,'(z) 
	+
	{\theta'(z)}^2 \left[
		E - \tilde{V}(\theta(z))
	\right]
	=
	0 \, 
\end{align}
Provided $\tilde{V}(z)$ is a rational function of $z$, cf. Eq. \eqref{eq:qhj-pot_transformed}, $\tilde{p}_f(z)$ 
 must be rational as well and can therefore be decomposed into partial fractions
 $\tilde{p}_{f,j}$, each of which evaluated at one of the poles $z_j \in \{\infty,0,1\}$,
\begin{align} \label{eq:qmf-fixed_poles-decomp}
	\tilde{p}_f(z) = \sum_{j=0}^{\#\{z_j\}-1} \tilde{p}_{f,j}(z)  \, 
\end{align}
with $\#\{z_j\}$ the number of
\textit{fixed poles} $z_j$.

For $j=1,2$, i.e., at $z_{1,2} \in \{0,1\}$, each of the components $\tilde{p}_{f,j}$
can be expanded in terms of a Laurent series
\begin{align} \label{eq:qmf-fixed_poles-decomp2}
	\tilde{p}_{f,j}(z) = \sum_{k=-d}^{\infty} c_{j k} (z - z_j)^k 
	\end{align}
with $d \in \mathbb{N}_{\geq 0}$ a
sufficiently large finite boundary.
  
In order to find the coefficients $c_{j k}$ of
Eq. \eqref{eq:qmf-fixed_poles-decomp2}, we plug each $\tilde{p}_{f,j}$ into Eq.  \eqref{eq:qhj-eq_transf3b} separately, 
\begin{align}\notag
	{}& \tilde{p}_{f,j}(z)^2 + \sqrt{B} \,\tilde{p}_{f,j}\,'(z)
	+\frac{E}{z-z^2}+\frac{\eta  (1-2 z)}{(z-1) z}
	-  \frac{\zeta (1-2 z)^2  }{(z-1) z}\\ \label{eq:qhj-eq_transf_at-poles-spec}
	+&
	B \left(
		\frac{K^2 (-4 r (z-1) z-1)+2 K M
		(2 z-1)-M^2+1}{4 (z-1)^2 z^2}
	\right)
	=0 \, 
\end{align}

At $z_0 = \infty$, however, the evaluation of $\tilde{p}_{f,0}$ from \eqref{eq:qmf-fixed_poles-decomp} requires 
another extension of the domain: as illustrated in Fig. \ref{fig:qhj_compactif2},
we have to add a point at infinity by introducing a copy of the complex plane
$\mathbb{C}$ with coordinates $w=1/z$
, which we denote as $\mathcal{D}_c(w)$. This compactifies
$\mathbb{C}$, so that $\mathcal{D}_c(1/w)
\cup \mathcal{D}_c(w)$ becomes a cover of the
Riemann sphere $\hat{\mathbb{C}} \cong \mathbb{C} \cup \{\infty\} $ \cite{frankel2011}. 
The inversion $w = 1/z$ allows for a reciprocal mapping between the elements of
$\mathcal{D}_c(z)$ and $\mathcal{D}_c(w)$. By introducing
$\hat{p}_{f}(w(z)) = \tilde{p}_{f}(z) $, we
 obtain the Laurent series expansion of the component $\tilde{p}_{f,0}(z)$ in Eq.\eqref{eq:qmf-fixed_poles-decomp}: 
\begin{align} \label{eq:qmf-fixed_poles-decomp3}
\tilde{p}_{f,0}(z) = \hat{p}_{f,0}(w(z)) = \sum_{k=-d}^{\infty} c_{0 k}
w(z)^k
\end{align}
Similarily to the case of the coefficients $c_{1k}$ and $c_{2k}$ above, the coefficients $c_{0 k}$ can be found by inserting
$\hat{p}_{f,0}(w)$ into the Riccati equation
 \eqref{eq:qhj-eq_transf3b} on $\mathcal{D}_c(w)\footnote{Here, we
 	bypass the \textit{normal form} of the Riccati equation, since $z \mapsto 1/z$
 	is a M\"obius transformation and hence does not yield any non-rational terms.}$
\begin{align}\notag
	{}&\hat{p}_{f,0}(w)^2-\sqrt{B}\, w^2\, \hat{p}_{f,0}\,'(w)
   -\frac{\eta\,  w (w-2)}{w-1}
   +\frac{ \zeta (w-2)^2 }{w-1}
   +\frac{E\, w^2}{w-1}\\
	\label{eq:qhj-eq_transf_at-pole-infty-spec} 
-&\frac{B w^2 \left(K^2 \left(w^2-4 \rho 
   (w-1)\right)+2 K M (w-2) w+\left(M^2-1\right) w^2\right)}{4 (w-1)^2}  = 0\, 
\end{align}
with $\lim_{z \to \infty}
 \tilde{p}_{f,0}(z) = \lim_{w \to 0} \hat{p}_{f,0}(w) = const$. 
 From the domain compactification we find that $\tilde{p}_{f,0}(z)$ on $\mathcal{D}_c(z)$ is bounded by $\hat{p}_{f,0}(w)$ at $w=0$
 on $\mathcal{D}_c(w)$.
 From Liouville's
 theorem of complex analysis \cite{ahlfors1966}, it then follows that
 $\tilde{p}_{f,0}(z) = const$. As a result, Eq. \eqref{eq:qmf-fixed_poles-decomp} 
 boils down to
\begin{align}\label{eq:qmf-fixed-term_reform}
	\tilde{p}_f(z) = \hat{p}_{f,0}(0)  +
	\sum_{j=1}^{\#\{z_j\}-1} \tilde{p}_{f,j}(z) \, 
\end{align}
Finally, by inserting the series expansion \eqref{eq:qmf-fixed_poles-decomp2} of
$\tilde{p}_{f,1}$ and $\tilde{p}_{f,2}$ at the poles $z_1=0$ and $z_2=1$
into Eq. \eqref{eq:qhj-eq_transf_at-poles-spec}  and
likewise the series expansion \eqref{eq:qmf-fixed_poles-decomp3} of $\hat{p}_{f,0}(w(z))$ at the pole at $z_0 = \infty$ into Eq.
\eqref{eq:qhj-eq_transf_at-pole-infty-spec} completely determines $\tilde{p}_{f}$ in
\eqref{eq:qmf-fixed-term_reform} as the exact and closed-form solution of the quantum Hamilton-Jacobi equation
\eqref{eq:qhj-eq_transf3b}.


Due to the squared $\tilde{p}_{f,j}$ term in
\eqref{eq:qhj-eq_transf_at-poles-spec} and
\eqref{eq:qhj-eq_transf_at-pole-infty-spec}, 
a branching and hence at least $2$ possible solutions 
$\tilde{p}_{f,j}$ can be expected for each $z_j$ resulting in $2^{\#\{z_j\}}$ combinations and as many
solutions $\tilde{p}_{f}$.
And indeed, for the coordinate choice $z=(\cos \theta +1)/2$ we find exactly $2^3$
algebraic sectors (see  Table \ref{table:qmfs}).
 \begin{table}[h]
\caption{\label{table:qmfs} 
	Quantum momentum function terms $\tilde{p}_{f}$ and $\hat{p}_{f,0}$ for each algebraic sector.
	}\vspace{1mm}
\begin{tabularx}{.75\linewidth}{l l l}
	\hline \noalign{\smallskip}
			Algebraic sector
	 	&	$\tilde{p}_f(z)$
	 	&	$\hat{p}_{f,0}(w)$
	\\ \hline\hline \noalign{\smallskip}
			$1_\pm$
		&	$\sqrt{B}\left(\frac{-K+M+1}{2 (z-1)}+\frac{K+M+1}{2 z}\right) \pm 2
		\sqrt{\zeta}$ 
		&	$\mp \frac{\eta\, w }{2  \sqrt{\zeta } } \pm 2
		\sqrt{\zeta}$ 
	\\\noalign{\smallskip}
		$2_\pm$
		&	$\sqrt{B}\left(\frac{K-M+1}{2 (z-1)}+\frac{K+M+1}{2 z}\right) \pm 2
		\sqrt{\zeta }$ 
		&	$\mp \frac{\eta\, w }{2 \sqrt{\zeta } } \pm 2
		\sqrt{\zeta }$ 
	 \\\noalign{\smallskip}
		$3_\pm$
		&	$\sqrt{B}\left(\frac{-K+M+1}{2 (z-1)} + \frac{-K-M+1}{2 z} \right) \pm 2
		\sqrt{\zeta }$ 
		&	$\mp \frac{\eta\, w }{2 \sqrt{\zeta }} \pm 2
		\sqrt{\zeta }$ 
	 \\\noalign{\smallskip}
		$4_\pm$
		&	$\sqrt{B}\left( \frac{K-M+1}{2 (z-1)} + \frac{-K-M+1}{2 z}\right) \pm 2
		\sqrt{\zeta }$ 
		&	$\mp \frac{\eta\, w }{2 \sqrt{\zeta }} \pm 2
		\sqrt{\zeta }$ 
	\\\noalign{\smallskip}\hline
\end{tabularx}
\end{table}
  
  
\subsection{Conditions of quasi-solvability}
\label{subsec:qs-conds}

We will now derive the conditions for quasi-solvability (cf. Ref. \cite{geojo2003}) and  with their help the polynomial
multipliers $\phi_{m,i}(z)$ of the seed functions in the factorized algebraic
wavefunctions $\tilde{\psi}_{i}(z)$ via Eq. \eqref{eq:new-wavefunc_factoriz}.

\begin{figure}
	\includegraphics[width=.6\textwidth]{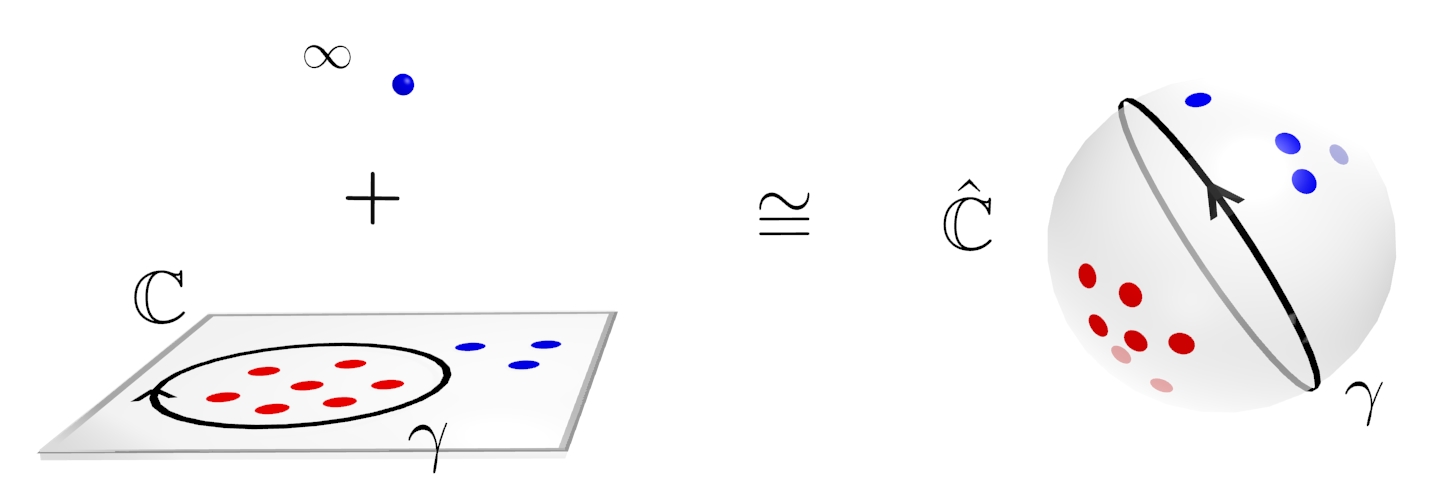}
	\caption{(a) Extension of the domain $\mathcal{D}_p(z)$ of the  potential $\tilde{V}_t(z)$, Eq. \eqref{eq:qhj-pot_transformed}, to the complex plane $\mathbb{C}$; fixed poles
	are depicted by the blue region and the $n$ moving poles by the red region.
	(b) Compactification of the complex plane to the Riemann sphere $\hat{\mathbb{C}}$ by adding a point at
	infinity shown in panel (a). 
	A contour integral of $\tilde{p}_m(z)$ along the path $\gamma$ enclosing
	all {\it moving} poles (red) is equal to the contour integral of $\tilde{p}_f(z)$
	along $\gamma$ in the opposite direction enclosing all fixed poles (blue).} 
	\label{fig:qhj_compactif2}
\end{figure}

The complexification of $z$, introduced in Subsection \ref{subsec:fixed-qmf-constr}, allows  to define the contour integral of the part $\tilde{p}_m(z) =
\sqrt{B}\,\phi_{m,i}'(z)/\phi_{m,i}(z)$ of the quantum momentum function $\tilde{p}(z)$ of Eq. \eqref{eq:qmf_decomp2} along the  curve
$\gamma$ enclosing the $n$ moving
poles (cf. the red dots in Fig. \ref{fig:qhj_compactif2}). 
By  Cauchy's argument
principle \cite{ahlfors1966}, this is equivalent to the sum of the corresponding
residues,
\begin{align} \label{eq:qs-cond_arg-principle} 
	\frac{1}{2 \pi i}\oint_\gamma
		\tilde{p}_m
 	\mathrm{d}z 
 	=  
 	\sqrt{B} \sum_{j=0}^{n}\mathrm{Res}\left(\frac{\phi_{m,i}'(z)}{\phi_{i}(z)},z_j \right)
 	=
 	\sqrt{B} \, n \, 
\end{align}
where $\mathrm{Res}(f(z),z_j)$ stands for the residue of a function $f(z)$ at its pole $z_j$. 
Let $z$ be an element of the compactified complex plane $\hat{\mathbb{C}}$.  This allows to redefine
the contour
 integral \eqref{eq:qs-cond_arg-principle} on the Riemann sphere (see
 Fig.
\ref{fig:qhj_compactif2})
and 
to identify it with the contour integral of $\tilde{p}_f(z)$ and
$\hat{p}_{f,0}(1/z)$ along $\gamma$ in the opposite direction (see Theorem 2.2 of Ref.
\cite{boothby1986}) enclosing all \textit{fixed} poles, $z_j \in
\{\infty,0,1\}$ and
represented by the blue region in Fig.
\ref{fig:qhj_compactif2}:  
\begin{align} \label{eq:qs-cond_general}
	-\frac{1}{2 \pi i }\oint_\gamma \left[ \tilde{p}_f(z) + \hat{p}_{f,0}(1/z) \right ]
	\,\mathrm{d}z = -
	\mathrm{Res}(\hat{p}_{f,0}(1/z),0)
	-\sum_{j=1}^{\#\{z_j\}-1}\mathrm{Res}(\tilde{p}_{f,j},z_j) =
	\sqrt{B}\, n \, 
\end{align}
with $\mathrm{Res}(\hat{p}_{f,0}(1/z),0) =
-\mathrm{Res}(\frac{1}{w^2} \hat{p}_{f,0}(w),0)$.


Eq. \eqref{eq:qs-cond_general}, which can be regarded as a ``quantization condition,'' leads ultimately to the conditions of quasi-solvability. All we need to do is to evaluate the residues $\mathrm{Res}(\tilde{p}_f,0)$,
$\mathrm{Res}(\tilde{p}_f,1)$, and $\mathrm{Res}(\hat{p}_{f,0},0)$ from the quantum momentum functions listed in Table \ref{table:qmfs}. The conditions of quasi-solvability then follow via \eqref{eq:qs-cond_general}
and are listed in Table \ref{table:qmfs_residues}. One can see that these fix the admissible ratios of the interaction parameters $\eta$ and $\zeta$, conveniently expressed in terms of the \textit{topological index}, $\kappa = \frac{\eta}{\sqrt{B\,\zeta}}$,  introduced in our earlier work
\cite{schmidt2014b}.
As discussed in Section
 \ref{sec:examples} and shown in Figs. \ref{fig:spectrum_num-alg_sectors_m2k1},
 \ref{fig:spectrum_num-alg_sectors_m1k1},
 \ref{fig:spectrum_num-alg_sectors_m1k0} and
 \ref{fig:spectrum_num-alg_sectors_m0p5k0}, the topological index allows to label the
 loci of the level crossings as well as the closed-form solutions obtained.

We note that the  residues and quasi-solvability conditions for the \textit{hyperbolic top} with $z =
(\cosh\theta+1)/2$ coincide with those for the \textit{trigonometric top} 
with $z = (\cos\theta+1)/2$. 
 \begin{table}[h]
\caption{\label{table:qmfs_residues} 
	The residues of the quantum momentum function and the quasi-solvability (QS) conditions for each algebraic sector 
	for the trigonometric and hyperbolic top with  coordinates $z=(\cos \theta + 1)/2$ and $z=(\cosh \theta + 1)/2$, respectively.
	}\vspace{1mm}
\begin{tabularx}{1.\linewidth}{l l l l l}
	\hline \noalign{\smallskip}
			sector
	 	&	$\mathrm{Res}(\hat{p}_f,\infty)$
	 	&	$\mathrm{Res}(\tilde{p}_f,0)$ 
	 	&	$\mathrm{Res}(\tilde{p}_f,1)$
	 	&	QS conditions
	\\ \noalign{\smallskip}\hline\hline \noalign{\smallskip}
			$1_\pm$
		&	$\mp \frac{\eta }{2 \sqrt{\zeta }}$
		&	$\frac{1}{2}\sqrt{B}\, (K+M+1)$
		&	$\frac{1}{2}\sqrt{B}\, (-K+M+1)$
		&	$\eta = \pm 2 \sqrt{B}\,\left(M +n +1\right)\sqrt{\zeta }$
	\\\noalign{\smallskip}
		$2_\pm$
		&	$\mp \frac{\eta }{2 \sqrt{\zeta }}$
		&	$\frac{1}{2}\sqrt{B}\, (K+M+1)$
		&	$\frac{1}{2}\sqrt{B}\, (K-M+1)$
		&	$\eta = \pm 2 \sqrt{B}\,\left(K +n +1\right)\sqrt{\zeta }$
	 \\\noalign{\smallskip}
		$3_\pm$
		&	$\mp \frac{\eta }{2 \sqrt{\zeta }}$
		&	$\frac{1}{2}\sqrt{B}\, (-K-M+1)$
		&	$\frac{1}{2}\sqrt{B}\, (-K+M+1)$
		&	$\eta = \pm 2 \sqrt{B}\,\left(-K +n +1\right)\sqrt{\zeta }$ 
	 \\\noalign{\smallskip}
		$4_\pm$
		&	$\mp \frac{\eta }{2 \sqrt{\zeta }}$
		&	$\frac{1}{2}\sqrt{B}\, (-K-M+1)$
		&	$\frac{1}{2}\sqrt{B}\, (K-M+1)$
		&	$\eta = \pm 2 \sqrt{B}\,\left(-M +n +1\right)\sqrt{\zeta }$	
	\\\noalign{\smallskip}\hline
\end{tabularx}
\end{table}


\subsection{Matrix elements}\label{subsec:step_matrix-els}  

By substituting the decomposition \eqref{eq:qmf_decomp2} into Eq. \eqref{eq:qhj-eq_transf-compact} 
and  multiplying by $1/\theta'(z)^2$, one obtains the
following eigenproblem
\begin{align} \notag
		{}&(T_{t} +E_i)\phi_{m,i}(z) \\ \notag
	={}&
		B \,\frac{1}{\theta'(z)^2} \, \left[\phi_{m,i}''(z)
		+2\sqrt{B}\, \tilde{p}_f(z)\, \phi_{m,i}'(z)\right]
	+
	\left[
		\frac{1}{\theta'(z)^2}
		\left(
			\tilde{p}_f(z)^2 + \sqrt{B}\,\tilde{p}_f'(z) 
		\right)
		+
		E_i - \tilde{V}_t(\theta(z))
	\right] \phi_{m,i}(z)\\\label{eq:sturm-liouville_general} 
	={}& 0 \, 
\end{align}
with the operator $T_t$ defined by Eq. \eqref{eq:ham_gauged2} and  fulfilling
$T_t = - T_h$.
Substituting for the quantum momentum functions $\tilde{p}_f$ from Table
\ref{table:qmfs_residues} then yields eight self-adjoint
operators $T_{t}$, each corresponding to a particular algebraic sector. The same applies to the hyperbolic analog $T_{h}$.
	
The matrix elements  
\begin{align}
	\left(T_t\right)_{k \ell} = \langle z^{k  } | T_t | z^{\ell  } \rangle\,   
\end{align}
of the Sturm-Liouville operator $T_t$ in the basis set of the monomials in $z$, Eq. \eqref{eq:polyn-span}, can be constructed either directly or via the residues
\begin{align} \label{eq:matrix-els-res}
	\langle z^{k  } | T_t | z^{\ell  } \rangle =
	\mathrm{Res}\left(\frac{1}{z}\times \frac{1}{z^k} T_t (z^\ell),0\right)	\,
\end{align}
which provide the constant part of $(1/z^k) T_t(z^\ell)$. 

Hence the eight Sturm-Liouville operators $T_t$ are each represented by a tri-diagonal matrix  whose elements are listed in 
Tables \ref{table:matrix_els1} and \ref{table:matrix_els2}.
Note that the integer $n$ in Table \ref{table:qmfs_residues} determines
the \textit{cutoff} dimension at which a given matrix can be decomposed 
into an (upper) finite $n \times n$ block and a (lower) infinite-dimensional block
\cite{b_friedrich2017}. Hence, the closed-form solutions can be constructed for $k, \ell \leq n$. Solutions beyond the $n-$dimensional block can only be determined numerically and demand multi-precision computations.

 \begin{table}[h]
\caption{\label{table:matrix_els1} 
	Matrix elements on the super- and subdiagonal for each algebraic
	sector}\vspace{1mm}
\setlength{\tabcolsep}{1.5em}
\begin{tabularx}{.7\linewidth}{l l l}
	\hline \noalign{\smallskip}
			Sector
	 	&	$\langle z^{\ell-1  } | T_t | z^{\ell  } \rangle$
	 	&	$\langle z^{\ell  } | T_t | z^{\ell-1  } \rangle$
	\\ \hline\hline \noalign{\smallskip}
			$1_\pm$
		&	$B \ell (K+M+\ell)$
		&	$2 \eta \mp 4 \sqrt{B}\, \sqrt{\zeta } (M+\ell)$
	\\\noalign{\smallskip}
		$2_\pm$
		&	$B \ell (K+M+\ell)$
		&	$2 \eta \mp 4 \sqrt{B}\,  \sqrt{\zeta } (K+\ell)$
	 \\\noalign{\smallskip}
		$3_\pm$
		&	$B \ell (-K-M+\ell)$ 
		&	$2 \eta \mp 4 \sqrt{B}\,  \sqrt{\zeta } (-K+\ell)$ 
	 \\\noalign{\smallskip}
		$4_\pm$
		&	$B \ell (-K-M+\ell)$
		&	$2 \eta \mp 4 \sqrt{B}\,  \sqrt{\zeta } (-M+\ell)$
	 \\\noalign{\smallskip}\hline
\end{tabularx}
\end{table}

 \begin{table}[h]
\caption{\label{table:matrix_els2} 
	Matrix elements on the main diagonal for each algebraic sector}\vspace{1mm}
\begin{tabularx}{.95\linewidth}{l l }
	\hline \noalign{\smallskip}
			Sector
	 	&	$\langle z^{\ell  } | T_t | z^{\ell  } \rangle$
	\\ \hline\hline \noalign{\smallskip}
			$1_{\pm}$
		&	$-\eta -B \left( M^2+2 M
		\ell+M+\ell (\ell+1) -\rho\, K^2 \right) \pm 2 \sqrt{B}\,  \sqrt{\zeta }
		(K+M+2 \ell+1)+\zeta$ \\\noalign{\smallskip}
		$2_{\pm}$
		&	$-\eta -B \left(K^2+2 K
		\ell+K+\ell (\ell+1) -\rho\, K^2 \right) \pm  2 \sqrt{B}\,  \sqrt{\zeta } (K+M+2
		\ell+1)+\zeta$ \\\noalign{\smallskip} 
		$3_{\pm}$
		&	$-\eta -B \left(K^2 - 2 K \ell-K+\ell (\ell+1)-\rho\, K^2 +\right)
		\pm 2 \sqrt{B}\,  \sqrt{\zeta } (-K-M+2 \ell+1)+\zeta$
		\\\noalign{\smallskip} $4_{\pm}$
		&	$-\eta -B \left(M^2-2 M \ell-M+\ell (\ell+1) -\rho\, K^2 \right)
		\pm 2 \sqrt{B}\,  \sqrt{\zeta } (-K-M+2 \ell+1)+\zeta$
\\\noalign{\smallskip}\hline
\end{tabularx}
\end{table}


\subsection{Closed-form wavefunctions} \label{subsec:alg-sols}

The lowest-order closed-form wave functions, i.e., the seed functions
 for $n=0$, are found by making use of Eqs. \eqref{eq:alg-sols_first-order_transf-new-old},
\eqref{eq:new-wavefunc_factoriz} and the algebraic quantum momentum functions listed in Table \ref{table:qmfs}:
\begin{align} \label{eq:direct-seed-from-new-qmf} 
	\psi_{t,0}(\theta) 
	\propto
	\left. \phi_{m,0}(z) \phi_{f}(z)
	\sqrt{\partial_z \theta(z) }
	\right|_{z=z(\theta)}
	\quad \text{with} \quad
	 \phi_f(z) = e^{\frac{1}{\sqrt{B}} \int^z \tilde{p}_f(y) \mathrm{d}y}
	 \; \quad \text{and} \quad
	 \phi_{m,0} = const.\,  
\end{align} 
Then the explicit expressions for the wavefunctions $\hat{\psi}_{t,0} (\theta)
= \psi_{t,0} (\theta)/\sqrt{\sin \theta}$, Eq. \eqref{eq:wavefunc-3d-gauge}, read
\begin{align} \label{eq:direct-seed-from-new-qmf_expl} 
	\hat{\psi}_{t,0}(\theta) 
	=
	e^{\pm \sqrt{\zeta/B } \cos (\theta)} \sin
    				^{\pm_1 K+ \mp_1 M}\left(\frac{\theta}{2}\right) \cos
    				^{\pm_2 K+ \pm_2 M}\left(\frac{\theta}{2}\right) \,  ,
\end{align}
where the signs are to be set with respect to the algebraic sectors.
Note that the dimension of an algebraic solution subspace is fixed to $n+1$,
so that $i$ labels the $n+1$ available algebraic solutions per
algebraic sector.

 A list of all closed-form wavefunctions 
\begin{align} \label{eq:alg-sols-all-levels}
	\hat{\psi}_{t,i}(\theta)=  \left. \hat{\psi}_{t,0}(\theta)
	\phi_{m,i}(z) \right|_{z=z(\theta)}	
\end{align}
which solve Schr\"odinger's Eq. \eqref{eq:ham_org-phys_symm-top_3-dim_separ} for the trigonometric symmetric top determined via
Eqs. \eqref{eq:new-wavefunc_factoriz} and \eqref{eq:direct-seed-from-new-qmf_expl}
(the coefficients in $\phi_{m,i}(z)$ are fixed by the eigenvectors of the matrix elements in  Tables \ref{table:matrix_els1} and \ref{table:matrix_els2}) is given in Table
\ref{tab:sols1} for $n=0,1$. Note that wavefunctions pertaining to sectors $1_-$ to $4_-$ differ from those pertaining to sectors $1_+$ to
$4_+$ by the change of branch, $\sqrt{\zeta} \mapsto
-\sqrt{\zeta}$.
According to the Abel-Ruffini theorem \cite{zoladek2000}, the existence of
algebraic solutions is not ensured for the characteristic polynomials of degree higher than $4$, although for $n > 4$ certain classes of characteristic
polynomials may yield further exact, but not necessarily closed-form,
solutions \cite{king2009,hagedorn2000}.

The construction of the solutions  $\hat{\psi}_{h,i}(\theta)$ of Eq.
\eqref{eq:ham_org-phys_symm-top_1-dim_hyperb} for the hyperbolic top via the Quantum Hamilton-Jacobi theory amounts to applying the
general mappings\footnote{These are not to be confused with the anti-isospectral transformation, which  yields a phase shift
such that
$\hat{\psi}_{t,i}^{}(i \theta) = e^{i \frac{\pi}{2}(\pm K \mp M )}
	\hat{\psi}_{h,i}(\theta) \, .$}
\begin{align} \label{eq:sols-trig-hyperb-mapping}
	\cos(\cdot) \mapsto \cosh(\cdot) 
	\quad \text{and} \quad
	\sin(\cdot) \mapsto \sinh(\cdot) \,,
\end{align}
to the trigonometric solutions $\hat{\psi}_{t,i}(\theta) $ in Eq. \eqref{eq:alg-sols-all-levels} 
and Table \ref{tab:sols1}.
The existence of the solutions $\hat{\psi}_{h,i}(\theta) $ is guaranteed by the
choice of the coordinate $z = (\cosh \theta)/2$ (or its M\"obius transformations).

The sufficient conditions for the normalizability
of the closed-form solutions $\hat{\psi}_{t,i}(\theta)$ of the
trigonometric and hyperbolic top are
summarized in Table \ref{table:alg-sols_normalizability}.

Furthermore, we find that the seed functions $\hat{\psi}_{t,0}(\theta)$ or
 $\hat{\psi}_{h,0}(\theta)$ which  satisfy one of the 
 normalizability conditions cause all higher-order solutions
 in the same sector, i.e., $\hat{\psi}_{t,i}(\theta)$ or $\hat{\psi}_{h,i}(\theta)$ for
 $n>0$, to be normalizable as well. However, the converse holds only for the hyperbolic top.
But if for the trigonometric top a higher order solution
$\hat{\psi}_{t,i}(\theta)$ for $n>0$ turns out to be normalizable,
then either the seed function of the same sector satisfies one of 
the above normalizability conditions or it is, up to the phase
$e^{ i\pi}$, identical  with a
solution of another sector whose seed function is normalizable. This situation is illustrated  in  Figures
 \ref{fig:spectrum_num-alg_sectors_m2k1},
 \ref{fig:spectrum_num-alg_sectors_m1k1} and
 \ref{fig:spectrum_num-alg_sectors_m1k0} for the cases $(K,M) = (1,2)$, $(K,M) =
 (1,1)$ and $(K,M) = (0,1)$ of Section
 \ref{sec:examples}.
 
%
\begin{sidewaystable}\footnotesize
	\caption{Algebraic solutions $\hat{\psi}_{t,i}(\theta)$ for the
	trigonometric top, Eq. 
	\eqref{eq:ham_org-phys_symm-top_3-dim_separ}, determined via Eq.
	\eqref{eq:alg-sols-all-levels}.
	Constructing the algebraic hyperbolic top solutions
	$\hat{\psi}_{h,i}(\theta)$, Eq. 
	\eqref{eq:ham_org-phys_symm-top_1-dim_hyperb}, requires the substitutions $\cos(\cdot)
	\mapsto \cosh(\cdot)
	\quad
	\text{and} \quad \sin(\cdot) \mapsto \sinh(\cdot)$ in the corresponding $\hat{\psi}_{t,i}(\theta)$.
	The dimension of the eigenspace for each algebraic sector is fixed by $n+1$, so
	that $i$ labels the available states in it.}
		\begin{tabularx}{1.02\textwidth}{ l l l X l }  
	    	\hline \noalign{\smallskip}
	    			sector
	   	 		&	
	   	 			$n$
	   	 		& 
	    			level $i$
	    		&  
	    			$E_i $ 
	    		& 
	    			$\hat{\psi}_{t,i}(\theta)$ 		
	    	\\ \noalign{\smallskip}\hline \hline \noalign{\smallskip}
    				\multirow{3}{*}{$1_\pm$} 
   	 			&	
    				0 
    			& 
    				0 
    			& 
    				$B(M^2+M-\rho K^2) \mp 2 \sqrt{B}\, K \sqrt{\zeta
    				}-\zeta $ 
    			& 
    				$e^{\pm\sqrt{\zeta/B } \cos (\theta)} \sin
    				^{(-K+M)}\left(\frac{\theta}{2}\right) \cos
    				^{(K+M)}\left(\frac{\theta}{2}\right)$ 
    		\\
    				\noalign{\smallskip}\cline{2-5}
    			& 
    				\multirow{2}{*}{1} 
    			& 
    				0
    			&  
    				$
    				B(M+1)^2- B\rho K^2 \mp 2 \sqrt{B}\,K
    				\sqrt{\zeta}-\zeta\newline 
    				 -\sqrt{B^2(M+1)^2 \pm 4 \sqrt{B}^3 K \sqrt{\zeta
    				}+ 4 B\zeta }$ 
    			&
    				$
    					 e^{\pm\sqrt{\zeta/B } \cos (\theta)}
	    				\sin^{(-K+M)}
	    				\left(\frac{\theta}{2}\right)
    					\cos^{(K+M)}
    					\left(\frac{\theta}{2}\right) 
    					 \left(
    						\frac{\sqrt{
    							B(M+1)^2 \pm 4 \sqrt{B}\, K
	    						\sqrt{\zeta }+4 \zeta 
	    					}
	    					+\sqrt{B} (M+1)
	    					}{
    					\pm 2 \sqrt{\zeta }
    				}
    				+ \cos (\theta)
    					\right)
    				$
    		\\ 	    
    		\noalign{\smallskip}\cline{3-5}	    			
   	 			&					
    			& 
    				1 
    			& 
    				$B (M+1)^2-B\rho K^2  \mp 2 \sqrt{B}\, K
    				\sqrt{\zeta}-\zeta \newline
    				+\sqrt{B^2 (M+1)^2 \pm 4 \sqrt{B}^3 K \sqrt{\zeta
    				}+4 B \zeta }$ 
    			&  
    				$e^{\pm\sqrt{\zeta/B } \cos (\theta)} \sin
    				^{(-K+M)}\left(\frac{\theta}{2}\right) \cos
    				^{(K+M)} \left(\frac{\theta}{2}\right) 
    				\left(\frac{-\sqrt{B (M+1)^2 \pm 4
    				\sqrt{B}\, K \sqrt{\zeta }+4 \zeta
    				}+\sqrt{B}(M+1)}{\pm 2 \sqrt{\zeta}}+\cos (\theta)\right)$
    				\\\noalign{\smallskip}\hline
		\multirow{3}{*}{$2_\pm$} 
   	 			&	
    				0 
    			& 
    				0 
    			& 
    				$B(K^2+K-\rho K^2) \mp 2 \sqrt{B}\, M \sqrt{\zeta
    				}-\zeta $ 
    			& 
    				$e^{\pm\sqrt{\zeta/B } \cos (\theta)} \sin
    				^{(K-M)}\left(\frac{\theta}{2}\right) \cos
    				^{(K+M)}\left(\frac{\theta}{2}\right)$
    		\\
			\noalign{\smallskip}\cline{2-5}
    			& 
    				\multirow{2}{*}{1} 
    			& 
    				0
    			&  
    				$B (K+1)^2- B \rho K^2 \mp 2
    				\sqrt{B}\, M \sqrt{\zeta }-\zeta \newline
    				-\sqrt{B^2 (K+1)^2 \pm 4 \sqrt{B}^3 M \sqrt{\zeta } + 4
    				B \zeta }$ 
    			&
    				$
    					e^{\pm\sqrt{\zeta/B } \cos (\theta)} 
    					\sin^{(K-M)} \left(\frac{\theta}{2}\right)
    					\cos^{(K+M)}\left(\frac{\theta}{2}\right)
    					\left(\frac{
    					\sqrt{B (K+1)^2 \pm 4 \sqrt{B}\, M \sqrt{\zeta }+4
    					\zeta }+ \sqrt{B} (K + 1) }{\pm 2
    					\sqrt{\zeta }} + \cos (\theta)\right)$
    		\\ 	    
    				\noalign{\smallskip}\cline{3-5}	    			
   	 			&					
    			& 
    				1 
    			& 
    				$B (K+1)^2- B \rho K^2
    				 \mp 2 \sqrt{B}\,M \sqrt{\zeta
    				}-\zeta\newline 
    				+\sqrt{B^2 (K+1)^2 \pm 4 \sqrt{B}^3 M \sqrt{\zeta }+4
    				B \zeta }$ 
    			&  
    				$e^{\pm\sqrt{\zeta/B } \cos (\theta)}
    				\sin^{(K-M)}\left(\frac{\theta}{2}\right)
    				\cos^{(K+M)}\left(\frac{\theta}{2}\right)
    				\left(\frac{
    				-\sqrt{B (K+1)^2 \pm 4 \sqrt{B}\, M \sqrt{\zeta }+4
    				\zeta }+\sqrt{B} ( K + 1) }{\pm 2
    				\sqrt{\zeta }} +\cos (\theta)\right)$
    				 \\\noalign{\smallskip}\hline
		\multirow{3}{*}{$3_\pm$} 
   	 			&	
    				0 
    			& 
    				0 
    			& 
    				$B(K^2-K-\rho K^2) \pm 2 \sqrt{B}\,M \sqrt{\zeta
    				}-\zeta $ 
    			& 
    				$e^{\pm \sqrt{\zeta/B } \cos (\theta)} \sin
    				^{(-K+M)}\left(\frac{\theta}{2}\right) \cos
    				^{(-K-M)}\left(\frac{\theta}{2}\right)$
    		\\
			\noalign{\smallskip}\cline{2-5}
    			& 
    				\multirow{2}{*}{1} 
    			& 
    				0
    			&  
    				$B (K-1)^2- B \rho K^2 \pm 2
    				\sqrt{B}\, M \sqrt{\zeta }-\zeta \newline
    				-\sqrt{B^2(K-1)^2 \mp 4 \sqrt{B}^3 M \sqrt{\zeta }+4
    				B \zeta }$ 
    			&
    				$
    					e^{\pm\sqrt{\zeta/B } \cos (\theta)}
    					\sin^{(-K+M)}\left(\frac{\theta}{2}\right) 
    					\cos^{(-K-M)}\left(\frac{\theta}{2}\right)
    					\left(\frac{ \sqrt{B (K-1)^2
    					\mp 4 \sqrt{B}\, M \sqrt{\zeta }+4 \zeta }
    					+\sqrt{B} (-K	+1)}{\pm 2 \sqrt{\zeta }} 
    					+ \cos (\theta) \right)$
    		\\ 	    
    				\noalign{\smallskip}\cline{3-5}	    			
   	 			&					
    			& 
    				1 
    			& 
    				$B (K-1)^2-B \rho K^2 \pm 2 M
    				\sqrt{B}\, \sqrt{\zeta }-\zeta \newline
    				+\sqrt{B^2 (K-1)^2 \mp 4 \sqrt{B}^3 M \sqrt{\zeta }+4
    				B \zeta }$ 
    			&  
    				$e^{\pm\sqrt{\zeta/B } \cos (\theta)}
    				\sin^{(-K+M)}\left(\frac{\theta}{2}\right) 
    				\cos^{(-K-M)}\left(\frac{\theta}{2}\right)
    				\left(\frac{ -\sqrt{B (K-1)^2
    				\mp 4 \sqrt{B}\, M \sqrt{\zeta }+4 \zeta
    				}+\sqrt{B}(-K+1)}{\pm2 \sqrt{\zeta }} + \cos (\theta)\right)$
			\\
					\noalign{\smallskip}\hline
    				\multirow{3}{*}{$4_\pm$} 
   	 			&	
    				0 
    			& 
    				0 
    			& 
    				$B(M^2-M-\rho K^2) \pm 2 \sqrt{B}\, K \sqrt{\zeta
    				}-\zeta $ 
    			& 
    				$e^{\pm\sqrt{\zeta/B } \cos (\theta)} \sin
    				^{(K-M)}\left(\frac{\theta}{2}\right) \cos
    				^{(-K-M)}\left(\frac{\theta}{2}\right)$
    		\\
    				\noalign{\smallskip}\cline{2-5}
    			& 
    				\multirow{2}{*}{1} 
    			& 
    				0
    			&  
    				$B (M-1)^2-B \rho K^2 \pm 2 \sqrt{B}\, K
    				\sqrt{\zeta }-\zeta \newline
    				-\sqrt{B^2(M-1)^2 \mp 4 \sqrt{B}^3 K \sqrt{\zeta
    				}+4 B \zeta }$ 
    			&
    				$
    					e^{\pm \sqrt{\zeta/B } \cos (\theta)} 
    					\sin^{(K-M)}\left(\frac{\theta}{2}\right)
    					\cos^{(-K-M)}\left(\frac{\theta}{2}\right)
    					\left(\frac{ \sqrt{B (M-1)^2 \mp 4
    					\sqrt{B}\, K \sqrt{\zeta }+4 \zeta }
    					+\sqrt{B}(-M+1)}{\pm 2 \sqrt{\zeta }} + \cos (\theta)\right)$
    		\\ 	    
    		\noalign{\smallskip}\cline{3-5}	    			
   	 			&					
    			& 
    				1 
    			& 
    				$B(M-1)^2-B \rho K^2 \pm 2 \sqrt{B}\, K
    				\sqrt{\zeta }-\zeta \newline
    				+\sqrt{B^2 (M-1)^2 \mp 4 \sqrt{B}^3 K \sqrt{\zeta
    				}+4 B \zeta }$ 
    			&  
    				$e^{\pm\sqrt{\zeta/B } \cos (\theta)}
    				\sin^{(K-M)}\left(\frac{\theta}{2}\right) 
    				\cos^{(-K-M)}\left(\frac{\theta}{2}\right)
    				\left(\frac{ -\sqrt{B (M-1)^2 \mp 4 \sqrt{B}\, K
    				\sqrt{\zeta }+4 \zeta }+\sqrt{B}(-M
    				+1)}{\pm 2 \sqrt{\zeta }}+ \cos(\theta)\right)$ 
    		\\\noalign{\smallskip}\hline 
	  \end{tabularx}	  \label{tab:sols1}	  
\end{sidewaystable}

 \begin{table}[h]
\caption{\label{table:alg-sols_normalizability} 
	Sufficient conditions for the normalizability of the wavefunctions
	$\hat{\psi}_{t,i}(\theta)$ and $\hat{\psi}_{h,i}(\theta)$ of the
	\textit{trigonometric} and \textit{hyperbolic top}, respectively.}\vspace{1mm}
\setlength{\tabcolsep}{1.5em}
\begin{tabularx}{.7\linewidth}{l l | l l}
	\hline 
			\multicolumn{2}{l|}{\textit{trigonometric top}}	 		
	 	&	\multicolumn{2}{l}{\textit{hyperbolic top}}
	\\   
			sector
	 	&	condition	 		
	 	&	sector
	 	&	condition
	\\ \hline\hline 
			$1_\pm$
		&	$K \leq M \wedge K \geq -M$
		&	$1_-$
		&	$K \leq M$
	\\
		$2_\pm$
		&	$K\geq M \wedge K\geq -M$
		&	$2_-$
		&	$K\geq M$
	 \\
		$3_\pm$
		&	$K \leq M \wedge K \leq -M$
		&	$3_-$
		&	$K \leq M$ 
	 \\
		$4_\pm$
		&	$K\geq M \wedge K \leq -M$
		&	$4_-$
		&	$K\geq M$
	 \\
	 \hline
\end{tabularx}
\end{table}

 
\subsection{Limit-point and Limit-circle classification}
\label{subsec:limit-point-circle_classif}  
\subsubsection{Trigonometric top}
In this section, we discuss the underlying physical and mathematical structure that leads to the existence/absence of (non-)normalizable closed-form solutions in Table \ref{tab:sols1}. In particular, we show that if an effective centrifugal potential, that we identify in Eq. \eqref{eq:Vcent}, remains bounded, then any solution to Eq. \eqref{eq:ham_org-phys_symm-top_1-dim} is square integrable. Conversely, there exist non-normalizable solutions to Eq. \eqref{eq:ham_org-phys_symm-top_1-dim} only, if this effective centrifugal potential is unbounded at the end-points $\theta=0$ or $\theta= \pi$. As shown in Table \ref{table:alg-sols_normalizability}, only some solutions found by the Quantum Hamilton-Jacobi method are normalizable and qualify therefore as physical states for the pendulum. In particular, we see that transitions $M=K$ or $M=-K$ are critical for the normalizability of the solutions presented in Table \ref{tab:sols1}. 
To understand these properties better, we start with some simple observations for Eq. \eqref{eq:ham_org-phys_symm-top_3-dim_separ} 
\begin{align} \notag
	\hat{\mathcal{H}}_t \hat{\psi}_t(\theta) 
	={}&
		B \left(
			-\partial_\theta^2 
			-
			\cot \theta\, \partial_\theta
			+
			\left(M^2 + K^2  \right)\csc^2 \theta 
			- 
			2 M K \csc \theta \, \cot \theta 
			-
			\rho K^2
			\right) \hat{\psi}_t(\theta)\\\notag
			&+
			\left(
			 -
			  \eta \cos{\theta} 
			 - \zeta 	
			\cos^2{\theta}
			\right) \hat{\psi}_t(\theta) 
		\\\label{eq:ham_org-phys_symm-top_3-dim_separ2}
	={}&
		E_t\, \hat{\psi}_t(\theta) \,  
\end{align} 
Note that normalizable solutions to
Eq. \eqref{eq:ham_org-phys_symm-top_3-dim_separ2} coincide with normalizable solutions to Eq. \eqref{eq:ham_org-phys_symm-top_1-dim} when gauged according to Eq. \eqref{eq:wavefunc-3d-gauge}.
The term 
\begin{align}\label{eq:Vcent}
V_{\text{cent,t}}(\theta)=\left(M^2 + K^2  \right)\csc^2 \theta - 2 M K
\csc \theta \, \cot \theta
\end{align}
is an effective centrifugal potential and is the decisive 
quantity for the presence or absence of normalizable solutions to 
Eq. \eqref{eq:ham_org-phys_symm-top_3-dim_separ2} as we explain now. In particular, 
the discussion of normalizability is independent of 
$\eta$ and $\zeta$ for the trigonometric problem because the
trigonometric potential, as defined in Eq. \eqref{eq:ext-pot}, is a bounded function.
We observe that exactly when $M \neq K$ the centrifugal potential $V_{\text{cent,t}}$ 
is \emph{confining} as $\theta \downarrow 0$, i.e.,
\begin{equation}
\label{eq:conf0}
\lim_{\theta \downarrow 0} V_{\text{cent,t}}(\theta) = \infty
\end{equation}
A similar result is true if $M \neq -K$ at $\theta=\pi$: Unless, $M =-K$, the centrifugal potential $V_{\text{cent,t}}$ is \emph{confining} as $\theta \uparrow \pi$, i.e.,
\begin{equation}
\label{eq:confpi}
\lim_{\theta \uparrow \pi} V_{\text{cent,t}}(\theta) = \infty
\end{equation}
    
Like any second order differential equation of Picard-Lindel\"of type, Eq. 
\eqref{eq:ham_org-phys_symm-top_3-dim_separ2}  possesses for any $E_t \in
\mathbb{C}$ two linearly independent solutions that we denote by $\hat{\psi}_{t}$ and $\hat{\varphi}_{t}$. Given one solution to Eq. \eqref{eq:ham_org-phys_symm-top_3-dim_separ2}, which we assume without loss of generality to be  $\hat{\psi}_{t},$ a linearly independent partner solution $\hat{\varphi}_{t}$ can be computed from \cite{teschl2011}
\begin{equation}
\label{eq:sec}
\hat{\varphi}_{t}(\theta)=\hat{\psi}_{t}(\theta) \int_{\frac{\pi}{2}}^{\theta} \frac{dt}{\sin(t) \hat{\psi}_{t}(t)^2}
\end{equation}
In other words, we have that $(\hat{\mathcal{H}}_t\hat{\psi}_{t})(\theta)=E_t 
\hat{\psi}_{t}(\theta)$ and $(\hat{\mathcal{H}}_t\hat{\varphi}_{t})(\theta)=E_t 
\hat{\varphi}_{t}(\theta)$ with $\hat{\psi}_{t},\hat{\varphi}_{t}$ linearly independent.
In particular, Eq. \eqref{eq:sec} allows us to compute a linearly independent second solution to the same energy for any closed-form solution shown in Table \ref{tab:sols1}. Yet, we will see that these solutions do not correspond to physical states of the system.

If $M \neq K$, i.e., confining centrifugal potential in the sense of Eq. \eqref{eq:conf0} at $\theta=0$, then one of the solutions $1_{+}$, $n=0$, $i=0$ and $2_{+}$, $n=0$, $i=0$ from Table \ref{tab:sols1} 
\begin{equation}
\label{eq:firstsol}
\hat{\psi}_{t,0}(\theta)=e^{\sqrt{\zeta/B}\cos(\theta)} \sin^{\pm
(M-K)}\left(\frac{\theta}{2} \right)\cos^{
M+K}\left(\frac{\theta}{2} \right)
\end{equation}
is not square integrable in any neighbourhood of $\theta=0.$ 
Similarly, if $M \neq -K,$ i.e., confining centrifugal potential at $\theta=\pi$ in the sense of Eq. \eqref{eq:confpi}, then one of the solutions $1_{+}$, $n=0$, $i=0$ and $3_{+}$, $n=0$, $i=0$ from Table \ref{tab:sols1} 
\begin{equation}
\label{eq: secondsol}
\hat{\psi}_{t,0}(\theta)=e^{\sqrt{\zeta/B}\cos(\theta)} \sin^{
M-K}\left(\frac{\theta}{2} \right)\cos^{\pm(M+K)}\left(\frac{\theta}{2} \right)
\end{equation}
is not square integrable in any neighbourhood of $\theta=\pi.$ The connection between the boundedness of the effective centrifugal potential Eq. \eqref{eq:Vcent} and the square integrability of solutions to Eq. \eqref{eq:ham_org-phys_symm-top_3-dim_separ2} can then be seen as follows:
By applying Eq. \eqref{eq:sec} to the solutions stated in Eq. \eqref{eq:firstsol} and Eq. \eqref{eq: secondsol}, one finds that Eq. \eqref{eq:firstsol}, Eq. \eqref{eq: secondsol}, and their respective linearly independent second solutions are not both square integrable at $\theta=0$ or $\theta=\pi$ exactly when the centrifugal potential is confining at that end-point. It is thus natural to expect that if the centrifugal potential at $\theta=0$ or $\theta=\pi$ remains bounded, any solution to Eq. \eqref{eq:ham_org-phys_symm-top_3-dim_separ2} is square integrable at that particular end-point. 
Hence, if $M=K=0$ then any solution to Eq. \eqref{eq:ham_org-phys_symm-top_3-dim_separ2} should be square integrable on the entire interval $(0,\pi).$ 

We now introduce a terminology from Sturm-Liouville theory to make this observation precise:
The operator $\hat{\mathcal{H}}_t$ is called {\it limit-circle} at $\theta=0$ or
$\theta=\pi$ if and only if for one value $E_t \in \mathbb{C}$ both linearly
independent solutions to $\hat{\mathcal{H}}_t\hat{\psi}_{t}=E_t \hat{\psi}_{t}$
are square integrable at $\theta=0$ or $\theta=\pi$, respectively. Otherwise, the operator $\hat{\mathcal{H}}_t$ is called {\it limit-point} at that end-point.

The rigorous footing for our previous argument is the Weyl alternative \cite{teschl2011} which states that the limit-circle property is independent of the energy $E_t$. Thus, if one verifies for a fixed energy $E_t$ that both solutions to Eq. \eqref{eq:ham_org-phys_symm-top_3-dim_separ2} are square integrable at one of the end-points, this will be the case for any other $E_t \in \mathbb{C}$ as well. 

We summarize our preceding discussion by observing that $\hat{\mathcal{H}}_t$ is
\begin{itemize}
\item limit-circle at both end-points if $M=K=0$, i.e., bounded centrifugal potential at both end-points and all solutions to Eq. \eqref{eq:ham_org-phys_symm-top_3-dim_separ2} are square integrable on the entire interval $(0,\pi),$
\item limit-circle at $\theta=0$ and limit-point at $\theta=\pi$ if $M=K \neq 0,$ i.e., bounded centrifugal potential at $\theta=0$, confining one at $\theta=\pi,$ and all solutions to Eq. \eqref{eq:ham_org-phys_symm-top_3-dim_separ2} are square integrable close to $\theta=0,$

\item limit-circle at $\theta=\pi$ and limit-point at $\theta=0$ if $M=-K \neq 0,$ i.e., bounded centrifugal potential at $\theta=\pi$, confining one at $\theta=0,$ and all solutions to Eq. \eqref{eq:ham_org-phys_symm-top_3-dim_separ2} are square integrable close to $\theta=\pi,$ and
\item limit-point at both end-points in any other case, i.e., confining centrifugal potential at both end-points. In this case, there do exist square integrable solutions but there do not exist two linearly independent solutions for a fixed energy that are both square integrable at the same end-point. 
\end{itemize}

Although we have obtained a rather complete description of when to expect normalizable or non-normalizable solutions, we still require a condition to exhibit the physical eigenstates among the normalizable ones. In fact, not every normalizable solution to Eq. \eqref{eq:ham_org-phys_symm-top_3-dim_separ2} is also an eigenstate in general. This is only true when $\hat{\mathcal{H}}_t$ is limit-point at both end-points. If $\hat{\mathcal{H}}_t$ is limit-circle at an end-point, then a normalizable solution $\psi_t$ to Eq. \eqref{eq:ham_org-phys_symm-top_3-dim_separ2} is an eigenfunction to $\hat{\mathcal{H}}_t$ if it satisfies at the limit-circle end-points
\begin{equation}
\label{eq:physicalcond}
\lim_{\theta \rightarrow 0, \pi } \sin(\theta)\hat{\psi}'_{t}(\theta)=0
\end{equation} 
The physical interpretation of the condition in Eq. \eqref{eq:physicalcond} 
is that eigenstates of the pendulum must have bounded wavefunctions and this condition filters out 
precisely the unbounded square integrable solutions. Mathematically, this condition is needed 
in order to obtain a self-adjoint operator $\hat{\mathcal{H}}_t $.

To illustrate our findings, we consider the limit-circle case at both end-points  $M=K=0$ and derive a solution using Eq. \eqref{eq:sec} that is not obtained by the Quantum Hamilton-Jacobi theory. In this case, the operator introduced in Eq. \eqref{eq:ham_org-phys_symm-top_3-dim_separ2} simplifies to
\begin{align}
\label{eq:simpler_form}
	\hat{\mathcal{H}}_t \hat{\psi}_t(\theta) 
	={}&
		B \left(
			- \frac{d^2}{d\theta^2}
			-
			\cot \theta\,\frac{d}{d\theta}
			 \right)\hat{\psi}_t(\theta)
			 -\left( \eta \cos{\theta} +\zeta 	
			\cos^2{\theta}
			\right) \hat{\psi}_t(\theta) =E_t\, \hat{\psi}_t(\theta) \,  
\end{align} 
Now, let $\eta= \pm 2 \sqrt{B} \sqrt{\zeta}$ and $E_t=-\xi$ then a
solution to \eqref{eq:simpler_form} is provided by the 
$1_{\pm},\ n=0 \text{ and } i=0$ solution from Table \ref{tab:sols1} which reads
\begin{equation}
\label{eq:firstsol2}
\hat{\psi}_{t,0}(\theta)=e^{\pm \sqrt{\zeta/B}\cos(\theta)}
\end{equation}
Applying the formula in Eq. \eqref{eq:sec} to this solution yields then another solution to Eq. \eqref{eq:simpler_form} for the same energy value that is not contained in Table \ref{tab:sols1}
\begin{equation}
\hat{\varphi}_{t,0}(\theta):=\frac{e^{\pm
\sqrt{\frac{\zeta}{B}}(\cos(\theta)-2)}}{2}\left[\operatorname{Ei}\left(\pm
4\sqrt{\frac{\zeta}{B}}\, \sin^2\left(\frac{\theta}{2} \right)\right)-e^{\pm4\sqrt{\frac{\zeta}{B}}}\operatorname{Ei}\left(\mp
4\sqrt{\frac{\zeta}{B}}\,\cos^2\left(\frac{\theta}{2} \right)\right)\right]
\end{equation}
where $\operatorname{Ei}$ is the exponential integral function. In other words, $\hat{\varphi}_{t,0}$ is a square integrable solution to Eq. \eqref{eq:simpler_form} that has not been found by the Quantum Hamilton-Jacobi theory, which does not satisfy the condition in Eq. \eqref{eq:physicalcond}, i.e.,
\begin{equation}
\lim_{\theta \rightarrow 0, \pi } \sin(\theta)\hat{\varphi}'_{t,0}(\theta)\neq 0
\end{equation} 
and is therefore not an eigenstate of the pendulum.


\subsubsection{Hyperbolic top}

After the comprehensive treatment of the trigonometric equation 
\eqref{eq:ham_org-phys_symm-top_3-dim_separ}, we just state the results for the 
hyperbolic potential \eqref{eq:ham_org-phys_symm-top_1-dim_hyperb} on 
$(0, \infty)$ for $\zeta > 0$. In this case, the centrifugal potential is
\begin{align}\label{Vcent_hyp}
V_{\text{cent,h}}(\theta)=-\left(M^2 + K^2  \right) \csch^2(\theta) + 2 M K
\coth(\theta)\csch(\theta)
\end{align}
Since the hyperbolic potential converges to $\infty$ as $\theta \rightarrow \infty$, this already implies that $\hat{\mathcal{H}}_h$ is limit-point at $\theta=\infty$. Thus, there is at most one square integrable solution for any energy value to $\hat{\mathcal{H}}_h$ at $\theta= \infty$ and it suffices to study the boundary value $\theta=0.$ For $\theta=0$, the centrifugal potential $V_{\text{cent,h}}$ is bounded only if $K=M$. The boundedness of the centrifugal potential is then, as for the trigonometric potential, equivalent to $\hat{\mathcal{H}}_h$ being limit-circle. Thus, exactly when $K=M$ all solutions to \eqref{eq:ham_org-phys_symm-top_1-dim_hyperb} for arbitrary energies $E$ are square integrable at $\theta=0$. Similar to the boundary condition in Eq. \eqref{eq:physicalcond} for the limit-circle case with trigonometric potential, we get an analogous boundary condition in the limit-circle case at $\theta=0$ for the hyperbolic potential, filtering out the physical eigenstates among the square integrable solutions,
\begin{equation}
\lim_{\theta \rightarrow 0 } \sinh(\theta)\hat{\psi}'_{h}(\theta)= 0
\end{equation}
A second linearly independent solution $\hat{\varphi}_{h}$ to some given solution $ \hat{\psi}_{h}$ for the same energy can be obtained from
\begin{equation}
\hat{\varphi}_{h}(\theta)=\hat{\psi}_{h}(\theta) \int_{1}^{\theta} \frac{dt}{\sinh(t) \hat{\psi}_{h}(t)^2}
\end{equation}

 
\section{Examples and observations} \label{sec:examples}
\begin{figure}
	\includegraphics[width=.7\textwidth]{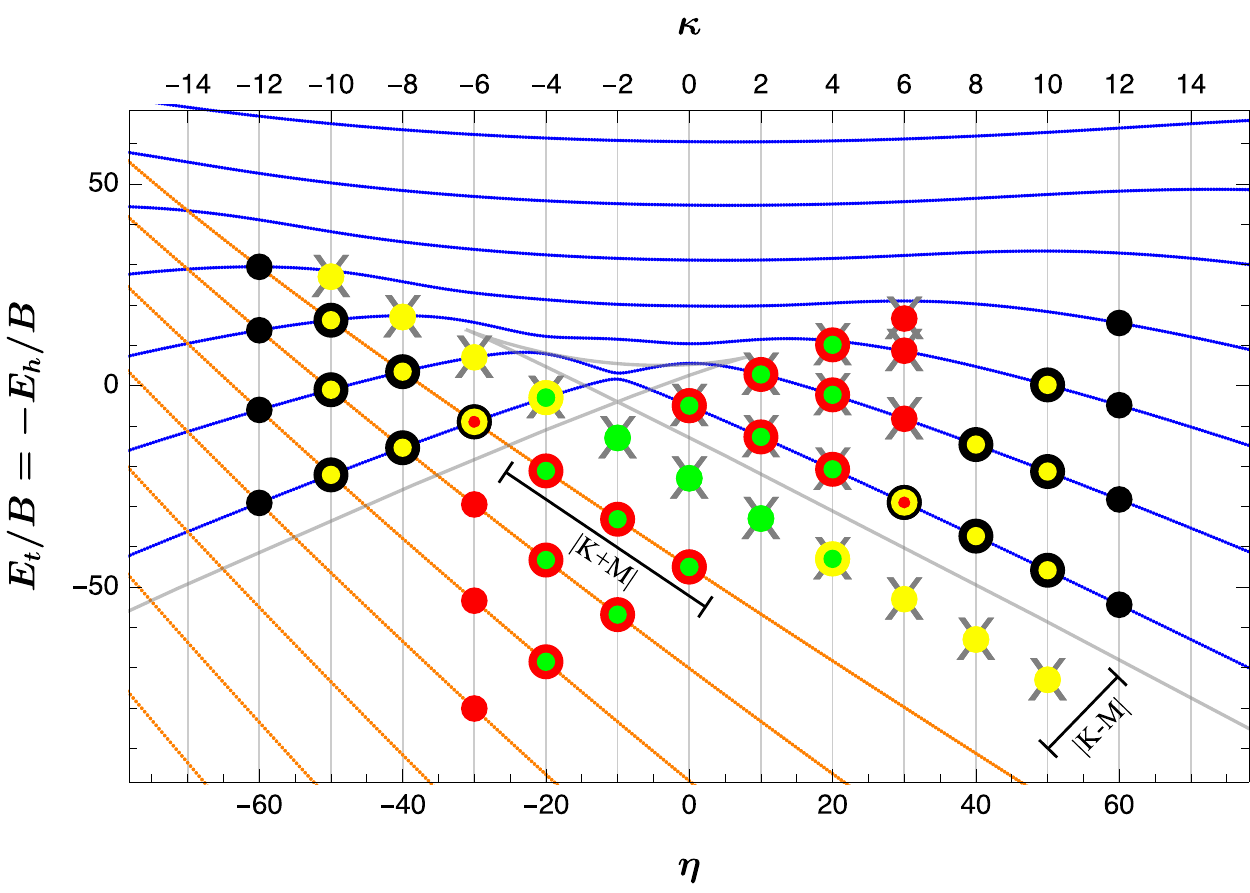}
	\caption{Numerical spectra for $M=2$, $K=1$ and $\zeta=25$, $\rho=0$:
	trigonometric top (blue), hyperbolic top (orange). 
	Black: sector $1_{+}$ (here $\kappa \geq 6$), sector $1_{-}$ ($\kappa \leq
	-6$); Yellow:
	sector $2_{+}$ ($\kappa \geq 4$), sector $2_{-}$ ($\kappa \leq -4$); 
	Red: sector $3_{+}$ ($\kappa \geq 0$), sector $3_{-}$ ($\kappa \leq 0$);
	Green:
	sector:
	$4_{+}$ ($\kappa \geq -2$), sector $4_{-}$ ($\kappa \leq 2$). Grey crosses mark non-normalizable solutions.
	Grey curves show local minima and maxima of the symmetric top potential, Eq. 
	\eqref{eq:org-effective-pot_1-dim}.  For computational details, see Appendix \ref{subsec:matrix-els-trig-top_for-num} and \ref{subsec:matrix-els-hyp-top_for-num}.}
	\label{fig:spectrum_num-alg_sectors_m2k1}
\end{figure}

\begin{figure}
	\centering
		\subfloat{ 
			\includegraphics[width=.35\textwidth]{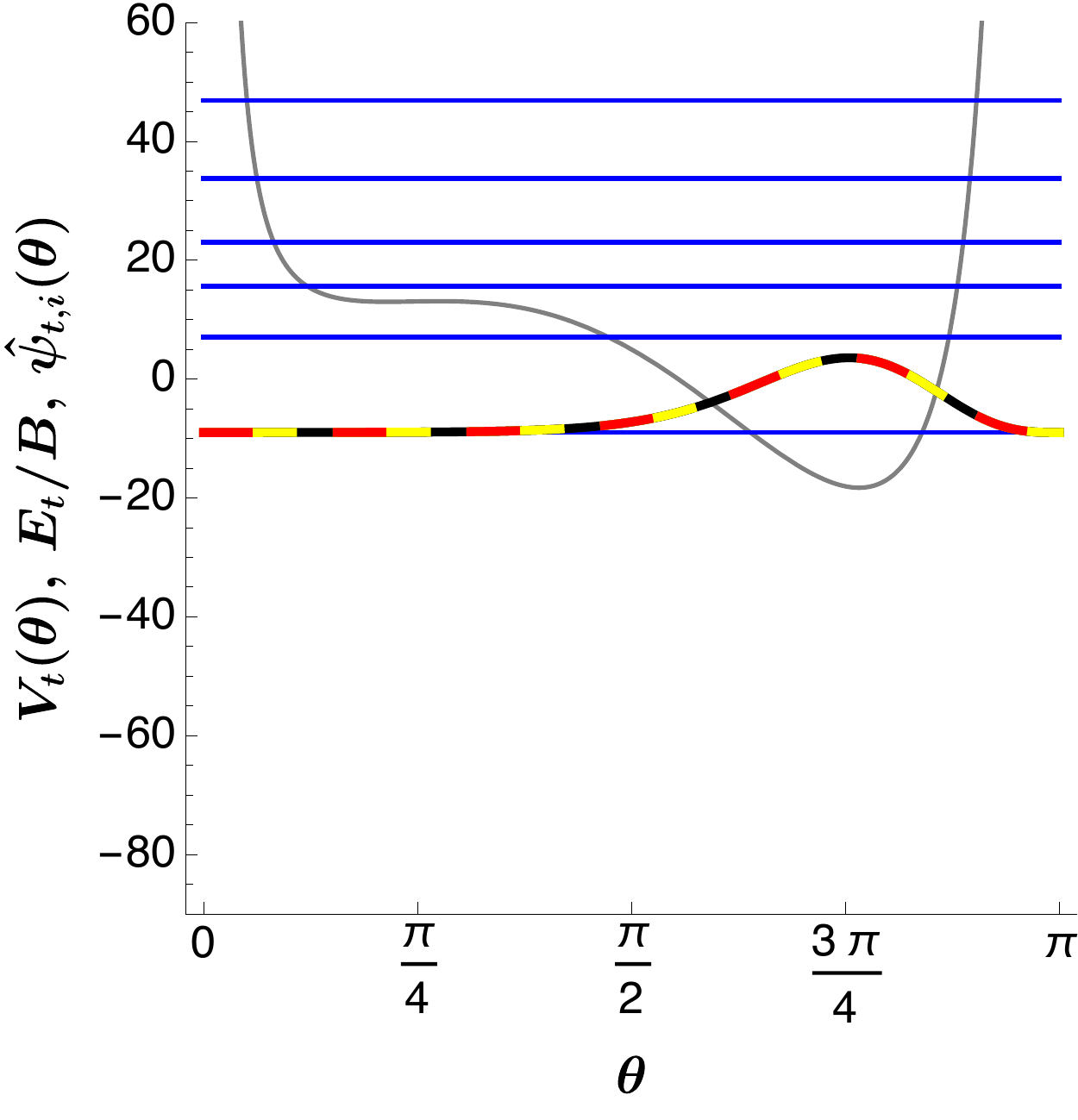}
			}
		\hspace{3mm}  
		\subfloat{ 
			\includegraphics[width=.35\textwidth]{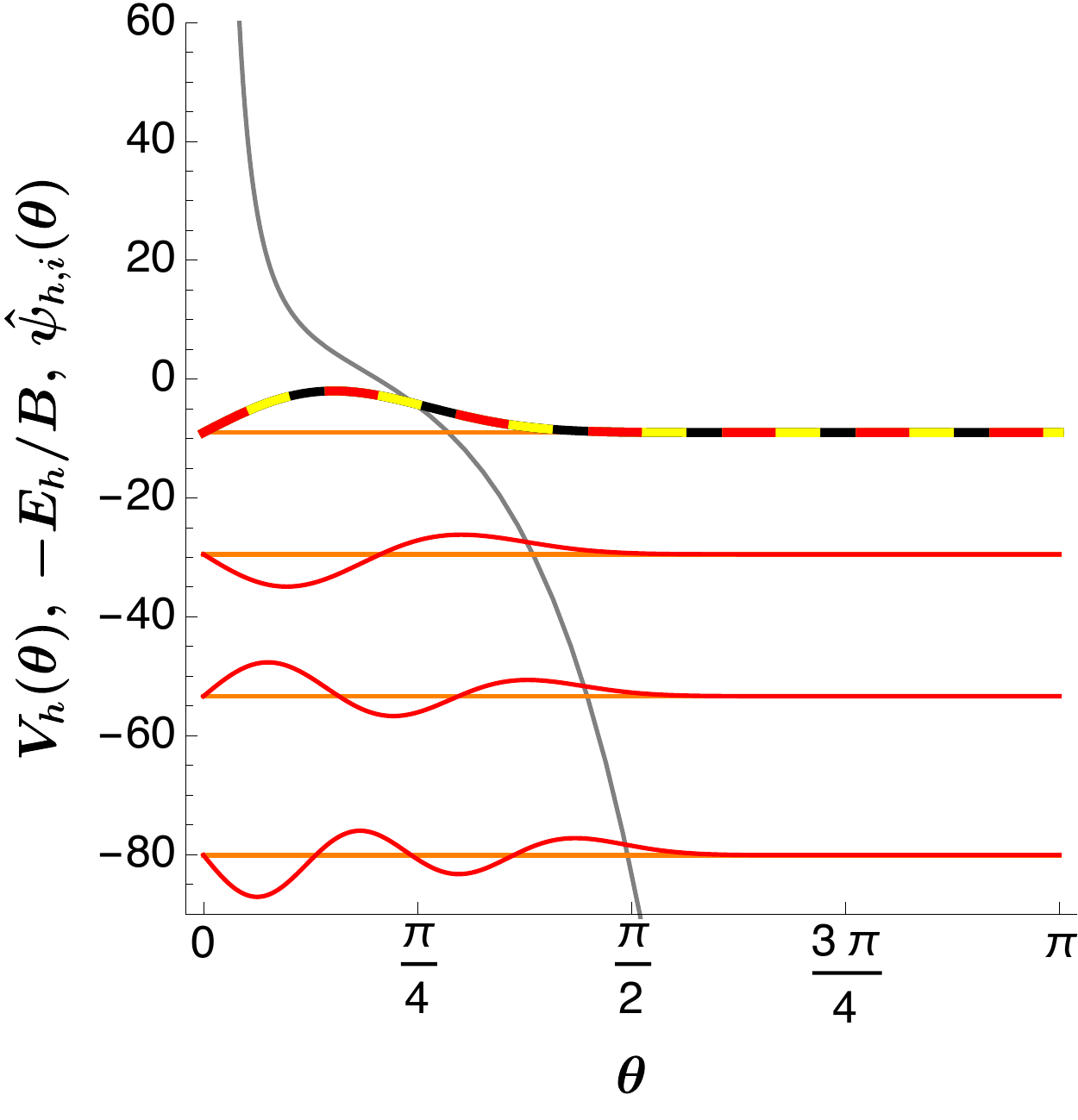}
			}
		\caption{Normalizable closed-form sample wavefunctons $\hat{\psi}_{t,i}
		(\theta)$ (left) and  $\hat{\psi}_{h,i}
		(\theta)$ (right). Black for sector $1_{\pm}$, yellow for sector $2_{\pm}$, and red for sector $3_{\pm}$. Blue: numerical eigenenergies for trigonometric
		top (left). Orange: numerical eigenenergies for hyperbolic top
		(right). Potentials shown in grey. Both plots for $M=2$, $K=1$, $\zeta=25$, $\eta=-30$.
		Same color-coding as in Fig.
		\ref{fig:spectrum_num-alg_sectors_m2k1} } 
		\label{fig:wavefuncs_m2-k1-zeta25-eta-40}
\end{figure}

In this section we take a closer look at the spectral patterns
in the cases $(K,M) = (1,2),\, (1,1),\, (0,1)\; \text{and}\; (0,1/2)$
and retrieve the symmetry-based special-case classifications of closed-form solutions from our earlier work
\cite{SchmiFri2015a,b_friedrich2017} for the
spherical pendulum, i.e., for $K=0$, as well as the planar pendulum and the
Razavy system, i.e., for $(K,M)=(0,1/2)$.

In Figs. \ref{fig:spectrum_num-alg_sectors_m2k1},
\ref{fig:spectrum_num-alg_sectors_m1k1}, \ref{fig:spectrum_num-alg_sectors_m1k0},
and \ref{fig:spectrum_num-alg_sectors_m0p5k0} below, we use the following color coding and graphical symbols: blue and orange curves show numerical
eigenenergies of the trigonometric and hyperbolic top obtained via the analytic
matrix elements of  Appendix \ref{subsec:matrix-els-trig-top_for-num} and overlap matrices 
of Appendix 
\ref{subsec:matrix-els-hyp-top_for-num}, respectively. The algebraic
energies of sectors $1_{\pm}$, $2_{\pm}$, $3_{\pm}$, and $4_{\pm}$ pertaining to
normalizbale wavefunctions are represented by black, yellow, red, and
green disks for $n \leq 3$. Nested disks represent eigenenergies of coinciding
closed-form solutions for different sectors.   
If the  closed-form
wavefunctions are not normalizable, the corresponding disks are furnished with a
grey cross. The extrema of the potential energy curves of the trigonometric top are shown by grey curves.
Note that level crossings (genuine and avoided) are marked  by the (integer) values of the topological
index $\kappa = \frac{\eta}{\sqrt{B\, \zeta}}$, which can be used as their label. 

In Figs. \ref{fig:wavefuncs_m2-k1-zeta25-eta-40} and
\ref{fig:wavefuncs_m1-k1-zeta25-eta-40} we show algebraic wavefunctions
colored according to same the scheme as described above for
Figs.
\ref{fig:spectrum_num-alg_sectors_m2k1},
\ref{fig:spectrum_num-alg_sectors_m1k1}, \ref{fig:spectrum_num-alg_sectors_m1k0}
and \ref{fig:spectrum_num-alg_sectors_m0p5k0}. Likewise, the numerical
eigenenergies of the trigonometric (left) and
hyperbolic top (right) are again shown in blue and orange,
respectively. Grey curves show the extrema of the trigonometric and hyperbolic top potentials.

As for the total number of  normalizable closed-form solutions obtained (for $n \leq 3$), we see in 
Figs. \ref{fig:spectrum_num-alg_sectors_m2k1},
\ref{fig:spectrum_num-alg_sectors_m1k1}, and
\ref{fig:spectrum_num-alg_sectors_m1k0} as well as in Table \ref{tab:sols1} that
for the trigonometric top it is
\begin{align} \label{eq:alg-sols-trig_nr}
	\#_{t}\{algebraic\,solutions\} = (n_{max}+1) (n_{max}+2) \,   
\end{align} 
Note, that the number of solutions 
takes into account all sectors for both branches $\pm \sqrt{\zeta}$. 
In the hyperbolic case we have
\begin{align} \notag
	{}&\#_{h}\{algebraic\,solutions\}\\ 
	={}& \begin{cases}
                (n_{max}+1) (n_{max}+2) -
               \frac{(n_{max}-|M+K|+1)(n_{max}-|M+K| + 2)}{2} \quad \text{for}
               \quad n_{max}-|M+K| \geq 0\\\label{eq:alg-sols-hyperb_nr}
               (n_{max}+1) (n_{max}+2) \quad \text{for} \quad
               n_{max}-|M+K| < 0
            \end{cases} \, 
\end{align}
Here, $K, M \in \mathbb{Z}$ and $n_{max}$ is the highest $n$ for which one
can derive closed-form solutions. As explained in Subsection \ref{subsec:alg-sols}, we
choose $n_{max} = 3$, which always yields $20$ closed-form solutions for each configuration
 $(K,M,\eta,\zeta)$ in the trigonometric top case for either of the 
 two branches $\pm \sqrt{\zeta}$.
On the other hand, for the hyperbolic top with $(M,K) = (0,0)$, $(1,1)$ or $(2,1)$,
we find $10$, $17$ or $19$ different closed-form solutions, whereas for $4-|M+K| > 0$ 
the total number becomes $30$ for each $(K,M,\eta,\zeta)$.


\mathversion{bold}
\subsection{$K=1,M=2$ } \label{subsec:exampl-k1m2}
\mathversion{normal}

We can read off from Fig. \ref{fig:spectrum_num-alg_sectors_m2k1} that
there are $10$ normalizable algebraic solutions for each of the sectors
$1_{\pm}$ (black). Out of these, sector $1_{+}$ ($\kappa \geq 6$) contains only
eigenvalues of the
trigonometric top, whereas the eigenvalues of the trigonometric 
and hyperbolic top coincide for sector $1_{-}$  ($\kappa \leq 6$). 
Furthermore, we have $6$ normalizable and $4$ non-normalizable closed-form
solutions for sectors $2_{+}$  ($\kappa \geq 4$) and $2_{-}$
($\kappa \leq 4$) each, shown in yellow. As all the normalizable
solutions coincide with sector $1_{\pm}$ solutions for $n \leq 2$, the
relative distribution of the solutions over the trigonometric and hyperbolic
top case is identical to that of sector $1_{\pm}$
solutions. 

Sector $3_{+}$ (red) contains $1$ normalizable solution at
$(\kappa,n) = (6,3)$ of the trigonometric top, which is identical to
the normalizable solution of sector $1_{+}$ at $(\kappa,n) = (6,0)$. 
A similar situation occurs for the highest excited sector $3_{-}$ (red) solution
at $(\kappa,n) = (-6,3)$ in that it coincides with the sector $1_{-}$ solution at
$(\kappa,n) = (6,0)$. While all other solutions of sector $3_{+}$ are
non-normalizable for the trigonometric and hyperbolic
top, every other sector $3_{-}$ solution is normalizable for the
hyperbolic top. There exist $6$ normalizable sector $4_{-}$ (green) solutions ($\kappa \leq 1$), overlapping with sector $3_{-}$ solutions, one for $n
\leq 2$. All other sector $4_{+}$ and $4_{-}$ (green) solutions are
non-normalizable. 

Thus we see that for all configurations with $K$ or $M \neq
0$ there emerge $3$  regions of normalizable solutions, which may partly
overlap:
firstly, one for the $\sqrt{\zeta}$ branch, which pertains to the
trigonometric top only; secondly, one for the $-\sqrt{\zeta}$ branch,
which pertains to the
hyperbolic top only and, thirdly,  one for the $-\sqrt{\zeta}$
branch that pertains to both the trigonometric and
hyperbolic top.
Ultimately, the cardinality of the overlap of the second and third region
($-\sqrt{\zeta}$ branch) as well as the spectral gap of
non-normalizability to the first region of the $\sqrt{\zeta}$ branch
suggest a vectorial measure for the distribution of these regions between the
trigonometric and hyperbolic numerical spectra, namely $|K+M|$ and
$|K-M|$. The former provides the number of the highest algebraic eigenenergies
of the hyperbolic top 
which are not coincident with any of the trigonometric top eigenenergies. The latter
gives the number of non-normalizable solutions in between the regions of
branch $\sqrt{\zeta}$ and $-\sqrt{\zeta}$, i.e., the above
mentioned spectral gap denoted by  grey crosses and running diagonally from the top
left to the bottom right.
As shown below for the configurations $(K,M)= (1,1), (1,0)$, this measure is valid for all $K,M \in
\mathbb{Z}$.

These observations are consistent with Eqs.
\eqref{eq:alg-sols-trig_nr} and
\eqref{eq:alg-sols-hyperb_nr}: the overall number of normalizable algebraic
solutions is $20$ for the trigonometric top and $19$ for the hyperbolic one. Besides, there are $20$ non-normalizable solutions for either case.

\mathversion{bold}
\subsection{$K=1,M=1$ }
\begin{figure}
	\includegraphics[width=.7\textwidth]{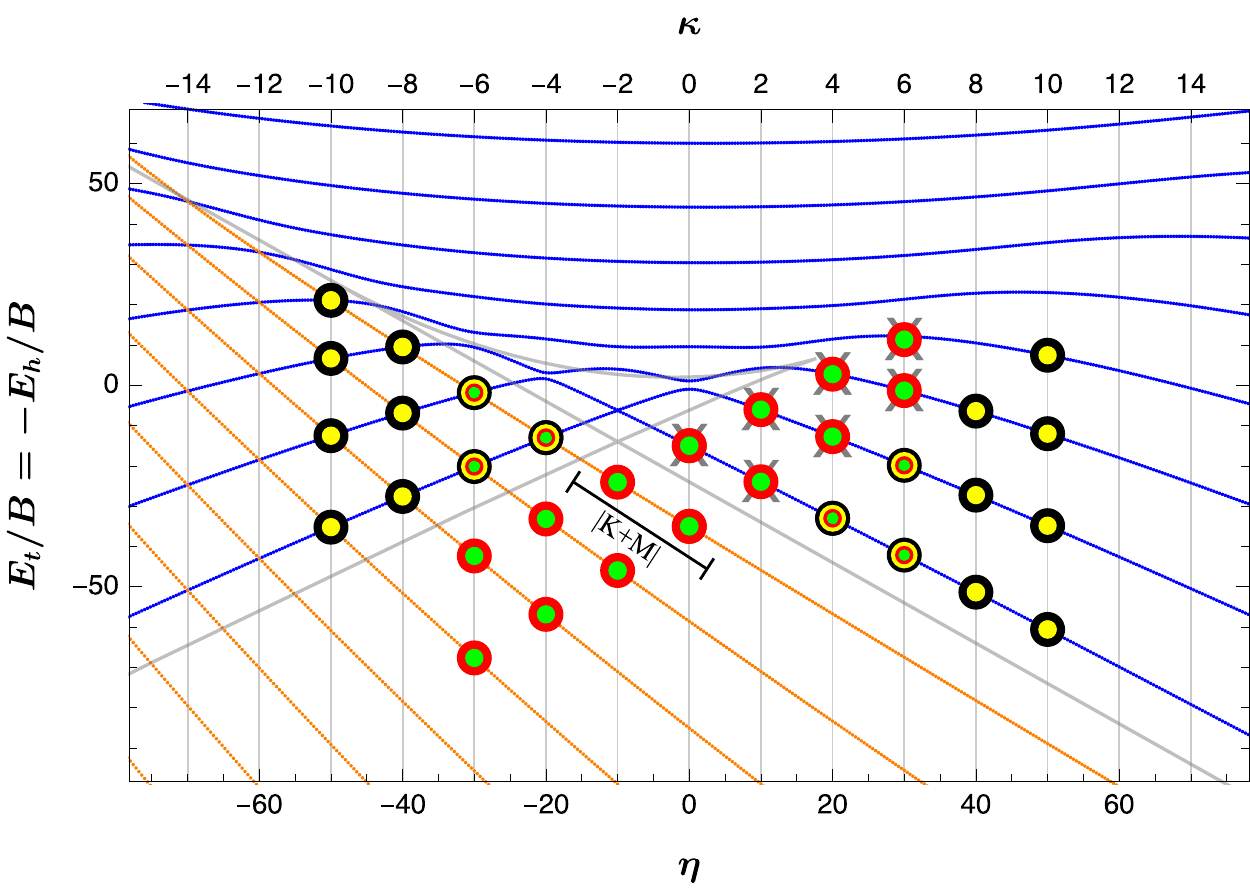}
	\caption{Numerical spectra for $M=1$, $K=1$ and $\zeta=25$, $\rho=0$:
	trigonometric top (blue), hyperbolic top (orange). 
	Black: sector $1_{+}$ (here $\kappa \geq 4$), sector $1_{-}$ ($\kappa \leq
	-4$); Yellow:
	sector $2_{+}$ ($\kappa \geq 4$), sector $2_{-}$ ($\kappa \leq -4$); 
	Red: sector $3_{+}$ ($\kappa \geq 0$), sector $3_{-}$ ($\kappa \leq 0$);
	Green:
	sector:  
	$4_{+}$ ($\kappa \geq 0$), sector $4_{-}$ ($\kappa \leq 0$). Grey crosses mark non-normalizable solutions.
Grey curve shows local minima and maxima of the symmetric top potential, Eq.
	\eqref{eq:org-effective-pot_1-dim}. For computational details, see Appendix \ref{subsec:matrix-els-trig-top_for-num} and \ref{subsec:matrix-els-hyp-top_for-num}.}
	\label{fig:spectrum_num-alg_sectors_m1k1}
\end{figure}

\begin{figure}
	\centering
		\subfloat{ 
			\includegraphics[width=.35\textwidth]{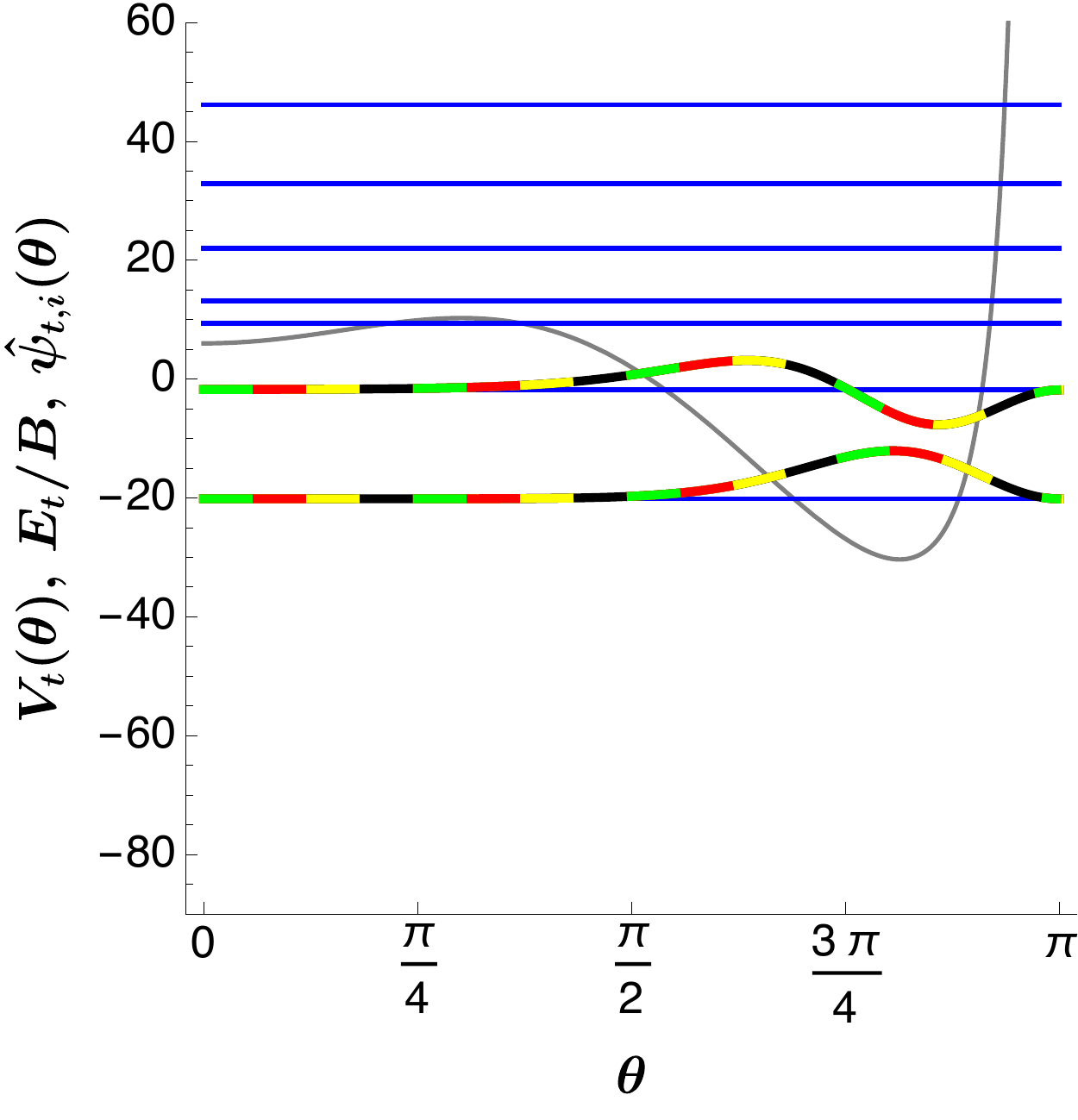}
			}
		\hspace{3mm} 
		\subfloat{ 
			\includegraphics[width=.35\textwidth]{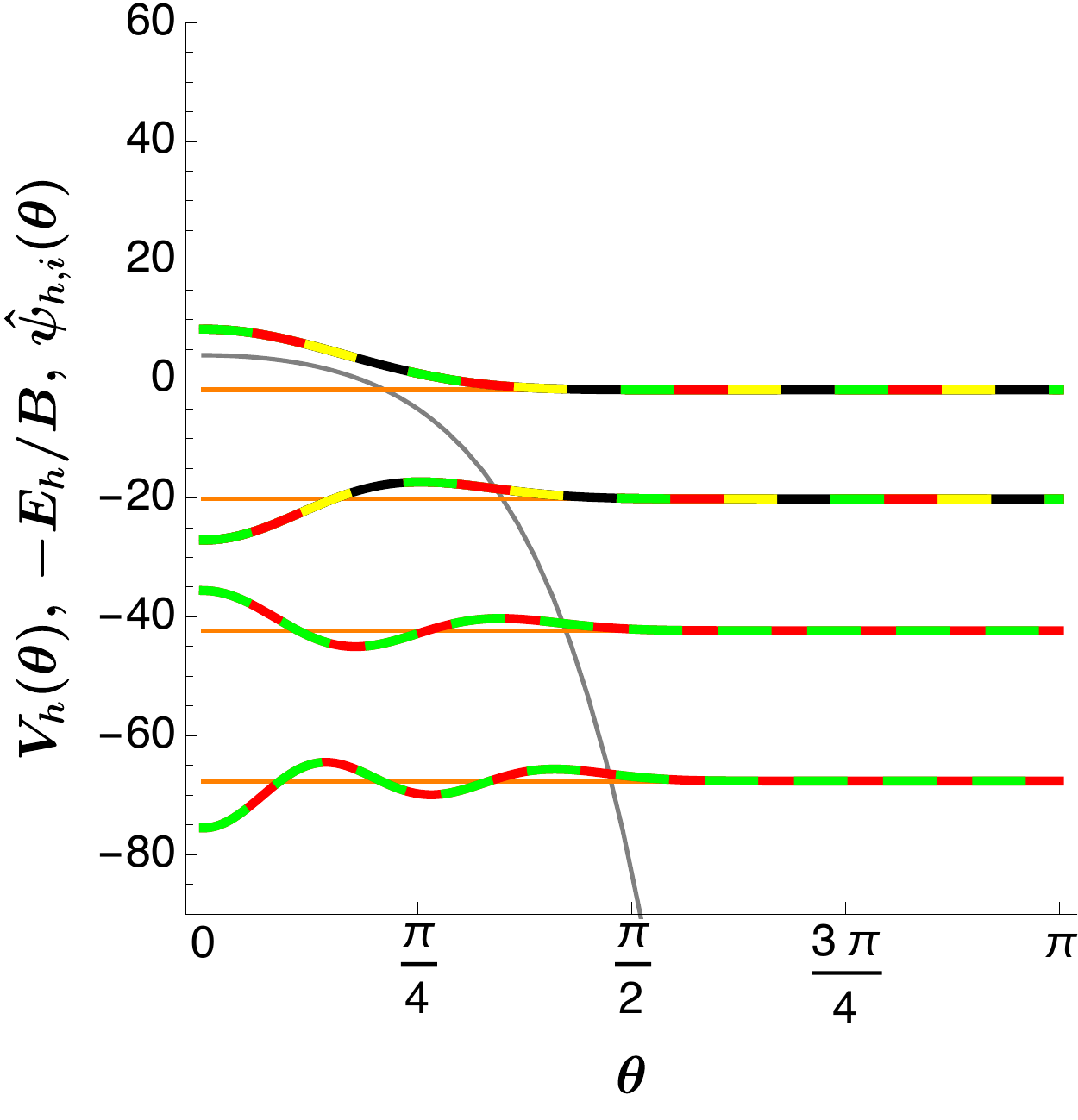}
			}
		\caption{Normalizable closed-form sample wavefunctons $\hat{\psi}_{t,i}
		(\theta)$ (left) and  $\hat{\psi}_{h,i}
		(\theta)$ (right). Black: sector $1_{\pm}$, yellow sector $2_{\pm}$, red sector $3_{\pm}$, and green sector
		$4_{\pm}$.  Blue:
		numerical eigenenergies for the trigonometric top (left). Orange: numerical eigenenergies for the hyperbolic top
		(right). Potentials shown in grey. Both plots: $M=1$, $K=1$, $\zeta=25$, $\eta=-30$.
		Same color-coding as in Fig.
		\ref{fig:spectrum_num-alg_sectors_m1k1}. }
		\label{fig:wavefuncs_m1-k1-zeta25-eta-40}
\end{figure}

\mathversion{normal}
The corresponding eigenenergies are shown in Fig.
\ref{fig:spectrum_num-alg_sectors_m1k1} and a sampling of the eigenfunctions in Fig. \ref{fig:wavefuncs_m1-k1-zeta25-eta-40}. There are $20$ closed-form
solutions for the trigonometric and $17$ for the hyperbolic top, in agreement with Eqs.  \eqref{eq:alg-sols-trig_nr} and
\eqref{eq:alg-sols-hyperb_nr}, respectively.  In addition,
we found $7$ non-normalizable closed-form solutions for the symmetric top.


%
\begin{figure}
	\includegraphics[width=.7\textwidth]{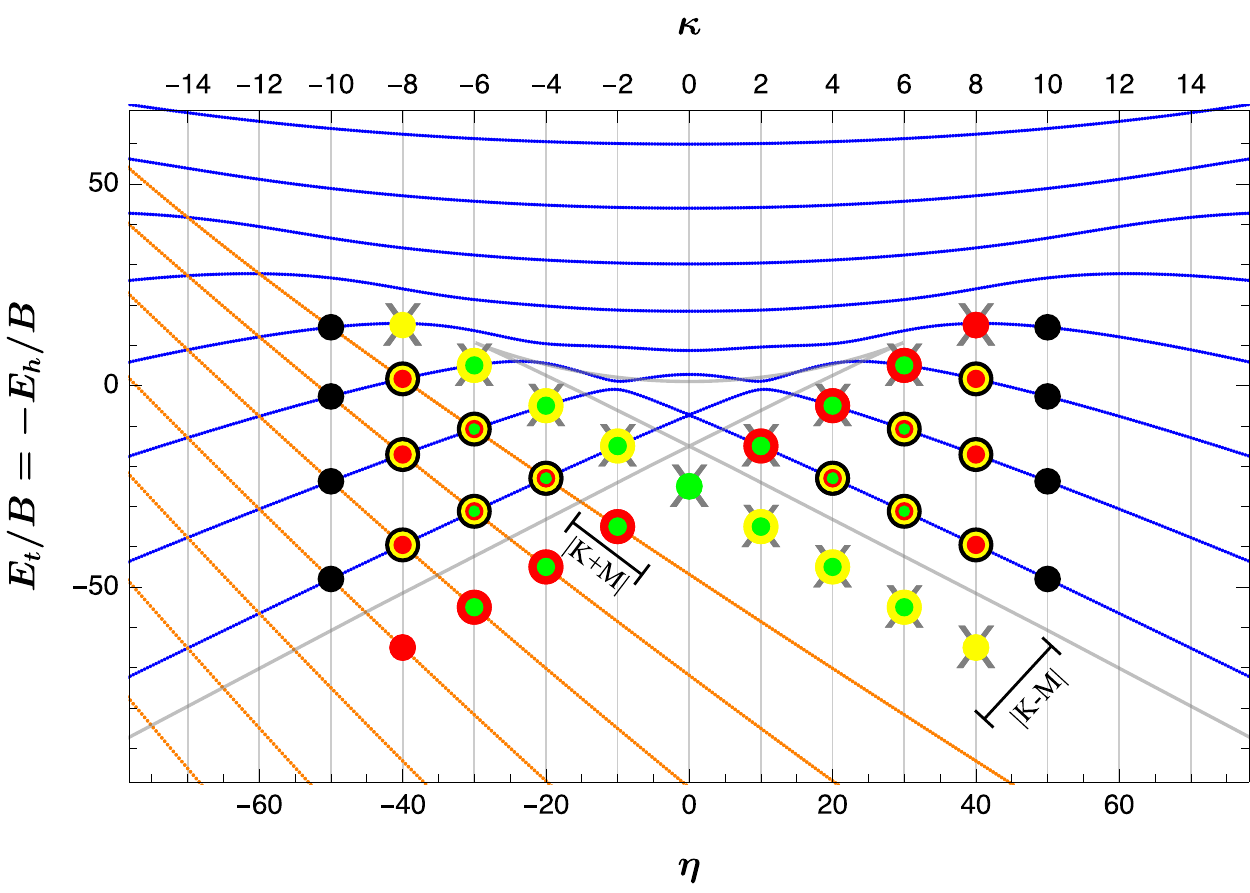}
	\caption{Numerical spectra for $M=1$, $K=0$ and $\zeta=25$, $\rho=0$:
	trigonometric top (blue), hyperbolic top (orange). 
	Black: sector $1_{+}$ (here $\kappa \geq 4$), sector $1_{-}$ ($\kappa \leq
	-4$); Yellow:
	sector $2_{+}$ ($\kappa \geq 2$), sector $2_{-}$ ($\kappa \leq -2$); 
	Red: sector $3_{+}$ ($\kappa \geq 2$), sector $3_{-}$ ($\kappa \leq -2$);
	Green:
	sector
	$4_{+}$ ($\kappa \geq 0$), sector $4_{-}$ ($\kappa \leq 0$). Grey crosses mark non-normalizable solutions.
Grey curves show local minima and maxima of the spherical pendulum potential, Eq. 
	\eqref{eq:org-effective-pot_1-dim}. For computational details, see Appendix \ref{subsec:matrix-els-trig-top_for-num} and \ref{subsec:matrix-els-hyp-top_for-num}.}
	\label{fig:spectrum_num-alg_sectors_m1k0} 
\end{figure} 

 \begin{table}[h]	
\caption{ \label{table:sectors-rel-spheric-pends-symms}
	Correlations between sectors and the algebraic
	seed-state cases obtained from SUSY QM in \cite{SchmiFri2015a} for the spherical pendulum ($K=0$). Note that in our earlier work the seed states $\hat{\psi}_{t,0}(\theta)$ were presented in gauged
	form, $\psi_{t,0}(\theta) = \hat{\psi}_{t,0}(\theta) \sqrt{\sin \theta}$,
	cf. Eq.
	\eqref{eq:wavefunc-3d-gauge}. }
\vspace{1mm}
\setlength{\tabcolsep}{.9em}
\begin{tabularx}{.7\linewidth}{l l l l }
	\hline 
			Sector 
		&  Case   
		& $\kappa$
		& $\psi_{t,0}(\theta) $ 
	\\
	\hline\hline
			$1_{+}$
		&   $1_-$ 
		& $2(M+1)$ 
		& $(\sin\theta)^{M+ \frac{1}{2}}
		\mathrm{e}^{\sqrt{\zeta/B}\cos\theta}$
	\\
		$2_{+}$  
		&  $2_+$ 
		& $2$ 
		& $(\sin\theta)^{ \frac{1}{2}}(\tan\theta)^{-M}
		\mathrm{e}^{\sqrt{\zeta/B}\cos\theta}$
	\\
		$3_{+}$
		&   $2_-$  
		& $2$ 
		& $(\sin\theta)^{ \frac{1}{2}}(\tan\theta)^{M}
		\mathrm{e}^{\sqrt{\zeta/B}\cos\theta}$ 
	\\
		$4_{+}$ 
		& $1_+$ 
		&  $2(1-M)$  
		& $(\sin\theta)^{-M+ \frac{1}{2}}
		\mathrm{e}^{\sqrt{\zeta/B}\cos\theta}$
	\\
	 \hline
\end{tabularx}
\end{table}

\begin{figure}
	\includegraphics[width=.7\textwidth]{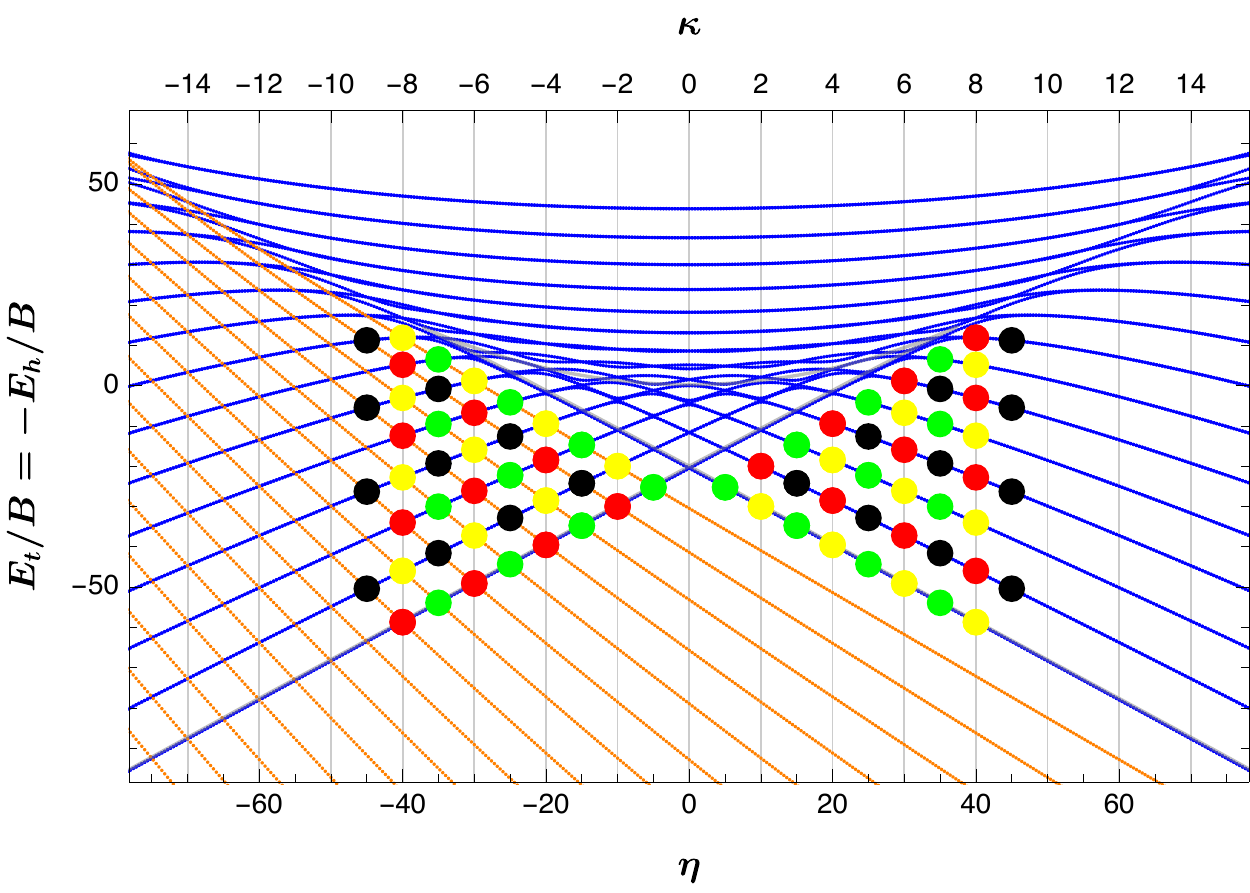}
	\caption{Numerical spectra for $M=1/2$, $K=0$ and $\zeta=25$, $\rho=0$:
	trigonometric top/planar pendulum (blue), hyperbolic top/Razavy system
	(orange).
	Black: sector $1_{+}$ (here $\kappa \geq 3$), sector $1_{-}$ ($\kappa \leq
	-3$); Yellow:
	sector $2_{+}$ ($\kappa \geq 2$), sector $2_{-}$ ($\kappa \leq -2$); 
	Red: sector $3_{+}$ ($\kappa \geq 2$), sector $3_{-}$ ($\kappa \leq -2$);
	Green:
	sector:
	$4_{+}$ ($\kappa \geq 1$), sector $4_{-}$ ($\kappa \leq -1$).
	Grey curves show the local minima and maxima of the trigonometric top potential, Eq.
	\eqref{eq:org-effective-pot_1-dim}. For computational details, see Appendix \ref{subsec:matrix-els-trig-top_for-num} and \ref{subsec:matrix-els-hyp-top_for-num}.}
	\label{fig:spectrum_num-alg_sectors_m0p5k0}  
\end{figure}
 
 \begin{table}[h]	
\caption{ \label{table:sectors-rel-planar-pends-symms}
Planar pendulum (left) and the Razavy system (right) as special cases of the trigonometric and hyperbolic tops, respectively, for $(K,M)=(0,1/2)$: correlations
between algebraic sectors and the irreducible representation $\Gamma_{t,h}$ of symmtery
groups from Ref. \cite{b_friedrich2017}. Note that in our earlier work the seed functions $\hat{\psi}_{t,0}(\theta)$ were presented in gauged
	form, $\psi_{t,0}(\theta) = \hat{\psi}_{t,0}(\theta) \sqrt{\sin \theta}$,
	cf.
	Eq. \eqref{eq:wavefunc-3d-gauge}.}
\vspace{1mm}
\setlength{\tabcolsep}{.85em}
\begin{tabularx}{.85\linewidth}{l l l l | l l l l}
	\hline 
			\multicolumn{4}{l|}{\text{Planar pendulum}}	 		
	 	&	\multicolumn{4}{l}{\text{Razavy system}}
	\\   
			Sector 
		&  $\Gamma_t$
		& $\kappa$ 
		& $\psi_{t,0}(\theta)$ 
		& Sector 
		& $\Gamma_h$
		& $\kappa$ 
		& $\psi_{h,0}(\theta)$ 
	\\
	\hline\hline
			$1_{+}$
		&   $A_2$ 
		& $3$ 
		& $\sin\theta\mathrm{e}^{\sqrt{\zeta/B}\cos\theta}$ 
		&
			\multirow{2}{*}{$3_{-}$} 
		& \multirow{2}{*}{$A''$} 
		& \multirow{2}{*}{2} 
		& \multirow{2}{*}{$\sinh\frac{\theta}{2}\mathrm{e}^{-\sqrt{\zeta/B}
		\cosh \theta}$}
	\\
		$2_{+}$  
		&  $B_1$ 
		& $2$ 
		& $\cos\frac{\theta}{2}\mathrm{e}^{\sqrt{\zeta/B}\cos\theta}$ 
		& 
		& 
		& 
		& 
	\\
		$3_{+}$
		&   $B_2$ 
		& $2$ 
		& $\sin\frac{\theta}{2}\mathrm{e}^{\sqrt{\zeta/B}\cos\theta}$ 
		& \multirow{2}{*}{$4_{-}$} 
		& \multirow{2}{*}{$A'$} 
		& \multirow{2}{*}{1}  
		& \multirow{2}{*}{$\mathrm{e}^{-\sqrt{\zeta/B} \cosh \theta}$} 
	\\
		$4_{+}$ 
		& $A_1$ 
		&  $1$  
		&  $\mathrm{e}^{\sqrt{\zeta/B} \cos\theta}$ 
		&  
		&  
		&   
		&
	\\
	 \hline
\end{tabularx}
\end{table}


\mathversion{bold}
\subsection{$K=0$}
\mathversion{normal}
For the trigonometric top with $n=0$ and
$\kappa = 2 k$, the case of $K=0$ coincides with that of the {\it spherical pendulum}, as investigated via SUSY QM in our previous work  \cite{LemMusKaisFriPRA2011,LemMusKaisFriNJP2011,SchmiFri2014a,SchmiFri2015a}.
The closed-form solutions listed in Ref. \cite{SchmiFri2015a} can now be directly retrieved from Table \ref{tab:sols1}. Table \ref{table:sectors-rel-spheric-pends-symms} provides their summary as well as a correlation between the algebraic sectors and the seed-state cases  introduced in Ref. \cite{SchmiFri2015a}.

Fig. \ref{fig:spectrum_num-alg_sectors_m1k0} displays the spectral
pattern for $(K,M)=(0,1)$ as the choice  example. We find, again, that, in agreement with Eqs. \eqref{eq:alg-sols-trig_nr} and
\eqref{eq:alg-sols-hyperb_nr}, there are $20$ normalizable
closed-form solutions for the trigonometric and $14$ for the
hyperbolic top. Note that for $n=0$ only the sector $1_+$ solution ($\kappa=4$) is
normalizable. Compared with our previous SUSY QM work as summarized in Ref. \cite{SchmiFri2015a}, the present work based on the QHJ theory delivers
ten times as many closed-form solutions (for $M>0$).


\mathversion{bold}
\subsection{$K=0,M=1/2$ }
\mathversion{normal}
The non-physical choice  of $M=1/2$ for a symmetric top makes it possible to  retrieve all our previous results concerning the quasi-solvability of the planar pendulum (in the trigonometric case) and the Razavy system (in the hyperbolic case), as reported in Refs. \cite{schmidt2014b,b_friedrich2017}. 
 
Our previous work 
\cite{schmidt2014b,b_friedrich2017} examined  the spatial symmetry classes and irreducible representations $\Gamma_{t,h}$ of  both the planar pendulum and the Razavy systems.  Table \ref{table:sectors-rel-planar-pends-symms} lists the closed-form solutions for $(K,M)=(0,1/2)$ as retrieved from Table \ref{tab:sols1} for $n \geq 0$ 
and provides a correlation between the algebraic sectors and the ireducible
representations of Ref. \cite{b_friedrich2017}. Note that in the pendulum case, the domain of the polar angle $\theta$ becomes $[0, 2\pi )$.

The energy levels of the planar pendulum and the Razavy systems together with the classification of the spectral patterns according to their  algebraic sectors are shown in Fig. \ref{fig:spectrum_num-alg_sectors_m0p5k0}. Compared with our previous work,  the figure includes  negative $\eta$ values.
Since for $(K,M)=(0,1/2)$
the topological index $\kappa$ can take both odd and even integer values, one has 
to multiply the number of solutions as given by Eq. \eqref{eq:alg-sols-trig_nr} by a factor of four.
Thus, for the planar pendulum, there are $40$ normalizable closed-form solutions in the
$\sqrt{\zeta}$ and $-\sqrt{\zeta}$ branch each and 
 $40$ normalizable closed-form solutions for the Razavy system.
%


\section{Conclusions}	

We have examined conditional quasi-solvability of a symmetric top subject to collinear orienting and aligning interactions -- the symmetric top quantum pendulum. We did so by invoking the Quantum Hamilton-Jacobi (QHJ) theory. Although the symmetric top quantum pendulum  represents a considerable generalization of the planar and spherical pendula, 
we have been able not only to retrieve the closed-form solutions identified previously for the special planar \cite{b_friedrich2017} and spherical  \cite{LemMusKaisFriPRA2011,LemMusKaisFriNJP2011,SchmiFri2015a} cases, but also to find a number of new solutions, including those for the generalized symmetric top case.  For the spherical pendulum case alone we have found ten times as many solutions as were  known from our previous work based on Supersymmetric Quantum Mechanics (SUSY QM).

Our previous work within the framework of SUSY QM  required an ``educated guess'' for the superpotential. A complexified superpotential makes also an appearance in the QHJ theory -- as the quantum momentum function (QMF). However, QHJ provides  a recipe for constructing the QMF, although it still needs ``an educated guess'' concerning the choice of suitable coordinates for expressing the closed-form solutions. This latter task is much less ad hoc than the former one and therefore easier to fulfil.

Apart form the trigonometric symmetric top, we have also tackled its anti-isospectral counterpart, the hyperbolic top, and examined the patterns of the algebraic and numerical spectra of both types of top for special choices of the projection quantum numbers $K$ and $M$.
We found that the distribution patterns of the algebraic
spectra from the different algebraic sectors can be charactereized by the quantum number
measures $| K + M |$ and $| K - M |$, see Figs.
\ref{fig:spectrum_num-alg_sectors_m2k1}, \ref{fig:spectrum_num-alg_sectors_m1k1}, and
 \ref{fig:spectrum_num-alg_sectors_m1k0}. An inspection of these figures reveals that the general pattern of the $4+4$ triangles (one for each of the four algebraic sectors/colors and type of top) is independent of $K$ and $M$. However, the triangles are shifted with respect to each other, with the shift depending  on both $|K-M|$ and $|K+M|$. These shifts in fact govern how many  and which solutions coincide, thus determining the total number of independent solutions. Moreover, as illustrated by the said figures, these shifts are characteristic for the transformation from the  symmetric to the spherical to the planar top. 
 Note that in none of the Figs. \ref{fig:spectrum_num-alg_sectors_m2k1}, \ref{fig:spectrum_num-alg_sectors_m1k1}, and
 \ref{fig:spectrum_num-alg_sectors_m1k0} are there more than four normalizable solutions at a given $\kappa$. But for the special case of $M=1/2$, there are up to $8$ solutions at a given $\kappa$, see Fig. \ref{fig:spectrum_num-alg_sectors_m0p5k0}.
 
In order to evaluate the physical relevance (i.e.,  normalizability) of the solutions found, we applied the limit-point and limit-circle classification and found a  condition that identifies bounded square integrable solutions. 
	
As noted in our previous work, the topological index $\kappa=\frac{\eta}{\sqrt{B\,\zeta}}$  labels the genuine and avoided crossings of the eigenenergy levels. 
According to, e.g., Eq. \eqref{eq:qmf-fixed-term_reform} and Table \ref{table:qmfs}, the
  	$\eta$-spacing between the level crossings, i.e., $2 \sqrt{\zeta}$ for $K,M$
  		integer and $\sqrt{\zeta}$ for $K,M$ half-integer, is identical
  		to the boundary at infinity of the rationalized QMF $\tilde{p}_f(z)$ on the
		  complexified domain $\mathcal{D}_c(z)$ and, respectively, to the boundary at
		  zero of the likewise rationalized QMF $\hat{p}_f(w)$ on $\mathcal{D}_c(w)$
		  with the mapping $z \mapsto w = 1/z$ between the domains. 
Like the quasi-solvability conditions, this boundary stems from the
		  domain compactification described in Subsection \ref{subsec:fixed-qmf-constr}.
		  It would be interesting to see whether this is a general feature of all
		  periodic Hamiltonians and their ``dual'' anti-isospectral hyperbolic partners. Of interest is also the question of whether  one could find any rules 
for determining the number of accessible exact solutions for $n_{max} > 3$.

The spectral properties at integer values of $\kappa$ -- and hence the quasi-exact solvability -- may be difficult to verify experimentally, mainly due to the uncertainties in setting the strengths of the electrostatic ($\eta$) and radiative ($\zeta$) fields. The interaction parameters for representative molecules are listed  in Table I of Ref. \cite{SchmiFri2014a} while Table II of Ref. \cite{SchmiFri2014a} provides the conversion factors needed to obtain the dimensionless reduced parameters from the molecular parameters expressed in customary units.

However, there are other features in the numerically calculated spectra of Figs. \ref{fig:spectrum_num-alg_sectors_m2k1}, \ref{fig:spectrum_num-alg_sectors_m1k1},
 \ref{fig:spectrum_num-alg_sectors_m1k0}, and
 \ref{fig:spectrum_num-alg_sectors_m0p5k0} that may be amenable to experimental testing. While all energy levels of the hyperbolic top are monotonous functions  of $\eta$, the lower curves for the trigonometric top exhibit extrema.  These become more numerous -- and also arise for the higher lying levels -- at larger values of $\zeta$. While these extrema are less pronounced for $K\neq 0$ (see Figs. \ref{fig:spectrum_num-alg_sectors_m2k1} and \ref{fig:spectrum_num-alg_sectors_m1k1}), they become more conspicuous in the spherical pendulum case, cf. Fig. \ref{fig:spectrum_num-alg_sectors_m1k0}. As detailed in our previous work \cite{SchmiFri2014a,SchmiFri2015a}, the loci of the extrema,  where all states energetically higher than the analytic solutions exhibit avoided crossings for a large-enough $\zeta$, coincide with the conditions of quasi-solvability (even integer values of $\kappa$).
A similar situation arises for the planar pendulum, Fig. \ref{fig:spectrum_num-alg_sectors_m0p5k0}, where again the conditions of quasi-solvability (integer values of $\kappa$) coincide with the loci of genuine (odd $\kappa$) or avoided (even $\kappa$) intersections of the higher states; note that the latter occur only for a large-enough $\zeta$ \cite{schmidt2014b,b_friedrich2017}.

The spectra of the spherical and  planar pendula are always such that the bottom part of the spectra for the values of $\kappa$ fulfilling the QS condition consists of single states whereas all higher states are found to form genuine or avoided crossings for large $\zeta$. This general pattern could in principle be confirmed in spectroscopic experiments -- by scanning  the $\eta$ or $\zeta$ parameters of our trigonometric model Hamiltonian, Eq. \eqref{eq:ham_org-phys_symm-top_3-dim_separ}. 

We note that the algebraic solutions found may serve as benchmarks for numerical analysis and the polynomial Ansatz they suggest could be useful for numerical calculations beyond quasi-solvability, i.e., for non-integer values of the topological index $\kappa$.
		  
In our ongoing work, we deal with the case of symmetric tops subject to non-collinear orienting and aligning interactions -- a much harder problem, both analytically and numerically.

	
\section*{Appendix}

\subsection{Quantum Hamilton-Jacobi equation in rational form}
\label{subsec:rational}

In order to be able to make use of the Laurent series expansion and the residue calculus for solving the quantum Hamilton-Jacobi equation \eqref{eq:qhj-eq}, we need to ensure that all terms besides those which contain the quantum momentum function are rational.

To this end, we first  transform the variable $\theta$ into a new variable $z = z(\theta)$ such that only rational terms in the new variable remain. The sought transformation is induced by the mapping
(cf. Ref. \cite{bank1981})
\begin{align} \label{eq:qmf_transf}
	p(\theta) = 
	\left.-\frac{i}{\theta'(z)} \left[
		\tilde{p}(z) 
		+ 
		\frac{1}{2} \sqrt{B}\, \ln \left(  \theta'(z) 
		\right) '
	\right]\right|_{z=z(\theta)}
\end{align}
of the new meromorphic quantum momentum function $\tilde{p}(z)$ and leads to 
\begin{align} \label{eq:qhj-eq_transf}
	\tilde{p}(z)^2 + \sqrt{B}\tilde{p}'(z) 
	+ 
	\frac{1}{2} B \left[
		 \frac{\theta'''(z)}{\theta'(z)}
		-
		\frac{3}{2}\, \frac{{\theta''(z)}^2}{{\theta'(z)}^2}				
	\right] 
	+
	{\theta'(z)}^2 \left(
		E - V(\theta(z))
	\right)
	=
	0 \, 
\end{align}
which can be viewed as a \textit{normal form} \cite{bank1981} of the Riccati
equation \eqref{eq:qhj-eq} with the inverse transformation of
coordinates\footnote{Multivaluedness, if existing, has to be taken in to account, unless it only arises from a phase shift of the form $e^{i \pi \nu}$ with $\nu$ integer.} $\theta = \theta(z)$. 

Eq.\eqref{eq:qhj-eq_transf} can be recast in terms of the Schwarzian
derivative,
\begin{align}
	\mathrm{S}(\theta)(z) = \left(
		\frac{\theta''(z)}{\theta'(z)}
	\right)'
	-
	\frac{1}{2} \left(
		\frac{\theta''(z)}{\theta'(z)}
	\right)^2 \, 
\end{align}
with the result
\begin{align} \label{eq:qhj-eq_transf2}
	\tilde{p}(z)^2 + \sqrt{B}\, \tilde{p}'(z) 
	+
	\frac{1}{2} B\, \mathrm{S}(\theta)(z)
	+
	{\theta'(z)}^2 \left(
		E - V(\theta(z))
	\right)
	=
	0 \, 
\end{align}
According to, e.g., Refs. \cite{lopez1994,ullate2007}, all coordinate transformations
$z=z(\theta)$ are admissible for which $r(y)$ in 
\begin{align}
	\theta(z) = \int^z \frac{\mathrm{d}y}{\sqrt{r(y)}}
\end{align}
are polynomials in $y$ of degree smaller or equal to $4$.
Thus all terms $\theta'(z)^2$ and
$\theta^{(n)}(z)/\theta^{(m)}(z)$, with $n$, $m$ the orders of the derivatives, 
must be rational functions, including $\mathrm{S}(\theta)(z)$.
By ensuring that also
$V_t(\theta(z))$ be a rational function via a suitable choice of $z(\theta)$, 
we obtain rational solutions $\tilde{p}(z)$ of Eq.
\eqref{eq:qhj-eq_transf2}.

We note that any M\"obius transformation $\frac{\alpha z(\theta) +\beta}{\gamma
z(\theta) + \delta}$ of the new coordinate $z$ will leave the terms in Eq.
\eqref{eq:qhj-eq_transf2} rational, since the Schwarzian derivative is invariant
under M\"obius transformations.

By making the \textit{Ansatz} 
\begin{align}
	\tilde{\psi}(z)= e^{ \frac{1}{\sqrt{B}} \int^z
	\tilde{p}(y) \mathrm{d}y}
\end{align}
for the wavefunction of the transformed problem, we can 
derive from Eq. \eqref{eq:qhj-eq_transf2} a new Schr\"odinger-type equation,
\begin{align} \label{eq:new-schrod-type-eq_general}
	- B \frac{1}{\theta'(z)^2}\tilde{\psi}''(z)
	+ \left[
		-\frac{1}{2} B\, \frac{\mathrm{S}(\theta)(z)}{{\theta'(z)}^2}
		+
		V(\theta(z))
	\right] \tilde{\psi}(z)
	=
	E \tilde{\psi}(z)	 \, 
\end{align}
The old wavefunction, $\psi(\theta)$, is then related to the new one,
$\tilde{\psi}(z)$, via
\begin{align} \label{eq:alg-sols_first-order_transf-new-old}
	\psi(\theta) = 
		\left. \tilde{\psi}(z) \sqrt{ \theta'(z) }
				\right|_{z=z(\theta)}		\, 	
\end{align}
which is equivalent to Eq. \eqref{eq:qmf_transf}
in that it induces the transformation from the original Schr\"odinger equation
\eqref{eq:ham_org-phys_symm-top_1-dim} to the new equation.
\eqref{eq:new-schrod-type-eq_general} with purely rational terms
 like Eq. \eqref{eq:qmf_transf} induces the transformation from the 
 original QHJ equation \eqref{eq:qhj-eq} to its normal Riccati form
 \eqref{eq:qhj-eq_transf2}.

By identifying the rational function
\begin{align} \label{eq:new-qhj-pot_general}
\tilde{V}(z) = 
	V(\theta(z))
	- \frac{1}{2} B\,
	\frac{\mathrm{S}(\theta)(z)}{{\theta'(z)}^2} 
			 \, 
\end{align}
as the new potential, we can recast Eq. \eqref{eq:qhj-eq_transf2} in a compact, rational form:
\begin{align} \label{eq:qhj-eq_transf-compact2}
	\tilde{p}(z)^2 + \sqrt{B}\, \tilde{p}'(z) 
	+
	{\theta'(z)}^2 \left[
		E - \tilde{V}(\theta(z))
	\right]
	=
	0 \, 
\end{align}



\subsection{Matrix elements for determining the numerical spectra of the trigonometric top}
\label{subsec:matrix-els-trig-top_for-num}

The numerical spectra in
Figs. \ref{fig:spectrum_num-alg_sectors_m1k1},
 \ref{fig:spectrum_num-alg_sectors_m1k0}, and
 \ref{fig:spectrum_num-alg_sectors_m0p5k0} were determined by diagonalizing the matrix representation of  Hamiltonian \eqref{eq:ham_org-phys_symm-top} in the symmetric top basis set 
\begin{align}
	| J K M \rangle = (-1)^{M-K} \sqrt{\frac{2 J+1}{8 \pi^2}} D_{-M -K}^J
\end{align}
with $D_{M
K}^{J}$  the Wigner D-matrices \cite{haertelt2008}. Note that
 $J \geq \mathrm{max} \{K, M\}$.
 
The potential energy terms, Eq. \eqref{eq:ext-pot}, can be
recast  in terms of the Wigner D-matrices as
\begin{align}\label{eq:cos_wigner-d-mat}
	\cos \theta ={}& D_{0 0}^1
\end{align}
and
\begin{align}\label{eq:cos2_wigner-d-mat}
	\cos^2 \theta ={}& \frac{2}{3} D_{0 0}^2 +\frac{1}{3} D_{0 0}^0 \, 
\end{align}

The matrix elements of the $\cos \theta$ and $\cos^2\theta$ operators, Eqs. \eqref{eq:cos_wigner-d-mat} and 
\eqref{eq:cos2_wigner-d-mat}, were obtained via the Gaunt-integral in terms of the $3$-$j$ symbols
\begin{align} \notag
	& \int_0^{2 \pi}\int_0^{\pi}\int_0^{2 \pi} D_{M_3
	K_3}^{J_3}(\varphi,\theta,\chi) D_{M_2 K_2}^{J_2}(\varphi,\theta,\chi) D_{M_1
	K_1}^{J_1}(\varphi,\theta,\chi)\,\mathrm{d}\phi\, \sin \theta\,
	\mathrm{d}\theta\, \mathrm{d}\chi\\
	={}& 8 \pi^2 \left(
	\begin{array}{ccc}
		J_1 & J_2 & J_3 \\
		M_1 & M_2 & M_3 
	\end{array} 
	\right)
	\left( 
	\begin{array}{ccc}
		J_1 & J_2 & J_3 \\
		K_1 & K_2 & K_3 
	\end{array} 
	\right)
\end{align}
with the result:
\begin{align}\notag
\langle J' K' M' | \cos \theta | J K M \rangle ={}& 
	\sum_{j = -1}^1
	(-1)^{M-K} \frac{\sqrt{(2 J+1)(2 (J + j) + 1)}}{8 \pi^2}\\
	& \times \left( 
	\begin{array}{ccc}
		J & 1 & J + j \\
		-M & 0 & M 
	\end{array} 
	\right)
	\left( 
	\begin{array}{ccc}
		J & 1 & J + j \\
		-K & 0 & K 
	\end{array} 
	\right)	
	\delta_{J', (J+j)}\delta_{M', M}\delta_{K', K}
\end{align}
and
\begin{align} \notag
\langle J' K' M' | \cos^2 \theta | J K M \rangle ={}& 
	\sum_{j = -2}^2
	(-1)^{M-K}
	\frac{\sqrt{(2 J+1)(2 (J + j) + 1)}}{8 \pi^2}\\\notag
	&\times
	\left[
		\frac{2}{3} 
		\left( 
		\begin{array}{ccc}
			J & 2 & J + j \\
			-M & 0 & M 
		\end{array} 
		\right)
		\left( 
		\begin{array}{ccc}
			J & 2 & J + j \\
			-K & 0 & K 
		\end{array} 
		\right)		\right.\\  \notag
		& \hspace*{5mm} \left.\vphantom{}	
		+
		\frac{1}{3} 
		\left(
		\begin{array}{ccc}
			J & 0 & J + j \\
			-M & 0 & M 
		\end{array} 
		\right)
		\left( 
		\begin{array}{ccc}
			J & 0 & J + j \\
			-K & 0 & K 
		\end{array} 
		\right)
	\right]\\
	&\times \delta_{J', (J+j)}\delta_{M', M}\delta_{K', K} \, 
\end{align}
Ultimately, the matrix elements of  Hamiltonian $\mathcal{H}$ of
\eqref{eq:ham_org-phys_symm-top} take the analytic form
\begin{align} \notag
\langle J' K' M' | \mathcal{H} | J K M \rangle ={}& 
	B (J (J+1) +\rho K^2) \delta_{J', J} \delta_{M', M}\delta_{K', K}\\
	&-\eta  \langle J' K' M' | \cos \theta | J K M \rangle
	- \zeta \langle J' K' M' | \cos^2 \theta | J K M \rangle \, 
\end{align}  
  
  
\subsection{Calculation of the numerical spectra of the hyperbolic top} 
\label{subsec:matrix-els-hyp-top_for-num}
For the numerical evaluation of the spectrum of the hyperbolic Hamiltonian of Eq. \eqref{eq:ham_org-phys_symm-top_1-dim_hyperb} on  $(0,\infty)$ for $K=M=1$, a  rapidly decaying sequence of basis functions is desirable. We circumvent this by making the substitution $x=\cosh(\theta)$ that leads to an equation on $(1,\infty)$ and a more robust numerical approximation of the spectrum. 

It turned out to be advantageous to choose a set of linearly independent trial functions $\varphi_i(x):=\frac{e^{-\sqrt{\zeta}x}}{1+x} x^i$ whose exponential decay coincided with the one of the known eigenfunctions.
With these, we were able to construct a symmetric matrix approximation $H_h^N$ of the Hamiltonian by choosing a finite cut-off $N \in \mathbb{N}$ such that for $i,j \in \left\{1,...,N \right\}$ the matrix elements read
\begin{equation}
\langle i \vert H_h^N \vert j \rangle=\int_{0}^{\infty} \overline{ \varphi_i(\cosh(\theta))} H_h \varphi_j(\cosh(\theta)) \sinh(\theta) d\theta 
\end{equation}
Such matrix elements can be computed using standard numerical integrators. Due to our particular choice of basis functions, 
the singularity of the potential at $\theta=0$ was removed and the integral 
was evaluated over smooth functions only.
The price to pay was that the functions $\varphi_i$ do not form an orthonormal basis. Thus, an additional symmetric positive-definite overlap matrix $\mathcal{O}$ with matrix elements 
\begin{equation}
\langle i \vert \mathcal O \vert j \rangle=\int_{1}^{\infty} \overline{ \varphi_i(x)} \varphi_j(x) dx 
\end{equation}
had to be computed numerically.
The energies of the hyperbolic pendulum were then obtained as the eigenvalues of the generalized problem
\begin{equation}
H_h^N v=\lambda \mathcal{O}v
\end{equation}
which can be solved with a standard software package.


\bibliography{QHJ_1}

\begin{thebibliography}{78}%
\makeatletter
\providecommand \@ifxundefined [1]{%
 \@ifx{#1\undefined}
}%
\providecommand \@ifnum [1]{%
 \ifnum #1\expandafter \@firstoftwo
 \else \expandafter \@secondoftwo
 \fi
}%
\providecommand \@ifx [1]{%
 \ifx #1\expandafter \@firstoftwo
 \else \expandafter \@secondoftwo
 \fi
}%
\providecommand \natexlab [1]{#1}%
\providecommand \enquote  [1]{``#1''}%
\providecommand \bibnamefont  [1]{#1}%
\providecommand \bibfnamefont [1]{#1}%
\providecommand \citenamefont [1]{#1}%
\providecommand \href@noop [0]{\@secondoftwo}%
\providecommand \href [0]{\begingroup \@sanitize@url \@href}%
\providecommand \@href[1]{\@@startlink{#1}\@@href}%
\providecommand \@@href[1]{\endgroup#1\@@endlink}%
\providecommand \@sanitize@url [0]{\catcode `\\12\catcode `\$12\catcode
  `\&12\catcode `\#12\catcode `\^12\catcode `\_12\catcode `\%12\relax}%
\providecommand \@@startlink[1]{}%
\providecommand \@@endlink[0]{}%
\providecommand \url  [0]{\begingroup\@sanitize@url \@url }%
\providecommand \@url [1]{\endgroup\@href {#1}{\urlprefix }}%
\providecommand \urlprefix  [0]{URL }%
\providecommand \Eprint [0]{\href }%
\providecommand \doibase [0]{http://dx.doi.org/}%
\providecommand \selectlanguage [0]{\@gobble}%
\providecommand \bibinfo  [0]{\@secondoftwo}%
\providecommand \bibfield  [0]{\@secondoftwo}%
\providecommand \translation [1]{[#1]}%
\providecommand \BibitemOpen [0]{}%
\providecommand \bibitemStop [0]{}%
\providecommand \bibitemNoStop [0]{.\EOS\space}%
\providecommand \EOS [0]{\spacefactor3000\relax}%
\providecommand \BibitemShut  [1]{\csname bibitem#1\endcsname}%
\let\auto@bib@innerbib\@empty
\bibitem [{\citenamefont {Aoiz}\ \emph {et~al.}(1998)\citenamefont {Aoiz},
  \citenamefont {Friedrich}, \citenamefont {Herrero}, \citenamefont {Rabanos},\
  and\ \citenamefont {Verdasco}}]{CPL-1998-Aoiz-Fri-Rab-Ver-Herr-H-DCl}%
  \BibitemOpen
  \bibfield  {author} {\bibinfo {author} {\bibfnamefont {F.}~\bibnamefont
  {Aoiz}}, \bibinfo {author} {\bibfnamefont {B.}~\bibnamefont {Friedrich}},
  \bibinfo {author} {\bibfnamefont {V.}~\bibnamefont {Herrero}}, \bibinfo
  {author} {\bibfnamefont {V.~S.}\ \bibnamefont {Rabanos}}, \ and\ \bibinfo
  {author} {\bibfnamefont {J.}~\bibnamefont {Verdasco}},\ }\href@noop {}
  {\bibfield  {journal} {\bibinfo  {journal} {Chemical Physics Letters}\
  }\textbf {\bibinfo {volume} {289}},\ \bibinfo {pages} {132 } (\bibinfo {year}
  {1998})}\BibitemShut {NoStop}%
\bibitem [{\citenamefont {H\"artelt}\ and\ \citenamefont
  {Friedrich}(2008)}]{haertelt2008}%
  \BibitemOpen
  \bibfield  {author} {\bibinfo {author} {\bibfnamefont {M.}~\bibnamefont
  {H\"artelt}}\ and\ \bibinfo {author} {\bibfnamefont {B.}~\bibnamefont
  {Friedrich}},\ }\href {\doibase http://dx.doi.org/10.1063/1.2929850}
  {\bibfield  {journal} {\bibinfo  {journal} {The Journal of Chemical Physics}\
  }\textbf {\bibinfo {volume} {128}},\ \bibinfo {eid} {224313} (\bibinfo {year}
  {2008}),\ http://dx.doi.org/10.1063/1.2929850}\BibitemShut {NoStop}%
\bibitem [{\citenamefont {Friedrich}\ and\ \citenamefont
  {Herschbach}(1999{\natexlab{a}})}]{friedrich1999}%
  \BibitemOpen
  \bibfield  {author} {\bibinfo {author} {\bibfnamefont {B.}~\bibnamefont
  {Friedrich}}\ and\ \bibinfo {author} {\bibfnamefont {D.}~\bibnamefont
  {Herschbach}},\ }\href {\doibase 10.1021/jp992131w} {\bibfield  {journal}
  {\bibinfo  {journal} {The Journal of Physical Chemistry A}\ }\textbf
  {\bibinfo {volume} {103}},\ \bibinfo {pages} {10280} (\bibinfo {year}
  {1999}{\natexlab{a}})}\BibitemShut {NoStop}%
\bibitem [{\citenamefont {Lemeshko}\ \emph {et~al.}(2013)\citenamefont
  {Lemeshko}, \citenamefont {Krems}, \citenamefont {Doyle},\ and\ \citenamefont
  {Kais}}]{LemKreDoyKais:MP2013}%
  \BibitemOpen
  \bibfield  {author} {\bibinfo {author} {\bibfnamefont {M.}~\bibnamefont
  {Lemeshko}}, \bibinfo {author} {\bibfnamefont {R.~V.}\ \bibnamefont {Krems}},
  \bibinfo {author} {\bibfnamefont {J.~M.}\ \bibnamefont {Doyle}}, \ and\
  \bibinfo {author} {\bibfnamefont {S.}~\bibnamefont {Kais}},\ }\href@noop {}
  {\bibfield  {journal} {\bibinfo  {journal} {Mol. Phys.}\ }\textbf {\bibinfo
  {volume} {111}},\ \bibinfo {pages} {1648Ð1682} (\bibinfo {year}
  {2013})}\BibitemShut {NoStop}%
\bibitem [{\citenamefont {Cooper}\ \emph {et~al.}(1995)\citenamefont {Cooper},
  \citenamefont {Khare},\ and\ \citenamefont
  {Sukhatme}}]{SUSYQM_Cooper_Khare_Sukhatme1995}%
  \BibitemOpen
  \bibfield  {author} {\bibinfo {author} {\bibfnamefont {F.}~\bibnamefont
  {Cooper}}, \bibinfo {author} {\bibfnamefont {A.}~\bibnamefont {Khare}}, \
  and\ \bibinfo {author} {\bibfnamefont {U.}~\bibnamefont {Sukhatme}},\
  }\href@noop {} {\bibfield  {journal} {\bibinfo  {journal} {Physics Reports}\
  }\textbf {\bibinfo {volume} {251}},\ \bibinfo {pages} {267} (\bibinfo {year}
  {1995})}\BibitemShut {NoStop}%
\bibitem [{\citenamefont {Brooks}(1976)}]{Brooks1976Science}%
  \BibitemOpen
  \bibfield  {author} {\bibinfo {author} {\bibfnamefont {P.}~\bibnamefont
  {Brooks}},\ }\href@noop {} {\bibfield  {journal} {\bibinfo  {journal}
  {Science}\ }\textbf {\bibinfo {volume} {193}},\ \bibinfo {pages} {11}
  (\bibinfo {year} {1976})}\BibitemShut {NoStop}%
\bibitem [{\citenamefont {Loesch}\ and\ \citenamefont
  {Remscheid}(1990)}]{LoeschRem1990}%
  \BibitemOpen
  \bibfield  {author} {\bibinfo {author} {\bibfnamefont {H.~J.}\ \bibnamefont
  {Loesch}}\ and\ \bibinfo {author} {\bibfnamefont {A.}~\bibnamefont
  {Remscheid}},\ }\href@noop {} {\bibfield  {journal} {\bibinfo  {journal} {J.
  Chem. Phys.}\ }\textbf {\bibinfo {volume} {93}},\ \bibinfo {pages} {4779}
  (\bibinfo {year} {1990})}\BibitemShut {NoStop}%
\bibitem [{\citenamefont {Friedrich}\ and\ \citenamefont
  {Herschbach}(1991)}]{FriHer1991Nature}%
  \BibitemOpen
  \bibfield  {author} {\bibinfo {author} {\bibfnamefont {B.}~\bibnamefont
  {Friedrich}}\ and\ \bibinfo {author} {\bibfnamefont {D.}~\bibnamefont
  {Herschbach}},\ }\href@noop {} {\bibfield  {journal} {\bibinfo  {journal}
  {Nature}\ }\textbf {\bibinfo {volume} {353}},\ \bibinfo {pages} {412}
  (\bibinfo {year} {1991})}\BibitemShut {NoStop}%
\bibitem [{\citenamefont {Ortigoso}\ \emph {et~al.}(1999)\citenamefont
  {Ortigoso}, \citenamefont {Rodr\'{i}guez}, \citenamefont {Gupta},\ and\
  \citenamefont {Friedrich}}]{Ortigoso1999}%
  \BibitemOpen
  \bibfield  {author} {\bibinfo {author} {\bibfnamefont {J.}~\bibnamefont
  {Ortigoso}}, \bibinfo {author} {\bibfnamefont {M.}~\bibnamefont
  {Rodr\'{i}guez}}, \bibinfo {author} {\bibfnamefont {M.}~\bibnamefont
  {Gupta}}, \ and\ \bibinfo {author} {\bibfnamefont {B.}~\bibnamefont
  {Friedrich}},\ }\href@noop {} {\bibfield  {journal} {\bibinfo  {journal} {J.
  Chem. Phys.}\ }\textbf {\bibinfo {volume} {110}},\ \bibinfo {pages} {3870}
  (\bibinfo {year} {1999})}\BibitemShut {NoStop}%
\bibitem [{\citenamefont {Seideman}(1999)}]{Seideman1999}%
  \BibitemOpen
  \bibfield  {author} {\bibinfo {author} {\bibfnamefont {T.}~\bibnamefont
  {Seideman}},\ }\href@noop {} {\bibfield  {journal} {\bibinfo  {journal}
  {Phys. Rev. Lett.}\ }\textbf {\bibinfo {volume} {83}},\ \bibinfo {pages}
  {4971} (\bibinfo {year} {1999})}\BibitemShut {NoStop}%
\bibitem [{\citenamefont {Larsen}\ \emph {et~al.}(2000)\citenamefont {Larsen},
  \citenamefont {Hald}, \citenamefont {Bjerre}, \citenamefont {Stapelfeldt},\
  and\ \citenamefont {Seideman}}]{Larsen:00a}%
  \BibitemOpen
  \bibfield  {author} {\bibinfo {author} {\bibfnamefont {J.}~\bibnamefont
  {Larsen}}, \bibinfo {author} {\bibfnamefont {K.}~\bibnamefont {Hald}},
  \bibinfo {author} {\bibfnamefont {N.}~\bibnamefont {Bjerre}}, \bibinfo
  {author} {\bibfnamefont {H.}~\bibnamefont {Stapelfeldt}}, \ and\ \bibinfo
  {author} {\bibfnamefont {T.}~\bibnamefont {Seideman}},\ }\href@noop {}
  {\bibfield  {journal} {\bibinfo  {journal} {Phys. Rev. Lett.}\ }\textbf
  {\bibinfo {volume} {85}},\ \bibinfo {pages} {2470} (\bibinfo {year}
  {2000})}\BibitemShut {NoStop}%
\bibitem [{\citenamefont {Cai}\ and\ \citenamefont
  {Friedrich}(2001)}]{CCCC2001}%
  \BibitemOpen
  \bibfield  {author} {\bibinfo {author} {\bibfnamefont {L.}~\bibnamefont
  {Cai}}\ and\ \bibinfo {author} {\bibfnamefont {B.}~\bibnamefont
  {Friedrich}},\ }\href@noop {} {\bibfield  {journal} {\bibinfo  {journal}
  {Coll. Czech Chem. Commun.}\ }\textbf {\bibinfo {volume} {66}},\ \bibinfo
  {pages} {991} (\bibinfo {year} {2001})}\BibitemShut {NoStop}%
\bibitem [{\citenamefont {Averbukh}\ and\ \citenamefont
  {Arvieu}(2001)}]{Averbukh:01a}%
  \BibitemOpen
  \bibfield  {author} {\bibinfo {author} {\bibfnamefont {I.}~\bibnamefont
  {Averbukh}}\ and\ \bibinfo {author} {\bibfnamefont {R.}~\bibnamefont
  {Arvieu}},\ }\href@noop {} {\bibfield  {journal} {\bibinfo  {journal} {Phys.
  Rev. Lett.}\ }\textbf {\bibinfo {volume} {87}},\ \bibinfo {pages} {163601}
  (\bibinfo {year} {2001})}\BibitemShut {NoStop}%
\bibitem [{\citenamefont {Leibscher}\ \emph {et~al.}(2003)\citenamefont
  {Leibscher}, \citenamefont {Averbukh},\ and\ \citenamefont
  {Rabitz}}]{Leibscher:03a}%
  \BibitemOpen
  \bibfield  {author} {\bibinfo {author} {\bibfnamefont {M.}~\bibnamefont
  {Leibscher}}, \bibinfo {author} {\bibfnamefont {I.}~\bibnamefont {Averbukh}},
  \ and\ \bibinfo {author} {\bibfnamefont {H.}~\bibnamefont {Rabitz}},\
  }\href@noop {} {\bibfield  {journal} {\bibinfo  {journal} {Phys. Rev. Lett.}\
  }\textbf {\bibinfo {volume} {90}},\ \bibinfo {pages} {213001} (\bibinfo
  {year} {2003})}\BibitemShut {NoStop}%
\bibitem [{\citenamefont {Leibscher}\ \emph {et~al.}(2004)\citenamefont
  {Leibscher}, \citenamefont {Averbukh},\ and\ \citenamefont
  {Rabitz}}]{Leibscher:04a}%
  \BibitemOpen
  \bibfield  {author} {\bibinfo {author} {\bibfnamefont {M.}~\bibnamefont
  {Leibscher}}, \bibinfo {author} {\bibfnamefont {I.}~\bibnamefont {Averbukh}},
  \ and\ \bibinfo {author} {\bibfnamefont {H.}~\bibnamefont {Rabitz}},\
  }\href@noop {} {\bibfield  {journal} {\bibinfo  {journal} {Phys. Rev. A}\
  }\textbf {\bibinfo {volume} {69}},\ \bibinfo {pages} {13402} (\bibinfo {year}
  {2004})}\BibitemShut {NoStop}%
\bibitem [{\citenamefont {Toennies}(1964)}]{Toennies1964}%
  \BibitemOpen
  \bibfield  {author} {\bibinfo {author} {\bibfnamefont {J.}~\bibnamefont
  {Toennies}},\ }\href@noop {} {\bibfield  {journal} {\bibinfo  {journal} {Z.
  Phys.}\ }\textbf {\bibinfo {volume} {177}},\ \bibinfo {pages} {84} (\bibinfo
  {year} {1964})}\BibitemShut {NoStop}%
\bibitem [{\citenamefont {Stolte}(1982)}]{StolteBerichte1982}%
  \BibitemOpen
  \bibfield  {author} {\bibinfo {author} {\bibfnamefont {S.}~\bibnamefont
  {Stolte}},\ }\href@noop {} {\bibfield  {journal} {\bibinfo  {journal}
  {Berichte Bunsen. Ges. Phys. Chem.}\ }\textbf {\bibinfo {volume} {413}},\
  \bibinfo {pages} {84} (\bibinfo {year} {1982})}\BibitemShut {NoStop}%
\bibitem [{\citenamefont {Stapelfeldt}\ \emph {et~al.}(1997)\citenamefont
  {Stapelfeldt}, \citenamefont {Sakai}, \citenamefont {Constant},\ and\
  \citenamefont {Corkum}}]{StapCorkumPRA1997}%
  \BibitemOpen
  \bibfield  {author} {\bibinfo {author} {\bibfnamefont {H.}~\bibnamefont
  {Stapelfeldt}}, \bibinfo {author} {\bibfnamefont {H.}~\bibnamefont {Sakai}},
  \bibinfo {author} {\bibfnamefont {E.}~\bibnamefont {Constant}}, \ and\
  \bibinfo {author} {\bibfnamefont {P.}~\bibnamefont {Corkum}},\ }\href@noop {}
  {\bibfield  {journal} {\bibinfo  {journal} {Phys. Rev. A}\ }\textbf {\bibinfo
  {volume} {79}},\ \bibinfo {pages} {2787} (\bibinfo {year}
  {1997})}\BibitemShut {NoStop}%
\bibitem [{\citenamefont {Kim}\ \emph {et~al.}(2016)\citenamefont {Kim},
  \citenamefont {Lee}, \citenamefont {Kim}, \citenamefont {Kwak}, \citenamefont
  {Friedrich},\ and\ \citenamefont {Zhao}}]{BumSukFriPRA2016}%
  \BibitemOpen
  \bibfield  {author} {\bibinfo {author} {\bibfnamefont {L.~Y.}\ \bibnamefont
  {Kim}}, \bibinfo {author} {\bibfnamefont {J.~H.}\ \bibnamefont {Lee}},
  \bibinfo {author} {\bibfnamefont {H.~A.}\ \bibnamefont {Kim}}, \bibinfo
  {author} {\bibfnamefont {S.~K.}\ \bibnamefont {Kwak}}, \bibinfo {author}
  {\bibfnamefont {B.}~\bibnamefont {Friedrich}}, \ and\ \bibinfo {author}
  {\bibfnamefont {B.~S.}\ \bibnamefont {Zhao}},\ }\href@noop {} {\bibfield
  {journal} {\bibinfo  {journal} {Phys. Rev. A}\ }\textbf {\bibinfo {volume}
  {94}},\ \bibinfo {pages} {013428} (\bibinfo {year} {2016})}\BibitemShut
  {NoStop}%
\bibitem [{\citenamefont {Truppe}\ \emph {et~al.}(2017)\citenamefont {Truppe},
  \citenamefont {Williams}, \citenamefont {Hambach}, \citenamefont {Caldwell},
  \citenamefont {Fitch}, \citenamefont {Hinds}, \citenamefont {Sauer},\ and\
  \citenamefont {Tarbutt}}]{TruppeTarbutt2017NaturePhysics}%
  \BibitemOpen
  \bibfield  {author} {\bibinfo {author} {\bibfnamefont {S.}~\bibnamefont
  {Truppe}}, \bibinfo {author} {\bibfnamefont {H.}~\bibnamefont {Williams}},
  \bibinfo {author} {\bibfnamefont {M.}~\bibnamefont {Hambach}}, \bibinfo
  {author} {\bibfnamefont {L.}~\bibnamefont {Caldwell}}, \bibinfo {author}
  {\bibfnamefont {N.}~\bibnamefont {Fitch}}, \bibinfo {author} {\bibfnamefont
  {E.}~\bibnamefont {Hinds}}, \bibinfo {author} {\bibfnamefont
  {B.}~\bibnamefont {Sauer}}, \ and\ \bibinfo {author} {\bibfnamefont
  {M.}~\bibnamefont {Tarbutt}},\ }\href@noop {} {\bibfield  {journal} {\bibinfo
   {journal} {Nature Physics}\ }\textbf {\bibinfo {volume} {13}},\ \bibinfo
  {pages} {1173} (\bibinfo {year} {2017})}\BibitemShut {NoStop}%
\bibitem [{\citenamefont {Bernstein}\ \emph {et~al.}(1987)\citenamefont
  {Bernstein}, \citenamefont {Herschbach},\ and\ \citenamefont
  {Levine}}]{BernHerschLevine1989}%
  \BibitemOpen
  \bibfield  {author} {\bibinfo {author} {\bibfnamefont {R.}~\bibnamefont
  {Bernstein}}, \bibinfo {author} {\bibfnamefont {D.}~\bibnamefont
  {Herschbach}}, \ and\ \bibinfo {author} {\bibfnamefont {R.}~\bibnamefont
  {Levine}},\ }\href@noop {} {\bibfield  {journal} {\bibinfo  {journal} {J.
  Phys. Chem.}\ }\textbf {\bibinfo {volume} {91}},\ \bibinfo {pages} {5365}
  (\bibinfo {year} {1987})}\BibitemShut {NoStop}%
\bibitem [{\citenamefont {Krems}\ \emph {et~al.}(2009)\citenamefont {Krems},
  \citenamefont {Stwalley},\ and\ \citenamefont
  {Friedrich}}]{KremsStwalleyFriedrich2009}%
  \BibitemOpen
  \bibfield  {author} {\bibinfo {author} {\bibfnamefont {R.}~\bibnamefont
  {Krems}}, \bibinfo {author} {\bibfnamefont {W.~C.}\ \bibnamefont {Stwalley}},
  \ and\ \bibinfo {author} {\bibfnamefont {B.}~\bibnamefont {Friedrich}},\
  }\href@noop {} {\emph {\bibinfo {title} {{Cold Molecules: Theory, Experi-
  ment, Applications}}}}\ (\bibinfo  {publisher} {CRC Press},\ \bibinfo
  {address} {Boca Raton, FL},\ \bibinfo {year} {2009})\BibitemShut {NoStop}%
\bibitem [{\citenamefont {Dulieu}\ and\ \citenamefont
  {Osterwalder}(2018)}]{DulieuOsterwalder2018}%
  \BibitemOpen
  \bibfield  {author} {\bibinfo {author} {\bibfnamefont {O.}~\bibnamefont
  {Dulieu}}\ and\ \bibinfo {author} {\bibfnamefont {A.}~\bibnamefont
  {Osterwalder}},\ }\href@noop {} {\emph {\bibinfo {title} {{Cold Chemistry:
  Molecular Scattering and Reactivity Near Absolute Zero}}}}\ (\bibinfo
  {publisher} {The Royal Society of Chemistry},\ \bibinfo {address}
  {Cambridge},\ \bibinfo {year} {2018})\BibitemShut {NoStop}%
\bibitem [{\citenamefont {Liptay}(1974)}]{Liptay1974}%
  \BibitemOpen
  \bibfield  {author} {\bibinfo {author} {\bibfnamefont {W.}~\bibnamefont
  {Liptay}},\ }\href@noop {} {\bibfield  {journal} {\bibinfo  {journal}
  {Berichte der Bunsengesellschaft Physikalische Chemie}\ }\textbf {\bibinfo
  {volume} {80}},\ \bibinfo {pages} {207} (\bibinfo {year} {1974})}\BibitemShut
  {NoStop}%
\bibitem [{\citenamefont {Friedrich}\ \emph {et~al.}(1994)\citenamefont
  {Friedrich}, \citenamefont {Slenczka},\ and\ \citenamefont
  {Herschbach}}]{FriSlenHer1994}%
  \BibitemOpen
  \bibfield  {author} {\bibinfo {author} {\bibfnamefont {B.}~\bibnamefont
  {Friedrich}}, \bibinfo {author} {\bibfnamefont {A.}~\bibnamefont {Slenczka}},
  \ and\ \bibinfo {author} {\bibfnamefont {D.}~\bibnamefont {Herschbach}},\
  }\href@noop {} {\bibfield  {journal} {\bibinfo  {journal} {Can. J. Phys.}\
  }\textbf {\bibinfo {volume} {72}},\ \bibinfo {pages} {897} (\bibinfo {year}
  {1994})}\BibitemShut {NoStop}%
\bibitem [{\citenamefont {Slenczka}(1999)}]{Slenczka1999}%
  \BibitemOpen
  \bibfield  {author} {\bibinfo {author} {\bibfnamefont {A.}~\bibnamefont
  {Slenczka}},\ }\href@noop {} {\bibfield  {journal} {\bibinfo  {journal}
  {Chemistry-a European Journal}\ }\textbf {\bibinfo {volume} {5}},\ \bibinfo
  {pages} {1136} (\bibinfo {year} {1999})}\BibitemShut {NoStop}%
\bibitem [{\citenamefont {Friedrich}\ and\ \citenamefont
  {Herschbach}(1999{\natexlab{b}})}]{JCP1999FriHer}%
  \BibitemOpen
  \bibfield  {author} {\bibinfo {author} {\bibfnamefont {B.}~\bibnamefont
  {Friedrich}}\ and\ \bibinfo {author} {\bibfnamefont {D.}~\bibnamefont
  {Herschbach}},\ }\href@noop {} {\bibfield  {journal} {\bibinfo  {journal} {J.
  Chem. Phys.}\ }\textbf {\bibinfo {volume} {111}},\ \bibinfo {pages} {6157}
  (\bibinfo {year} {1999}{\natexlab{b}})}\BibitemShut {NoStop}%
\bibitem [{\citenamefont {Cai}\ \emph {et~al.}(2001)\citenamefont {Cai},
  \citenamefont {Marango},\ and\ \citenamefont {Friedrich}}]{CaiFri2001}%
  \BibitemOpen
  \bibfield  {author} {\bibinfo {author} {\bibfnamefont {L.}~\bibnamefont
  {Cai}}, \bibinfo {author} {\bibfnamefont {J.}~\bibnamefont {Marango}}, \ and\
  \bibinfo {author} {\bibfnamefont {B.}~\bibnamefont {Friedrich}},\ }\href@noop
  {} {\bibfield  {journal} {\bibinfo  {journal} {Phys. Rev. Lett.}\ }\textbf
  {\bibinfo {volume} {86}},\ \bibinfo {pages} {775} (\bibinfo {year}
  {2001})}\BibitemShut {NoStop}%
\bibitem [{\citenamefont {Sakai}\ \emph {et~al.}(2003)\citenamefont {Sakai},
  \citenamefont {Minemoto}, \citenamefont {Nanjo}, \citenamefont {Tanji},\ and\
  \citenamefont {Suzuki}}]{PRL2003Sakai}%
  \BibitemOpen
  \bibfield  {author} {\bibinfo {author} {\bibfnamefont {H.}~\bibnamefont
  {Sakai}}, \bibinfo {author} {\bibfnamefont {S.}~\bibnamefont {Minemoto}},
  \bibinfo {author} {\bibfnamefont {H.}~\bibnamefont {Nanjo}}, \bibinfo
  {author} {\bibfnamefont {H.}~\bibnamefont {Tanji}}, \ and\ \bibinfo {author}
  {\bibfnamefont {T.}~\bibnamefont {Suzuki}},\ }\href@noop {} {\bibfield
  {journal} {\bibinfo  {journal} {Phys. Rev. Lett.}\ }\textbf {\bibinfo
  {volume} {90}},\ \bibinfo {pages} {83001} (\bibinfo {year}
  {2003})}\BibitemShut {NoStop}%
\bibitem [{\citenamefont {Friedrich}\ \emph {et~al.}(2003)\citenamefont
  {Friedrich}, \citenamefont {Nahler},\ and\ \citenamefont
  {Buck}}]{JMO2003-Fri}%
  \BibitemOpen
  \bibfield  {author} {\bibinfo {author} {\bibfnamefont {B.}~\bibnamefont
  {Friedrich}}, \bibinfo {author} {\bibfnamefont {N.~H.}\ \bibnamefont
  {Nahler}}, \ and\ \bibinfo {author} {\bibfnamefont {U.}~\bibnamefont
  {Buck}},\ }\href@noop {} {\bibfield  {journal} {\bibinfo  {journal} {J. Mod.
  Opt.}\ }\textbf {\bibinfo {volume} {50}},\ \bibinfo {pages} {2677} (\bibinfo
  {year} {2003})}\BibitemShut {NoStop}%
\bibitem [{\citenamefont {Nielsen}\ \emph {et~al.}(2012)\citenamefont
  {Nielsen}, \citenamefont {Stapelfeldt}, \citenamefont {K\"{u}pper},
  \citenamefont {Friedrich}, \citenamefont {Omiste},\ and\ \citenamefont
  {Gonz\'alez-F\'erez}}]{RosarioPRL2012}%
  \BibitemOpen
  \bibfield  {author} {\bibinfo {author} {\bibfnamefont {J.~H.}\ \bibnamefont
  {Nielsen}}, \bibinfo {author} {\bibfnamefont {H.}~\bibnamefont
  {Stapelfeldt}}, \bibinfo {author} {\bibfnamefont {J.}~\bibnamefont
  {K\"{u}pper}}, \bibinfo {author} {\bibfnamefont {B.}~\bibnamefont
  {Friedrich}}, \bibinfo {author} {\bibfnamefont {J.~J.}\ \bibnamefont
  {Omiste}}, \ and\ \bibinfo {author} {\bibfnamefont {R.}~\bibnamefont
  {Gonz\'alez-F\'erez}},\ }\href@noop {} {\bibfield  {journal} {\bibinfo
  {journal} {Phys. Rev. Lett.}\ }\textbf {\bibinfo {volume} {108}},\ \bibinfo
  {pages} {193001} (\bibinfo {year} {2012})}\BibitemShut {NoStop}%
\bibitem [{\citenamefont {Schmidt}\ and\ \citenamefont
  {Friedrich}(2014{\natexlab{a}})}]{schmidt2014b}%
  \BibitemOpen
  \bibfield  {author} {\bibinfo {author} {\bibfnamefont {B.}~\bibnamefont
  {Schmidt}}\ and\ \bibinfo {author} {\bibfnamefont {B.}~\bibnamefont
  {Friedrich}},\ }\href {\doibase 10.3389/fphy.2014.00037} {\bibfield
  {journal} {\bibinfo  {journal} {Front. Phys.}\ }\textbf {\bibinfo {volume}
  {2}},\ \bibinfo {pages} {1} (\bibinfo {year} {2014}{\natexlab{a}})},\ \Eprint
  {http://arxiv.org/abs/1404.2243} {1404.2243} \BibitemShut {NoStop}%
\bibitem [{\citenamefont {Schmidt}\ and\ \citenamefont
  {Friedrich}(2015)}]{SchmiFri2015a}%
  \BibitemOpen
  \bibfield  {author} {\bibinfo {author} {\bibfnamefont {B.}~\bibnamefont
  {Schmidt}}\ and\ \bibinfo {author} {\bibfnamefont {B.}~\bibnamefont
  {Friedrich}},\ }\href@noop {} {\bibfield  {journal} {\bibinfo  {journal}
  {Phys. Rev. A}\ }\textbf {\bibinfo {volume} {91}},\ \bibinfo {pages} {022111}
  (\bibinfo {year} {2015})}\BibitemShut {NoStop}%
\bibitem [{\citenamefont {Bisgaard}\ \emph {et~al.}(2009)\citenamefont
  {Bisgaard}, \citenamefont {Clarkin}, \citenamefont {Wu}, \citenamefont {Lee},
  \citenamefont {Gessner}, \citenamefont {Hayden},\ and\ \citenamefont
  {Stolow}}]{Bisgaard2009}%
  \BibitemOpen
  \bibfield  {author} {\bibinfo {author} {\bibfnamefont {C.~Z.}\ \bibnamefont
  {Bisgaard}}, \bibinfo {author} {\bibfnamefont {O.~J.}\ \bibnamefont
  {Clarkin}}, \bibinfo {author} {\bibfnamefont {G.}~\bibnamefont {Wu}},
  \bibinfo {author} {\bibfnamefont {A.~M.~D.}\ \bibnamefont {Lee}}, \bibinfo
  {author} {\bibfnamefont {O.}~\bibnamefont {Gessner}}, \bibinfo {author}
  {\bibfnamefont {C.~C.}\ \bibnamefont {Hayden}}, \ and\ \bibinfo {author}
  {\bibfnamefont {A.}~\bibnamefont {Stolow}},\ }\href@noop {} {\bibfield
  {journal} {\bibinfo  {journal} {Science}\ }\textbf {\bibinfo {volume}
  {323}},\ \bibinfo {pages} {1464} (\bibinfo {year} {2009})}\BibitemShut
  {NoStop}%
\bibitem [{\citenamefont {Holmegaard}\ \emph {et~al.}(2010)\citenamefont
  {Holmegaard}, \citenamefont {Hansen}, \citenamefont {Kalh\o~j}, \citenamefont
  {{Louise Kragh}}, \citenamefont {Stapelfeldt}, \citenamefont {Filsinger},
  \citenamefont {K\"{u}pper}, \citenamefont {Meijer}, \citenamefont
  {Dimitrovski}, \citenamefont {Abu-samha}, \citenamefont {Martiny},\ and\
  \citenamefont {{Bojer Madsen}}}]{Holmegaard2010}%
  \BibitemOpen
  \bibfield  {author} {\bibinfo {author} {\bibfnamefont {L.}~\bibnamefont
  {Holmegaard}}, \bibinfo {author} {\bibfnamefont {J.~L.}\ \bibnamefont
  {Hansen}}, \bibinfo {author} {\bibfnamefont {L.}~\bibnamefont {Kalh\o~j}},
  \bibinfo {author} {\bibfnamefont {S.}~\bibnamefont {{Louise Kragh}}},
  \bibinfo {author} {\bibfnamefont {H.}~\bibnamefont {Stapelfeldt}}, \bibinfo
  {author} {\bibfnamefont {F.}~\bibnamefont {Filsinger}}, \bibinfo {author}
  {\bibfnamefont {J.}~\bibnamefont {K\"{u}pper}}, \bibinfo {author}
  {\bibfnamefont {G.}~\bibnamefont {Meijer}}, \bibinfo {author} {\bibfnamefont
  {D.}~\bibnamefont {Dimitrovski}}, \bibinfo {author} {\bibfnamefont
  {M.}~\bibnamefont {Abu-samha}}, \bibinfo {author} {\bibfnamefont {C.~P.~J.}\
  \bibnamefont {Martiny}}, \ and\ \bibinfo {author} {\bibfnamefont
  {L.}~\bibnamefont {{Bojer Madsen}}},\ }\href@noop {} {\bibfield  {journal}
  {\bibinfo  {journal} {Nature Phys.}\ }\textbf {\bibinfo {volume} {6}},\
  \bibinfo {pages} {428} (\bibinfo {year} {2010})}\BibitemShut {NoStop}%
\bibitem [{\citenamefont {Hansen}\ \emph {et~al.}(2011)\citenamefont {Hansen},
  \citenamefont {Stapelfeldt}, \citenamefont {Dimitrovski}, \citenamefont
  {Abu-samha}, \citenamefont {Martiny},\ and\ \citenamefont
  {Madsen}}]{Hansen2011}%
  \BibitemOpen
  \bibfield  {author} {\bibinfo {author} {\bibfnamefont {J.~L.}\ \bibnamefont
  {Hansen}}, \bibinfo {author} {\bibfnamefont {H.}~\bibnamefont {Stapelfeldt}},
  \bibinfo {author} {\bibfnamefont {D.}~\bibnamefont {Dimitrovski}}, \bibinfo
  {author} {\bibfnamefont {M.}~\bibnamefont {Abu-samha}}, \bibinfo {author}
  {\bibfnamefont {C.~P.~J.}\ \bibnamefont {Martiny}}, \ and\ \bibinfo {author}
  {\bibfnamefont {L.}~\bibnamefont {Madsen}},\ }\href@noop {} {\bibfield
  {journal} {\bibinfo  {journal} {Phys. Rev. Lett.}\ }\textbf {\bibinfo
  {volume} {106}},\ \bibinfo {pages} {073001} (\bibinfo {year}
  {2011})}\BibitemShut {NoStop}%
\bibitem [{\citenamefont {Landers}\ \emph {et~al.}(2001)\citenamefont
  {Landers}, \citenamefont {Weber}, \citenamefont {Ali}, \citenamefont
  {Cassimi}, \citenamefont {Hattass}, \citenamefont {Jagutzki}, \citenamefont
  {Nauert}, \citenamefont {Osipov}, \citenamefont {Staudte}, \citenamefont
  {Prior}, \citenamefont {Schmidt-B\"{o}cking}, \citenamefont {Cocke},\ and\
  \citenamefont {D\"{o}rner}}]{Landers2001}%
  \BibitemOpen
  \bibfield  {author} {\bibinfo {author} {\bibfnamefont {A.}~\bibnamefont
  {Landers}}, \bibinfo {author} {\bibfnamefont {T.}~\bibnamefont {Weber}},
  \bibinfo {author} {\bibfnamefont {I.}~\bibnamefont {Ali}}, \bibinfo {author}
  {\bibfnamefont {A.}~\bibnamefont {Cassimi}}, \bibinfo {author} {\bibfnamefont
  {M.}~\bibnamefont {Hattass}}, \bibinfo {author} {\bibfnamefont
  {O.}~\bibnamefont {Jagutzki}}, \bibinfo {author} {\bibfnamefont
  {A.}~\bibnamefont {Nauert}}, \bibinfo {author} {\bibfnamefont
  {T.}~\bibnamefont {Osipov}}, \bibinfo {author} {\bibfnamefont
  {A.}~\bibnamefont {Staudte}}, \bibinfo {author} {\bibfnamefont
  {M.}~\bibnamefont {Prior}}, \bibinfo {author} {\bibfnamefont
  {H.}~\bibnamefont {Schmidt-B\"{o}cking}}, \bibinfo {author} {\bibfnamefont
  {C.}~\bibnamefont {Cocke}}, \ and\ \bibinfo {author} {\bibfnamefont
  {R.}~\bibnamefont {D\"{o}rner}},\ }\href@noop {} {\bibfield  {journal}
  {\bibinfo  {journal} {Phys. Rev. Lett.}\ }\textbf {\bibinfo {volume} {87}},\
  \bibinfo {pages} {013002} (\bibinfo {year} {2001})}\BibitemShut {NoStop}%
\bibitem [{\citenamefont {Itatani}\ \emph {et~al.}(2004)\citenamefont
  {Itatani}, \citenamefont {Levesque}, \citenamefont {Zeidler}, \citenamefont
  {Niikura}, \citenamefont {P\'{e}pin}, \citenamefont {Kieffer}, \citenamefont
  {Corkum},\ and\ \citenamefont {Villeneuve}}]{Itatani:04a}%
  \BibitemOpen
  \bibfield  {author} {\bibinfo {author} {\bibfnamefont {J.}~\bibnamefont
  {Itatani}}, \bibinfo {author} {\bibfnamefont {J.}~\bibnamefont {Levesque}},
  \bibinfo {author} {\bibfnamefont {D.}~\bibnamefont {Zeidler}}, \bibinfo
  {author} {\bibfnamefont {H.}~\bibnamefont {Niikura}}, \bibinfo {author}
  {\bibfnamefont {H.}~\bibnamefont {P\'{e}pin}}, \bibinfo {author}
  {\bibfnamefont {J.~C.}\ \bibnamefont {Kieffer}}, \bibinfo {author}
  {\bibfnamefont {P.~B.}\ \bibnamefont {Corkum}}, \ and\ \bibinfo {author}
  {\bibfnamefont {D.~M.}\ \bibnamefont {Villeneuve}},\ }\href@noop {}
  {\bibfield  {journal} {\bibinfo  {journal} {Nature}\ }\textbf {\bibinfo
  {volume} {432}},\ \bibinfo {pages} {867} (\bibinfo {year}
  {2004})}\BibitemShut {NoStop}%
\bibitem [{\citenamefont {Jin}\ \emph {et~al.}(2012)\citenamefont {Jin},
  \citenamefont {Bertrand}, \citenamefont {Lucchese}, \citenamefont
  {W\"{o}rner}, \citenamefont {Corkum}, \citenamefont {Villeneuve},
  \citenamefont {Le},\ and\ \citenamefont {Lin}}]{CorkumHHG}%
  \BibitemOpen
  \bibfield  {author} {\bibinfo {author} {\bibfnamefont {C.}~\bibnamefont
  {Jin}}, \bibinfo {author} {\bibfnamefont {J.}~\bibnamefont {Bertrand}},
  \bibinfo {author} {\bibfnamefont {R.}~\bibnamefont {Lucchese}}, \bibinfo
  {author} {\bibfnamefont {H.}~\bibnamefont {W\"{o}rner}}, \bibinfo {author}
  {\bibfnamefont {P.}~\bibnamefont {Corkum}}, \bibinfo {author} {\bibfnamefont
  {D.}~\bibnamefont {Villeneuve}}, \bibinfo {author} {\bibfnamefont {A.-T.}\
  \bibnamefont {Le}}, \ and\ \bibinfo {author} {\bibfnamefont {C.}~\bibnamefont
  {Lin}},\ }\href@noop {} {\bibfield  {journal} {\bibinfo  {journal} {Phys.
  Rev. A}\ }\textbf {\bibinfo {volume} {85}},\ \bibinfo {pages} {013405}
  (\bibinfo {year} {2012})}\BibitemShut {NoStop}%
\bibitem [{\citenamefont {Yuan}\ and\ \citenamefont
  {Bandrauk}(2009)}]{BandraukHHG}%
  \BibitemOpen
  \bibfield  {author} {\bibinfo {author} {\bibfnamefont {K.-J.}\ \bibnamefont
  {Yuan}}\ and\ \bibinfo {author} {\bibfnamefont {A.}~\bibnamefont
  {Bandrauk}},\ }\href@noop {} {\bibfield  {journal} {\bibinfo  {journal}
  {Phys. Rev. A}\ }\textbf {\bibinfo {volume} {80}},\ \bibinfo {pages} {053404}
  (\bibinfo {year} {2009})}\BibitemShut {NoStop}%
\bibitem [{\citenamefont {Smirnova}\ \emph {et~al.}(2009)\citenamefont
  {Smirnova}, \citenamefont {Mairesse}, \citenamefont {Patchkovskii},
  \citenamefont {Dudovich}, \citenamefont {Villeneuve}, \citenamefont
  {Corkum},\ and\ \citenamefont {Ivanov}}]{Smirnova2009}%
  \BibitemOpen
  \bibfield  {author} {\bibinfo {author} {\bibfnamefont {O.}~\bibnamefont
  {Smirnova}}, \bibinfo {author} {\bibfnamefont {Y.}~\bibnamefont {Mairesse}},
  \bibinfo {author} {\bibfnamefont {S.}~\bibnamefont {Patchkovskii}}, \bibinfo
  {author} {\bibfnamefont {N.}~\bibnamefont {Dudovich}}, \bibinfo {author}
  {\bibfnamefont {D.}~\bibnamefont {Villeneuve}}, \bibinfo {author}
  {\bibfnamefont {P.}~\bibnamefont {Corkum}}, \ and\ \bibinfo {author}
  {\bibfnamefont {M.~Y.}\ \bibnamefont {Ivanov}},\ }\href@noop {} {\bibfield
  {journal} {\bibinfo  {journal} {Nature}\ }\textbf {\bibinfo {volume} {460}},\
  \bibinfo {pages} {972} (\bibinfo {year} {2009})}\BibitemShut {NoStop}%
\bibitem [{\citenamefont {Rupenyan}\ \emph {et~al.}(2012)\citenamefont
  {Rupenyan}, \citenamefont {Bertrand}, \citenamefont {Villeneuve},\ and\
  \citenamefont {W\"{o}rner}}]{Woerner}%
  \BibitemOpen
  \bibfield  {author} {\bibinfo {author} {\bibfnamefont {A.}~\bibnamefont
  {Rupenyan}}, \bibinfo {author} {\bibfnamefont {J.}~\bibnamefont {Bertrand}},
  \bibinfo {author} {\bibfnamefont {D.}~\bibnamefont {Villeneuve}}, \ and\
  \bibinfo {author} {\bibfnamefont {H.}~\bibnamefont {W\"{o}rner}},\
  }\href@noop {} {\bibfield  {journal} {\bibinfo  {journal} {Phys. Rev. Lett.}\
  }\textbf {\bibinfo {volume} {108}},\ \bibinfo {pages} {033903} (\bibinfo
  {year} {2012})}\BibitemShut {NoStop}%
\bibitem [{\citenamefont {Baranov}\ \emph {et~al.}(2012)\citenamefont
  {Baranov}, \citenamefont {Dalmonte}, \citenamefont {Pupillo},\ and\
  \citenamefont {Zoller}}]{Bar:12}%
  \BibitemOpen
  \bibfield  {author} {\bibinfo {author} {\bibfnamefont {M.~A.}\ \bibnamefont
  {Baranov}}, \bibinfo {author} {\bibfnamefont {M.}~\bibnamefont {Dalmonte}},
  \bibinfo {author} {\bibfnamefont {G.}~\bibnamefont {Pupillo}}, \ and\
  \bibinfo {author} {\bibfnamefont {P.}~\bibnamefont {Zoller}},\ }\href@noop {}
  {\bibfield  {journal} {\bibinfo  {journal} {Chem. Rev.}\ }\textbf {\bibinfo
  {volume} {112}},\ \bibinfo {pages} {5012} (\bibinfo {year}
  {2012})}\BibitemShut {NoStop}%
\bibitem [{\citenamefont {Manmana}\ \emph {et~al.}(2013)\citenamefont
  {Manmana}, \citenamefont {Stoudenmire}, \citenamefont {Hazzard},
  \citenamefont {Rey},\ and\ \citenamefont {Gorshkov}}]{Man:13}%
  \BibitemOpen
  \bibfield  {author} {\bibinfo {author} {\bibfnamefont {S.~R.}\ \bibnamefont
  {Manmana}}, \bibinfo {author} {\bibfnamefont {E.~M.}\ \bibnamefont
  {Stoudenmire}}, \bibinfo {author} {\bibfnamefont {K.~R.~A.}\ \bibnamefont
  {Hazzard}}, \bibinfo {author} {\bibfnamefont {A.~M.}\ \bibnamefont {Rey}}, \
  and\ \bibinfo {author} {\bibfnamefont {A.~V.}\ \bibnamefont {Gorshkov}},\
  }\href@noop {} {\bibfield  {journal} {\bibinfo  {journal} {Physical Review
  B}\ }\textbf {\bibinfo {volume} {87}},\ \bibinfo {pages} {081106 R} (\bibinfo
  {year} {2013})}\BibitemShut {NoStop}%
\bibitem [{\citenamefont {DeMille}(2002)}]{DeMille2002}%
  \BibitemOpen
  \bibfield  {author} {\bibinfo {author} {\bibfnamefont {D.}~\bibnamefont
  {DeMille}},\ }\href@noop {} {\bibfield  {journal} {\bibinfo  {journal} {Phys.
  Rev. Lett.}\ }\textbf {\bibinfo {volume} {88}},\ \bibinfo {pages} {067901}
  (\bibinfo {year} {2002})}\BibitemShut {NoStop}%
\bibitem [{\citenamefont {Schneider}\ \emph {et~al.}(2007)\citenamefont
  {Schneider}, \citenamefont {Gollub}, \citenamefont {Kompa},\ and\
  \citenamefont {de~Vivie-Riedle}}]{de6}%
  \BibitemOpen
  \bibfield  {author} {\bibinfo {author} {\bibfnamefont {B.}~\bibnamefont
  {Schneider}}, \bibinfo {author} {\bibfnamefont {C.}~\bibnamefont {Gollub}},
  \bibinfo {author} {\bibfnamefont {K.-L.}\ \bibnamefont {Kompa}}, \ and\
  \bibinfo {author} {\bibfnamefont {R.}~\bibnamefont {de~Vivie-Riedle}},\
  }\href@noop {} {\bibfield  {journal} {\bibinfo  {journal} {Chem. Phys.}\
  }\textbf {\bibinfo {volume} {338}},\ \bibinfo {pages} {291} (\bibinfo {year}
  {2007})}\BibitemShut {NoStop}%
\bibitem [{\citenamefont {Korff}\ \emph {et~al.}(2005)\citenamefont {Korff},
  \citenamefont {Troppmann}, \citenamefont {Kompa},\ and\ \citenamefont
  {de~Vivie-Riedle}}]{de7}%
  \BibitemOpen
  \bibfield  {author} {\bibinfo {author} {\bibfnamefont {B.~M.~R.}\
  \bibnamefont {Korff}}, \bibinfo {author} {\bibfnamefont {U.}~\bibnamefont
  {Troppmann}}, \bibinfo {author} {\bibfnamefont {K.-L.}\ \bibnamefont
  {Kompa}}, \ and\ \bibinfo {author} {\bibfnamefont {R.}~\bibnamefont
  {de~Vivie-Riedle}},\ }\href@noop {} {\bibfield  {journal} {\bibinfo
  {journal} {J. Chem. Phys.}\ }\textbf {\bibinfo {volume} {123}},\ \bibinfo
  {pages} {244509} (\bibinfo {year} {2005})}\BibitemShut {NoStop}%
\bibitem [{\citenamefont {Troppmann}\ \emph {et~al.}(2003)\citenamefont
  {Troppmann}, \citenamefont {Tesch},\ and\ \citenamefont
  {de~Vivie-Riedle}}]{de8}%
  \BibitemOpen
  \bibfield  {author} {\bibinfo {author} {\bibfnamefont {U.}~\bibnamefont
  {Troppmann}}, \bibinfo {author} {\bibfnamefont {C.~M.}\ \bibnamefont
  {Tesch}}, \ and\ \bibinfo {author} {\bibfnamefont {R.}~\bibnamefont
  {de~Vivie-Riedle}},\ }\href@noop {} {\bibfield  {journal} {\bibinfo
  {journal} {Chem. Phys. Lett.}\ }\textbf {\bibinfo {volume} {378}},\ \bibinfo
  {pages} {273} (\bibinfo {year} {2003})}\BibitemShut {NoStop}%
\bibitem [{\citenamefont {Wei}\ \emph {et~al.}(2011{\natexlab{a}})\citenamefont
  {Wei}, \citenamefont {Kais}, \citenamefont {Friedrich},\ and\ \citenamefont
  {Herschbach}}]{QiKaisFrieHer2011a}%
  \BibitemOpen
  \bibfield  {author} {\bibinfo {author} {\bibfnamefont {Q.}~\bibnamefont
  {Wei}}, \bibinfo {author} {\bibfnamefont {S.}~\bibnamefont {Kais}}, \bibinfo
  {author} {\bibfnamefont {B.}~\bibnamefont {Friedrich}}, \ and\ \bibinfo
  {author} {\bibfnamefont {D.}~\bibnamefont {Herschbach}},\ }\href@noop {}
  {\bibfield  {journal} {\bibinfo  {journal} {J. Chem. Phys.}\ }\textbf
  {\bibinfo {volume} {134}},\ \bibinfo {pages} {124107} (\bibinfo {year}
  {2011}{\natexlab{a}})}\BibitemShut {NoStop}%
\bibitem [{\citenamefont {Wei}\ \emph {et~al.}(2011{\natexlab{b}})\citenamefont
  {Wei}, \citenamefont {Kais}, \citenamefont {Friedrich},\ and\ \citenamefont
  {Herschbach}}]{QiKaisFrieHer2011b}%
  \BibitemOpen
  \bibfield  {author} {\bibinfo {author} {\bibfnamefont {Q.}~\bibnamefont
  {Wei}}, \bibinfo {author} {\bibfnamefont {S.}~\bibnamefont {Kais}}, \bibinfo
  {author} {\bibfnamefont {B.}~\bibnamefont {Friedrich}}, \ and\ \bibinfo
  {author} {\bibfnamefont {D.}~\bibnamefont {Herschbach}},\ }\href@noop {}
  {\bibfield  {journal} {\bibinfo  {journal} {J. Chem. Phys.}\ }\textbf
  {\bibinfo {volume} {135}},\ \bibinfo {pages} {154102} (\bibinfo {year}
  {2011}{\natexlab{b}})}\BibitemShut {NoStop}%
\bibitem [{\citenamefont {Karra}\ \emph {et~al.}(2016)\citenamefont {Karra},
  \citenamefont {Sharma}, \citenamefont {Friedrich}, \citenamefont {Kais},\
  and\ \citenamefont {Herschbach}}]{Karra2016}%
  \BibitemOpen
  \bibfield  {author} {\bibinfo {author} {\bibfnamefont {M.}~\bibnamefont
  {Karra}}, \bibinfo {author} {\bibfnamefont {K.}~\bibnamefont {Sharma}},
  \bibinfo {author} {\bibfnamefont {B.}~\bibnamefont {Friedrich}}, \bibinfo
  {author} {\bibfnamefont {S.}~\bibnamefont {Kais}}, \ and\ \bibinfo {author}
  {\bibfnamefont {D.}~\bibnamefont {Herschbach}},\ }\href@noop {} {\bibfield
  {journal} {\bibinfo  {journal} {J. Chem. Phys.}\ }\textbf {\bibinfo {volume}
  {144}},\ \bibinfo {pages} {094301} (\bibinfo {year} {2016})}\BibitemShut
  {NoStop}%
\bibitem [{\citenamefont {Lefebvre-Brion}\ and\ \citenamefont
  {Field}(2004)}]{Levefre-Field:2004}%
  \BibitemOpen
  \bibfield  {author} {\bibinfo {author} {\bibfnamefont {H.}~\bibnamefont
  {Lefebvre-Brion}}\ and\ \bibinfo {author} {\bibfnamefont {R.}~\bibnamefont
  {Field}},\ }\href@noop {} {\emph {\bibinfo {title} {{The spectra and dynamics
  of diatomic molecules}}}}\ (\bibinfo  {publisher} {Elsevier},\ \bibinfo
  {address} {Amsterdam},\ \bibinfo {year} {2004})\BibitemShut {NoStop}%
\bibitem [{\citenamefont {Leacock}\ and\ \citenamefont
  {Padgett}(1983)}]{leacock1983}%
  \BibitemOpen
  \bibfield  {author} {\bibinfo {author} {\bibfnamefont {R.~A.}\ \bibnamefont
  {Leacock}}\ and\ \bibinfo {author} {\bibfnamefont {M.~J.}\ \bibnamefont
  {Padgett}},\ }\href {\doibase 10.1103/PhysRevD.28.2491} {\bibfield  {journal}
  {\bibinfo  {journal} {Phys. Rev. D}\ }\textbf {\bibinfo {volume} {28}},\
  \bibinfo {pages} {2491} (\bibinfo {year} {1983})}\BibitemShut {NoStop}%
\bibitem [{\citenamefont {Geojo}\ \emph {et~al.}(2003)\citenamefont {Geojo},
  \citenamefont {Ranjani},\ and\ \citenamefont {Kapoor}}]{geojo2003}%
  \BibitemOpen
  \bibfield  {author} {\bibinfo {author} {\bibfnamefont {K.~G.}\ \bibnamefont
  {Geojo}}, \bibinfo {author} {\bibfnamefont {S.~S.}\ \bibnamefont {Ranjani}},
  \ and\ \bibinfo {author} {\bibfnamefont {A.~K.}\ \bibnamefont {Kapoor}},\
  }\href {http://stacks.iop.org/0305-4470/36/i=16/a=309} {\bibfield  {journal}
  {\bibinfo  {journal} {Journal of Physics A: Mathematical and General}\
  }\textbf {\bibinfo {volume} {36}},\ \bibinfo {pages} {4591} (\bibinfo {year}
  {2003})}\BibitemShut {NoStop}%
\bibitem [{\citenamefont {Lemeshko}\ \emph
  {et~al.}(2011{\natexlab{a}})\citenamefont {Lemeshko}, \citenamefont
  {Mustafa}, \citenamefont {Kais},\ and\ \citenamefont
  {Friedrich}}]{LemMusKaisFriPRA2011}%
  \BibitemOpen
  \bibfield  {author} {\bibinfo {author} {\bibfnamefont {M.}~\bibnamefont
  {Lemeshko}}, \bibinfo {author} {\bibfnamefont {M.}~\bibnamefont {Mustafa}},
  \bibinfo {author} {\bibfnamefont {S.}~\bibnamefont {Kais}}, \ and\ \bibinfo
  {author} {\bibfnamefont {B.}~\bibnamefont {Friedrich}},\ }\href@noop {}
  {\bibfield  {journal} {\bibinfo  {journal} {Phys. Rev. A}\ }\textbf {\bibinfo
  {volume} {83}},\ \bibinfo {pages} {043415} (\bibinfo {year}
  {2011}{\natexlab{a}})}\BibitemShut {NoStop}%
\bibitem [{\citenamefont {Lemeshko}\ \emph
  {et~al.}(2011{\natexlab{b}})\citenamefont {Lemeshko}, \citenamefont
  {Mustafa}, \citenamefont {Kais},\ and\ \citenamefont
  {Friedrich}}]{LemMusKaisFriNJP2011}%
  \BibitemOpen
  \bibfield  {author} {\bibinfo {author} {\bibfnamefont {M.}~\bibnamefont
  {Lemeshko}}, \bibinfo {author} {\bibfnamefont {M.}~\bibnamefont {Mustafa}},
  \bibinfo {author} {\bibfnamefont {S.}~\bibnamefont {Kais}}, \ and\ \bibinfo
  {author} {\bibfnamefont {B.}~\bibnamefont {Friedrich}},\ }\href@noop {}
  {\bibfield  {journal} {\bibinfo  {journal} {New J. Phys.}\ }\textbf {\bibinfo
  {volume} {13}},\ \bibinfo {pages} {063036} (\bibinfo {year}
  {2011}{\natexlab{b}})}\BibitemShut {NoStop}%
\bibitem [{\citenamefont {Becker}\ \emph {et~al.}(2017)\citenamefont {Becker},
  \citenamefont {Mirahmadi}, \citenamefont {Schmidt}, \citenamefont {Schatz},\
  and\ \citenamefont {Friedrich}}]{b_friedrich2017}%
  \BibitemOpen
  \bibfield  {author} {\bibinfo {author} {\bibfnamefont {S.}~\bibnamefont
  {Becker}}, \bibinfo {author} {\bibfnamefont {M.}~\bibnamefont {Mirahmadi}},
  \bibinfo {author} {\bibfnamefont {B.}~\bibnamefont {Schmidt}}, \bibinfo
  {author} {\bibfnamefont {K.}~\bibnamefont {Schatz}}, \ and\ \bibinfo {author}
  {\bibfnamefont {B.}~\bibnamefont {Friedrich}},\ }\href {\doibase
  10.1140/epjd/e2017-80134-6} {\bibfield  {journal} {\bibinfo  {journal} {The
  European Physical Journal D}\ }\textbf {\bibinfo {volume} {71}},\ \bibinfo
  {pages} {149} (\bibinfo {year} {2017})}\BibitemShut {NoStop}%
\bibitem [{\citenamefont {Krajewska}\ \emph {et~al.}(1997)\citenamefont
  {Krajewska}, \citenamefont {Ushveridze},\ and\ \citenamefont
  {Walczak}}]{Krajewska1997a}%
  \BibitemOpen
  \bibfield  {author} {\bibinfo {author} {\bibfnamefont {A.}~\bibnamefont
  {Krajewska}}, \bibinfo {author} {\bibfnamefont {A.}~\bibnamefont
  {Ushveridze}}, \ and\ \bibinfo {author} {\bibfnamefont {Z.}~\bibnamefont
  {Walczak}},\ }\href {\doibase 10.1142/S0217732397001242} {\bibfield
  {journal} {\bibinfo  {journal} {Mod. Phys. Lett. A}\ }\textbf {\bibinfo
  {volume} {12}},\ \bibinfo {pages} {1225} (\bibinfo {year}
  {1997})}\BibitemShut {NoStop}%
\bibitem [{\citenamefont {Sato}\ and\ \citenamefont {Tanaka}(2002)}]{sato2002}%
  \BibitemOpen
  \bibfield  {author} {\bibinfo {author} {\bibfnamefont {M.}~\bibnamefont
  {Sato}}\ and\ \bibinfo {author} {\bibfnamefont {T.}~\bibnamefont {Tanaka}},\
  }\href {\doibase http://dx.doi.org/10.1063/1.1485115} {\bibfield  {journal}
  {\bibinfo  {journal} {Journal of Mathematical Physics}\ }\textbf {\bibinfo
  {volume} {43}},\ \bibinfo {pages} {3484} (\bibinfo {year}
  {2002})}\BibitemShut {NoStop}%
\bibitem [{\citenamefont {de~Souza~Dutra}(1993)}]{dutra1993}%
  \BibitemOpen
  \bibfield  {author} {\bibinfo {author} {\bibfnamefont {A.}~\bibnamefont
  {de~Souza~Dutra}},\ }\href {\doibase 10.1103/PhysRevA.47.R2435} {\bibfield
  {journal} {\bibinfo  {journal} {Phys. Rev. A}\ }\textbf {\bibinfo {volume}
  {47}},\ \bibinfo {pages} {R2435} (\bibinfo {year} {1993})}\BibitemShut
  {NoStop}%
\bibitem [{\citenamefont {Gonz{\'a}lez-Lopez}\ \emph
  {et~al.}(1993)\citenamefont {Gonz{\'a}lez-Lopez}, \citenamefont {Kamran},\
  and\ \citenamefont {Olver}}]{lopez1993}%
  \BibitemOpen
  \bibfield  {author} {\bibinfo {author} {\bibfnamefont {A.}~\bibnamefont
  {Gonz{\'a}lez-Lopez}}, \bibinfo {author} {\bibfnamefont {N.}~\bibnamefont
  {Kamran}}, \ and\ \bibinfo {author} {\bibfnamefont {P.~J.}\ \bibnamefont
  {Olver}},\ }\href@noop {} {\bibfield  {journal} {\bibinfo  {journal}
  {Communications in Mathematical Physics}\ }\textbf {\bibinfo {volume}
  {153}},\ \bibinfo {pages} {117} (\bibinfo {year} {1993})}\BibitemShut
  {NoStop}%
\bibitem [{wik()}]{wiki-Closed-form_expression}%
  \BibitemOpen
  \href@noop {} {\enquote {\bibinfo {title} {Closed-form expression},}\
  }\bibinfo {howpublished}
  {\url{https://en.wikipedia.org/wiki/Closed-form_expression}},\ \bibinfo
  {note} {accessed February 12, 2018}\BibitemShut {NoStop}%
\bibitem [{\citenamefont {Vilenkin}\ and\ \citenamefont
  {Klimyk}(1995)}]{vilenkin1995}%
  \BibitemOpen
  \bibfield  {author} {\bibinfo {author} {\bibfnamefont {N.}~\bibnamefont
  {Vilenkin}}\ and\ \bibinfo {author} {\bibfnamefont {A.}~\bibnamefont
  {Klimyk}},\ }\enquote {\bibinfo {title} {Representation of {L}ie groups and
  special functions},}\ in\ \href {\doibase 10.1007/978-94-017-2885-0} {\emph
  {\bibinfo {booktitle} {Mathematics and Its Applications}}}\ (\bibinfo
  {publisher} {Springer Netherlands},\ \bibinfo {year} {1995})\BibitemShut
  {NoStop}%
\bibitem [{\citenamefont {Sezgin}\ \emph {et~al.}(1998)\citenamefont {Sezgin},
  \citenamefont {Verdiyev},\ and\ \citenamefont {Verdiyev}}]{sezgin1998}%
  \BibitemOpen
  \bibfield  {author} {\bibinfo {author} {\bibfnamefont {M.}~\bibnamefont
  {Sezgin}}, \bibinfo {author} {\bibfnamefont {A.~Y.}\ \bibnamefont
  {Verdiyev}}, \ and\ \bibinfo {author} {\bibfnamefont {Y.~A.}\ \bibnamefont
  {Verdiyev}},\ }\href {\doibase 10.1063/1.532268} {\bibfield  {journal}
  {\bibinfo  {journal} {Journal of Mathematical Physics}\ }\textbf {\bibinfo
  {volume} {39}},\ \bibinfo {pages} {1910} (\bibinfo {year}
  {1998})}\BibitemShut {NoStop}%
\bibitem [{\citenamefont {Schmidt}\ and\ \citenamefont
  {Friedrich}(2014{\natexlab{b}})}]{SchmiFri2014a}%
  \BibitemOpen
  \bibfield  {author} {\bibinfo {author} {\bibfnamefont {B.}~\bibnamefont
  {Schmidt}}\ and\ \bibinfo {author} {\bibfnamefont {B.}~\bibnamefont
  {Friedrich}},\ }\href {http://dx.doi.org/10.1063/1.4864465} {\bibfield
  {journal} {\bibinfo  {journal} {J. Chem. Phys.}\ }\textbf {\bibinfo {volume}
  {140}},\ \bibinfo {pages} {064317} (\bibinfo {year}
  {2014}{\natexlab{b}})}\BibitemShut {NoStop}%
\bibitem [{\citenamefont {Gangopadhyaya}\ \emph {et~al.}(2011)\citenamefont
  {Gangopadhyaya}, \citenamefont {Mallow},\ and\ \citenamefont
  {Rasinariu}}]{gangopadhyaya2011}%
  \BibitemOpen
  \bibfield  {author} {\bibinfo {author} {\bibfnamefont {A.}~\bibnamefont
  {Gangopadhyaya}}, \bibinfo {author} {\bibfnamefont {J.~V.}\ \bibnamefont
  {Mallow}}, \ and\ \bibinfo {author} {\bibfnamefont {C.}~\bibnamefont
  {Rasinariu}},\ }\href@noop {} {\emph {\bibinfo {title} {Supersymmetric
  Quantum Mechanics - An Introduction}}}\ (\bibinfo  {publisher} {World
  Scientific},\ \bibinfo {address} {Singapore},\ \bibinfo {year}
  {2011})\BibitemShut {NoStop}%
\bibitem [{\citenamefont {{Ranjani}}\ \emph {et~al.}(2004)\citenamefont
  {{Ranjani}}, \citenamefont {{Geojo}}, \citenamefont {{Kapoor}},\ and\
  \citenamefont {{Panigrahi}}}]{ranjani2004}%
  \BibitemOpen
  \bibfield  {author} {\bibinfo {author} {\bibfnamefont {S.~S.}\ \bibnamefont
  {{Ranjani}}}, \bibinfo {author} {\bibfnamefont {K.~G.}\ \bibnamefont
  {{Geojo}}}, \bibinfo {author} {\bibfnamefont {A.~K.}\ \bibnamefont
  {{Kapoor}}}, \ and\ \bibinfo {author} {\bibfnamefont {P.~K.}\ \bibnamefont
  {{Panigrahi}}},\ }\href {\doibase 10.1142/S0217732304013799} {\bibfield
  {journal} {\bibinfo  {journal} {Modern Physics Letters A}\ }\textbf {\bibinfo
  {volume} {19}},\ \bibinfo {pages} {1457} (\bibinfo {year} {2004})},\ \Eprint
  {http://arxiv.org/abs/quant-ph/0211168} {quant-ph/0211168} \BibitemShut
  {NoStop}%
\bibitem [{\citenamefont {{Bank}}\ \emph {et~al.}(1981)\citenamefont {{Bank}},
  \citenamefont {Gundersen},\ and\ \citenamefont {Laine}}]{bank1981}%
  \BibitemOpen
  \bibfield  {author} {\bibinfo {author} {\bibfnamefont {S.~B.}\ \bibnamefont
  {{Bank}}}, \bibinfo {author} {\bibfnamefont {G.~G.}\ \bibnamefont
  {Gundersen}}, \ and\ \bibinfo {author} {\bibfnamefont {I.}~\bibnamefont
  {Laine}},\ }\href@noop {} {\bibfield  {journal} {\bibinfo  {journal} {Annales
  Academiae Scientiarum Fennicae, Series A. I. Mathematica}\ }\textbf {\bibinfo
  {volume} {6}},\ \bibinfo {pages} {369} (\bibinfo {year} {1981})}\BibitemShut
  {NoStop}%
\bibitem [{\citenamefont {Gonz{\'a}lez-L{\'o}pez}\ \emph
  {et~al.}(1994)\citenamefont {Gonz{\'a}lez-L{\'o}pez}, \citenamefont
  {Kamran},\ and\ \citenamefont {Olver}}]{lopez1994}%
  \BibitemOpen
  \bibfield  {author} {\bibinfo {author} {\bibfnamefont {A.}~\bibnamefont
  {Gonz{\'a}lez-L{\'o}pez}}, \bibinfo {author} {\bibfnamefont {N.}~\bibnamefont
  {Kamran}}, \ and\ \bibinfo {author} {\bibfnamefont {P.~J.}\ \bibnamefont
  {Olver}},\ }\href@noop {} {\bibfield  {journal} {\bibinfo  {journal}
  {Contemp. Math}\ }\textbf {\bibinfo {volume} {160}},\ \bibinfo {pages} {113}
  (\bibinfo {year} {1994})}\BibitemShut {NoStop}%
\bibitem [{\citenamefont {G{\'o}mez-Ullate}\ \emph {et~al.}(2013)\citenamefont
  {G{\'o}mez-Ullate}, \citenamefont {Kamran},\ and\ \citenamefont
  {Milson}}]{ullate2013}%
  \BibitemOpen
  \bibfield  {author} {\bibinfo {author} {\bibfnamefont {D.}~\bibnamefont
  {G{\'o}mez-Ullate}}, \bibinfo {author} {\bibfnamefont {N.}~\bibnamefont
  {Kamran}}, \ and\ \bibinfo {author} {\bibfnamefont {R.}~\bibnamefont
  {Milson}},\ }\href {\doibase 10.1007/s10208-012-9128-6} {\bibfield  {journal}
  {\bibinfo  {journal} {Foundations of Computational Mathematics}\ }\textbf
  {\bibinfo {volume} {13}},\ \bibinfo {pages} {615} (\bibinfo {year}
  {2013})}\BibitemShut {NoStop}%
\bibitem [{\citenamefont {Frankel}(2011)}]{frankel2011}%
  \BibitemOpen
  \bibfield  {author} {\bibinfo {author} {\bibfnamefont {T.}~\bibnamefont
  {Frankel}},\ }\href {\doibase 10.1017/CBO9781139061377} {\emph {\bibinfo
  {title} {The Geometry of Physics: An Introduction}}},\ \bibinfo {edition}
  {3rd}\ ed.\ (\bibinfo  {publisher} {Cambridge University Press},\ \bibinfo
  {year} {2011})\BibitemShut {NoStop}%
\bibitem [{\citenamefont {Ahlfors}(1966)}]{ahlfors1966}%
  \BibitemOpen
  \bibfield  {author} {\bibinfo {author} {\bibfnamefont {L.~V.}\ \bibnamefont
  {Ahlfors}},\ }\href@noop {} {\emph {\bibinfo {title} {Complex Analysis}}},\
  \bibinfo {edition} {2nd}\ ed.\ (\bibinfo  {publisher} {MacGraw-Hill Book
  Company},\ \bibinfo {year} {1966})\BibitemShut {NoStop}%
\bibitem [{\citenamefont {Boothby}(1986)}]{boothby1986}%
  \BibitemOpen
  \bibfield  {author} {\bibinfo {author} {\bibfnamefont {W.~M.}\ \bibnamefont
  {Boothby}},\ }\href@noop {} {\emph {\bibinfo {title} {An Introduction to
  Differentiable Manifolds and Riemannian Geometry. Revised 2nd Ed.}}}\
  (\bibinfo  {publisher} {Academic Press Inc. (London) Ltd.},\ \bibinfo
  {address} {New York},\ \bibinfo {year} {1986})\BibitemShut {NoStop}%
\bibitem [{\citenamefont {Zoladek}(2000)}]{zoladek2000}%
  \BibitemOpen
  \bibfield  {author} {\bibinfo {author} {\bibfnamefont {H.}~\bibnamefont
  {Zoladek}},\ }\href {https://projecteuclid.org:443/euclid.tmna/1471875703}
  {\bibfield  {journal} {\bibinfo  {journal} {Topol. Methods Nonlinear Anal.}\
  }\textbf {\bibinfo {volume} {16}},\ \bibinfo {pages} {253} (\bibinfo {year}
  {2000})}\BibitemShut {NoStop}%
\bibitem [{\citenamefont {King}(2009)}]{king2009}%
  \BibitemOpen
  \bibfield  {author} {\bibinfo {author} {\bibfnamefont {B.~R.}\ \bibnamefont
  {King}},\ }\href {\doibase 10.1007/978-0-8176-4849-7} {\emph {\bibinfo
  {title} {Beyond the Quartic Equation}}}\ (\bibinfo  {publisher}
  {Birkh{\"a}user Boston},\ \bibinfo {year} {2009})\BibitemShut {NoStop}%
\bibitem [{\citenamefont {Hagedorn}(2000)}]{hagedorn2000}%
  \BibitemOpen
  \bibfield  {author} {\bibinfo {author} {\bibfnamefont {T.~R.}\ \bibnamefont
  {Hagedorn}},\ }\href {\doibase http://dx.doi.org/10.1006/jabr.2000.8428}
  {\bibfield  {journal} {\bibinfo  {journal} {Journal of Algebra}\ }\textbf
  {\bibinfo {volume} {233}},\ \bibinfo {pages} {704 } (\bibinfo {year}
  {2000})}\BibitemShut {NoStop}%
\bibitem [{\citenamefont {Teschl}(2014)}]{teschl2011}%
  \BibitemOpen
  \bibfield  {author} {\bibinfo {author} {\bibfnamefont {G.}~\bibnamefont
  {Teschl}},\ }\href@noop {} {\emph {\bibinfo {title} {Mathematical Methods in
  Quantum Mechanics}}},\ \bibinfo {edition} {2nd}\ ed.\ (\bibinfo  {publisher}
  {American Mathematical Society},\ \bibinfo {year} {2014})\BibitemShut
  {NoStop}%
\bibitem [{\citenamefont {G{\'o}mez-Ullate}\ \emph {et~al.}(2007)\citenamefont
  {G{\'o}mez-Ullate}, \citenamefont {Kamran},\ and\ \citenamefont
  {Milson}}]{ullate2007}%
  \BibitemOpen
  \bibfield  {author} {\bibinfo {author} {\bibfnamefont {D.}~\bibnamefont
  {G{\'o}mez-Ullate}}, \bibinfo {author} {\bibfnamefont {N.}~\bibnamefont
  {Kamran}}, \ and\ \bibinfo {author} {\bibfnamefont {R.}~\bibnamefont
  {Milson}},\ }\href {http://stacks.iop.org/0266-5611/23/i=5/a=008} {\bibfield
  {journal} {\bibinfo  {journal} {Inverse Problems}\ }\textbf {\bibinfo
  {volume} {23}},\ \bibinfo {pages} {1915} (\bibinfo {year}
  {2007})}\BibitemShut {NoStop}%
\end{thebibliography}%


\end{document}